\documentclass{article}

\usepackage{amssymb,amsfonts,amsmath}
\usepackage{cite,enumerate,float}
\usepackage{color}

\usepackage{physics}

\usepackage{mathrsfs}

\usepackage{tikz}
\usetikzlibrary{arrows}
\usetikzlibrary{arrows.meta}
\usetikzlibrary{positioning}
\usetikzlibrary{shapes,snakes}
\usetikzlibrary{fit}
\usetikzlibrary{decorations.pathmorphing,decorations.pathreplacing,decorations.markings}
\usetikzlibrary{calc}

\usepackage{ytableau}

\usepackage{xcolor}
\definecolor{burgundy}{rgb}{0.5, 0.0, 0.13}
\definecolor{olive}{rgb}{0.50, 0.50, 0.0}
\usepackage[linktocpage=true,colorlinks=true,linkcolor=burgundy,citecolor=black!20!blue,urlcolor=black!40!violet]{hyperref}

\usepackage[most]{tcolorbox}

\usepackage{pdflscape}
\usepackage{array, longtable}
\newcolumntype{C}{>{$}c<{$}}

\def\be{\begin{eqnarray}}
\def\ee{\end{eqnarray}}

\def\Tr{{\rm Tr}\,}

\definecolor{red}{rgb}{1,0,0}
\definecolor{orange}{rgb}{1,0.5,0}
\definecolor{violet}{rgb}{0.7,0,1}

\def\CI {{\cal I}}

\def\CN {{\cal N}}
\def\CO {{\cal O}}

\def\CR {{\cal R}}

\def\CO {{\cal O}}

\def\CI {{{\cal I}}}

\def\CT {{\cal T}}

\def\IC{\mathbb{C}}

\def\IN{\mathbb{N}}

\def\IP{\mathbb{P}}
\def\IR{{\mathbb{R}}}

\def\IZ{{\mathbb{Z}}}

\def\ff{\mathfrak{f}}

\def\fg{\mathfrak{g}}

\def\fl{\mathfrak{l}}

\def\fo{\mathfrak{o}}

\def\fA{\mathfrak{A}}

\def\fQ{\mathfrak{Q}}

\def\lm{\limits}

\DeclareSymbolFont{bbsymbol}{U}{bbold}{m}{n}
\DeclareMathSymbol{\bbzero}{\mathbin}{bbsymbol}{"30}
\DeclareMathSymbol{\bbone}{\mathbin}{bbsymbol}{"31}
\DeclareMathSymbol{\bbtwo}{\mathbin}{bbsymbol}{"32}
\DeclareMathSymbol{\bbthree}{\mathbin}{bbsymbol}{"33}
\DeclareMathSymbol{\bbfour}{\mathbin}{bbsymbol}{"34}
\DeclareMathSymbol{\bbfive}{\mathbin}{bbsymbol}{"35}
\DeclareMathSymbol{\bbsix}{\mathbin}{bbsymbol}{"36}
\DeclareMathSymbol{\bbseven}{\mathbin}{bbsymbol}{"37}
\DeclareMathSymbol{\bbeight}{\mathbin}{bbsymbol}{"38}
\DeclareMathSymbol{\bbnine}{\mathbin}{bbsymbol}{"39}

\newcommand\sqbox[1]{{
		\setbox0=\hbox{\mbox{$\Box$}}
		\setbox1=\hbox{\mbox{\raisebox{0.35ex}{\small #1}}}
		\mbox{\raisebox{-0.2ex}{\rlap{\hbox to \wd0{\hss{\box1}\hss}}\box0}}
}}
\newcommand\ssqbox[1]{{
		\setbox0=\hbox{\mbox{$\scriptstyle\Box$}}
		\setbox1=\hbox{\mbox{\raisebox{0.35ex}{\tiny #1}}}
		\mbox{\raisebox{-0.2ex}{\rlap{\hbox to \wd0{\hss{\box1}\hss}}\box0}}
}}

\def\myblue{white!40!blue}
\def\graphcol{ white!30!orange}

\tcbset{boxrule=0.6pt,colback=white!97!blue}
\def\stile{\begin{tikzpicture}[scale=0.15]
		\draw (0,0) -- (1,0) -- (1,-1) -- (0,-1) -- cycle;
\end{tikzpicture}}
\def\shtile{\begin{tikzpicture}[scale=0.15]
		\draw (0,0) -- (1,0) -- (0,-1) -- cycle;
\end{tikzpicture}}
\def\shhtile{\begin{tikzpicture}[scale=0.15]
		\draw (1,-1) -- (1,0) -- (0,-1) -- cycle;
\end{tikzpicture}}

\def\ntile{\begin{tikzpicture}[scale=0.2]
		\draw (0,0) -- (1,0) -- (1,-1) -- (0,-1) -- cycle;
\end{tikzpicture}}
\def\nhtile{\begin{tikzpicture}[scale=0.2]
		\draw (0,0) -- (1,0) -- (0,-1) -- cycle;
\end{tikzpicture}}
\def\nhhtile{\begin{tikzpicture}[scale=0.2]
		\draw (1,-1) -- (1,0) -- (0,-1) -- cycle;
\end{tikzpicture}}

\hoffset= - 1.0in         % switch off for draft style
\numberwithin{equation}{section}
\begin{document}

\hfill MIPT/TH-07/24

\hfill ITEP/TH-10/24

\hfill IITP/TH-09/24

\vskip 1.5in
%\vskip 1cm
\begin{center}
	
{\bf\Large Wall-Crossing Effects on Quiver BPS Algebras}
	
	\vskip 0.2in
	\renewcommand{\thefootnote}{\fnsymbol{footnote}}
	{Dmitry Galakhov$^{2,3,4,}$\footnote[2]{e-mail: galakhov@itep.ru},  Alexei Morozov$^{1,2,3,4,}$\footnote[3]{e-mail: morozov@itep.ru} and Nikita Tselousov$^{1,2,4,}$\footnote[4]{e-mail: tselousov.ns@phystech.edu}}
	\vskip 0.2in
	\renewcommand{\thefootnote}{\roman{footnote}}
	{\small{
			\textit{$^1$MIPT, 141701, Dolgoprudny, Russia}
			\vskip 0 cm
			\textit{$^2$NRC “Kurchatov Institute”, 123182, Moscow, Russia}
			\vskip 0 cm
			\textit{$^3$IITP RAS, 127051, Moscow, Russia}
			\vskip 0 cm
			\textit{$^4$ITEP, Moscow, Russia}
	}}
\end{center}

\vskip 0.2in
\baselineskip 16pt

\centerline{ABSTRACT}

\bigskip

{\footnotesize
	BPS states in supersymmetric theories can admit additional algebro-geometric structures in their spectra,
	described as quiver Yangian algebras.
	Equivariant fixed points on the quiver variety are interpreted as vectors populating a representation module,
	and matrix elements for the generators are then defined as Duistermaat-Heckman integrals in the vicinity of these points.
	The well-known wall-crossing phenomena are that the fixed point spectrum
	establishes a dependence on the stability (Fayet-Illiopolous) parameters $\zeta$,  jumping abruptly across the walls of marginal stability,
	which divide the $\zeta$-space into a collection of stability chambers -- ``phases'' of the theory.
	The standard construction of the quiver Yangian algebra relies heavily on the molten crystal model,
	valid in a sole cyclic chamber where all the $\zeta$-parameters have the same sign.
	We propose to lift this restriction and investigate the effects of the wall-crossing phenomena
	on the quiver Yangian algebra and its representations --
	starting with the example of affine super-Yangian $\mathsf{Y}(\widehat{\mathfrak{gl}}_{1|1})$.
	In addition to the molten crystal construction more general atomic structures appear,
	in other non-cyclic phases (chambers of the $\zeta$-space).
	We call them {\it glasses} and also divide in a few different classes.
	For some of the new phases we manage to associate an algebraic structure again
	as a representation of the same affine Yangian $\mathsf{Y}(\widehat{\mathfrak{gl}}_{1|1})$.
	This observation supports an earlier conjecture that the BPS algebraic structures
	can be considered as new wall-crossing invariants.
	
}

\bigskip

\bigskip

\tableofcontents

\section{Introduction}

One of the most intriguing and physically beautiful concepts related to the Bogomol'nyi-Prasad-Sommerfeld (BPS) states in string theories and QFTs with supersymmetry is a \emph{wall-crossing} phenomenon (see \cite{Yamazaki:2010fz,Andriyash:2010qv,Cecotti:2009uf,Aganagic:2009kf,Kontsevich:2008fj,FelixKlein,Aganagic:2010qr,Bao:2022oyn,Sulkowski:2009rw} for reviews).
This phenomenon is the closest cousin to the phase transitions in physical systems.
Yet unlike the majority of physical systems without supersymmetry BPS states in supersymmetric theories have quantum numbers protected from quantum fluctuation corrections.
Therefore a process of crossing a \emph{marginal stability wall} that divide the parameter (moduli) space in stability chambers (different phases) where BPS spectra differ admits in many cases an analytic description.

In this note we would like to initiate a thorough consideration of such an object as a BPS algebra in non-canonical phases, and follow especially  its transformations under the wall-crossing.
As the basic model to proceed we choose BPS states in a system of D6/D4/D2/D0 branes in type IIA wrapping a resolution of a conifold singularity $\CO(-1)^{\otimes 2}\to\IC\IP^1$, and the respective algebra is an affine super-Yangian $\mathsf{Y}(\widehat{\fg\fl}_{1|1})$.
On one hand this algebra has rather rich representations that would make emerging phenomena accompanying the transition under the wall-crossing more transparent.
On the other hand the structure of the phase diagram for this model is not that complicated yet.
In particular, a large portion of the stability chambers is related to the canonical cyclic chamber by a sequence of \emph{mutations} (Seiberg dualities).

\bigskip

The BPS algebra was first defined in general in \cite{Harvey:1995fq, Harvey:1996gc}.
This rather generic definition has been transforming over recent year in applications to various specific models.
Here we follow a route proposed in \cite{Li:2020rij, Galakhov:2020vyb}:
\begin{enumerate}
	\item One considers a family of quiver $\mathfrak{Q}$ gauge theories describing an IR behavior of D-brane systems on toric Calabi-Yau 3-folds.
	\item Classical vacua in this theory are described by \emph{fixed points} on the quiver variety and depend explicitly on Fayet-Illiopolous stability parameters $\zeta$.
	A canonical definition follows a specific choice of the stability parameters defining a cyclic chamber (when all $\zeta$ have the same sign) on the phase diagram.
	\item Fixed points are enumerated by (some modification of) Young diagrams $\lambda$ (3d molten crystals in a more general setting).
	Then on diagrams we define matrix elements ${\bf E}_{\lambda,\lambda+\Box}$, ${\bf F}_{\lambda,\lambda-\Box}$ describing processes of \emph{adding/subtracting boxes} $\Box$ by corresponding equivariant integrals over quiver moduli spaces.
	\item We observe that matrix elements ${\bf E}_{\lambda,\lambda+\Box}$, ${\bf F}_{\lambda,\lambda-\Box}$ satisfy specific \emph{hysteresis} relations connecting different modification paths in the space of diagrams.
	\item From the hysteresis relations it follows naturally that matrices ${\bf E}_{\lambda,\lambda+\Box}$, ${\bf F}_{\lambda,\lambda-\Box}$ form \emph{a representation} of a quiver BPS algebra $\mathfrak{A}$.
\end{enumerate}

Schematically we could represent this route in the following way:
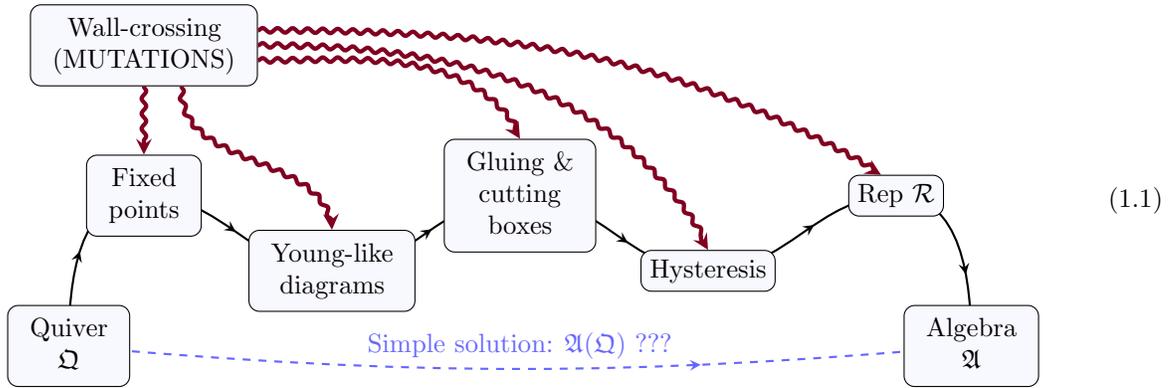
\begin{equation}\label{scheme}
	\begin{array}{c}
		\begin{tikzpicture}
			\draw[thick, postaction={decorate},
			decoration={markings, mark= at position 0.08 with {\arrow{stealth}}, mark= at position 0.24 with {\arrow{stealth}},mark= at position 0.42 with {\arrow{stealth}}, mark= at position 0.6 with {\arrow{stealth}}, mark= at position 0.77 with {\arrow{stealth}}, mark= at position 0.94 with {\arrow{stealth}}}] (0,0) to[out=90,in=180] (1,2) to[out=0,in=180] (3.5,1) to[out=0,in=180] (6,2) to[out=0,in=180] (8.5,1) to[out=0,in=180] (11,2) to[out=0,in=90] (12,0);
			\draw[thick, dashed, \myblue,postaction={decorate},
			decoration={markings, mark= at position 0.7 with {\arrow{stealth}}}] (0,0) to[out=355,in=185] node[pos=0.5,above] {Simple solution: $\fA(\fQ)$ ???} (12,0);
			\node[draw, rounded corners, fill=white!97!blue] (A) at (0,0) {$\begin{array}{c}
					\mbox{Quiver}\\
					\mathfrak{Q}
				\end{array}$};
			\node[draw, rounded corners, fill=white!97!blue] (B) at (1,2) {$\begin{array}{c}
					\mbox{Fixed}\\
					\mbox{points}
				\end{array}$};
			\node[draw, rounded corners, fill=white!97!blue] (C) at (3.5,1) {$\begin{array}{c}
					\mbox{Young-like}\\
					\mbox{diagrams}
				\end{array}$};
			\node[draw, rounded corners, fill=white!97!blue] (D) at (6,2) {$\begin{array}{c}
					\mbox{Gluing \&}\\
					\mbox{cutting}\\
					\mbox{boxes}
				\end{array}$};
			\node[draw, rounded corners, fill=white!97!blue] (E) at (8.5,1) {Hysteresis};
			\node[draw, rounded corners, fill=white!97!blue] (F) at (11,2) {Rep $\CR$};
			\node[draw, rounded corners, fill=white!97!blue] (G) at (12,0) {$\begin{array}{c}
					\mbox{Algebra}\\
					\mathfrak{A}
				\end{array}$};
			\node[draw, rounded corners, fill=white!97!blue] (H) at (1,4) {$\begin{array}{c}
					\mbox{Wall-crossing}\\
					\mbox{(MUTATIONS)}
				\end{array}$};
			\draw[ultra thick, burgundy, -stealth, decorate,decoration={snake,amplitude=.3mm,segment length=2mm,post length=1mm}] (H.south) to[out=270,in=90] (B.north);
			\draw[ultra thick, burgundy, -stealth, decorate,decoration={snake,amplitude=.3mm,segment length=2mm,post length=1mm}] ([shift={(0.5,0)}]H.south) to[out=270,in=90] (C.north);
			\draw[ultra thick, burgundy, -stealth, decorate,decoration={snake,amplitude=.3mm,segment length=2mm,post length=1mm}] ([shift={(0,-0.2)}]H.east) to[out=0,in=120] (D.north);
			\draw[ultra thick, burgundy, -stealth, decorate,decoration={snake,amplitude=.3mm,segment length=2mm,post length=1mm}] (H.east) to[out=0,in=120] (E.north);
			\draw[ultra thick, burgundy, -stealth, decorate,decoration={snake,amplitude=.3mm,segment length=2mm,post length=1mm}] ([shift={(0,0.2)}]H.east) to[out=0,in=150] ([shift={(-0.2,0)}]F.north);
		\end{tikzpicture}
	\end{array}
\end{equation}

Moving across a marginal stability wall forces a number and a classification of fixed points to jump abruptly.
Therefore we are in no position to predict a priory how these jumps will affect the whole reasoning chain \eqref{scheme} or if it is still valid to be followed at all.

The aim of this note is to follow scheme \eqref{scheme} from the first principles in cases when the stability parameters are outside the cyclic chamber and to observe modifications or their absence in the final element $\mathfrak{A}$, and its representations.

We should stress that there are a few hypothesis regarding a generic behavior of $\mathfrak{A}$ under the wall-crossing.
Despite its certain abstractness the whole procedure \eqref{scheme} leads to a rather simple (even somewhat paradoxical) conclusion that $\mathfrak{A}$ is defined on a set of Chevalley-like generators by relations depending solely on the shape of $\mathfrak{Q}$ \cite{Li:2023zub}.
Neither stability parameters nor even a superpotential contribute to the final expressions for the algebra relations.
This might lead to a conclusion that the original and the mutated after the wall-crossing algebras are simply isomorphic.
Another observation in favor of this hypothesis is related to cohomological Hall algebras (CoHA) \cite{Kontsevich:2010px} that might be considered as a Borel positive subalgebra of $\mathfrak{A}$.
A physical construction of the CoHA follows an effective theory localized on the dual Coulomb branch \cite{Galakhov:2018lta}.
The formulation of the CoHA does not involve the stability parameters either.
Moreover it allows one to reconstruct correctly the wall-crossing formulas for characters of the BPS states on two sides of marginal stability walls.
This observation indicates that the wall-crossing does not affect much neither the CoHA, nor $\mathfrak{A}$.

On the other hand transitions between rays of commutative families \cite{Mironov:2020pcd, Mironov:2023wga} in the DIM algebra \cite{Ding:1996mq, Miki1, DIM1, DIM2} (a quantum toroidal deformation of affine Yangian $\mathsf{Y}(\widehat{\fg\fl}_1)$), known as Miki automorphisms \cite{Miki1,Miki2}, are induced by transitions in stability parameters \cite{Smirnov:2018drm, Smirnov:2021cyf, Crew:2020psc} that might be interpreted further as a form of a wall-crossing.
Furthermore, another argument that the wall-crossing does not leave $\mathfrak{A}$ completely unchanged is the following.
In general, algebra $\mathsf{Y}(\widehat{\fg\fl}_{m|n})$ for sufficiently large $m$ and $n$ has more than one Dynkin diagram representation.
However all these representations are related by automorphisms \cite{BM} forming a version of the braid group.
In the physical language these automorphisms correspond to mutations on odd nodes of quiver $\mathfrak{Q}$.
Since even the simplest formulation of $\mathfrak{A}$ depends on $\mathfrak{Q}$ explicitly, and $\mathfrak{Q}$ varies in the above example, these authomorphisms act quite non-trivially.
Therefore, we conclude that \emph{mutations} of $\mathfrak{Q}$ are expected to act on $\mathfrak{A}$ by \emph{automorphisms}.
And there is no physical principle forbidding this action to be \emph{outer} with respect to $\mathfrak{A}$ and quite non-trivial.

This paper is organized as follows.
In sec.\ref{sec:summary} we discuss the motivation for our model choice and summarize results.
In sec.\ref{sec:Yang_rem} we give a brief reminder on the affine super-Yangian $\mathsf{Y}(\widehat{\fg\fl}_{1|1})$ and its semi-Fock representations based on super-partitions.
In sec.\ref{sec:atomic_plot} we discuss the BPS vacuum equations and introduce a pictorial language to describe them.
In sec.\ref{sec:lot_of_examples} we calculate a lot of explicit examples of fixed points on the quiver variety.
In sec.\ref{sec:phases} we discuss the structure of the phase portrait in general and establish dualities for some phases via mutations.
In sec.\ref{sec:wc_algebra} we construct representations of the BPS algebra in new phases and confirm that those are representations of $\mathsf{Y}(\widehat{\fg\fl}_{1|1})$.
Finally, in Appendix \ref{sec:Y(gl(m|n))} we describe some relevant mutations for more general algebra $\mathsf{Y}(\widehat{\fg\fl}_{1|1})$.

%%%%%%%%%%%%%%%%%%%%%%%%%%%%%%%%%%%%%%%%%%%%%%%%%%%%%%%%%%%%%%%%%%%
%%%%%%%%%%%%%%%%%%%%%%%%%%%%%%%%%%%%%%%%%%%%%%%%%%%%%%%%%%%%%%%%%%%
%%%%%%%%%%%%%%%%%%%%%%%%%%%%%%%%%%%%%%%%%%%%%%%%%%%%%%%%%%%%%%%%%%%
%%%%%%%%%%%%%%%%%%%%%%%%%%%%%%%%%%%%%%%%%%%%%%%%%%%%%%%%%%%%%%%%%%%

\section{Motivation \& result summary}\label{sec:summary}

The wall-crossing phenomenon is well-known for being a notoriously difficult problem.
BPS states -- fixed points on the quiver variety in question --  are usually defined quasi-classically as solutions to certain equations depending explicitly on parameters $\zeta$ called stability parameters.
Varying these parameters forces the BPS spectra to jump discontinuously over loci of co-dimension 1 called marginal stability (MS) walls.
MS walls divide the $\zeta$-space into chambers we call ``phases''.

Experiments indicate wall-crossing establishes exemplary traits of a chaotic behavior.
For example, we might acquire any type of counting fixed points:
\begin{enumerate}
	\item There might be no fixed points (supersymmetric BPS states) at all.
	\item There might be a finite set of solutions. For example, in the Argyres-Douglass AD$_3$ theory \cite{Argyres:1995wt}.
	\item There might be a countable infinite set of solutions. Moreover, counting patterns may differ drastically in various examples: some may be counted by $\IN$ (weak coupling regime of $\CN=2$ 4d $SU(2)$ SYM \cite{Alim:2011kw}), or by lattices and Young diagrams (e.g. 4d $\CN=4$ SYM \cite{Ryzhov:2001bp}).
	Generating functions of solution numbers in some cases may be Dedekind or Macmahon functions (our case of toric Calabi-Yau threefold) or solutions to algebraic equations \cite{Galakhov:2013oja}.
	\item Sometimes solutions to the BPS equations may develop their own moduli of various types \cite{Bena:2012hf,Lee:2012sc,Beaujard:2021fsk}.
	Those may contribute as uncountable infinite sets.
\end{enumerate}

Settings representing quiver Yangians has a miraculous advantage: for a specific window of stability parameters called a cyclic chamber there is a counting of fixed points by so called molten crystals \cite{Okounkov:2003sp,Yamazaki:2010fz}.
What behavior fixed points have outside this window is unknown in general.
However as far as we are aware there are no generic methods to define the counting pattern explicitly.
Molten crystals due to the melting rule may be identified with (some generalization of) Young diagrams \cite{Galakhov:2023mak, Galakhov:2024mbz}. 

Despite there were attempts \cite{Harvey:1995fq,Harvey:1996gc} to define an algebraic structure for systems of BPS states in general those methods may be inapplicable in certain situations \cite{Galakhov:2018lta}.
Quiver Yangian algebras we are after come from BPS states of D-brane systems on toric Calabi-Yau threefolds \cite{Rapcak:2020ueh,Li:2020rij}.
In practice, we do not construct an algebra itself, rather having enumerated the fixed points as partitions (Young diagrams) $\lambda$ we construct matrix elements ${\bf E}_{\lambda,\lambda+a}$, ${\bf F}_{\lambda,\lambda-a}$ corresponding to adding/subtracting an element to/from diagram.
Further we notice that the diagrams may by associated with vectors $|\lambda\rangle$ of the algebra representation space and matrices  ${\bf E}$, ${\bf F}$ form a representation of $\fA$ on this space.
This construction may be affected by the wall-crossing to the regions outside the cyclic phase in many ways as it is depicted in \eqref{scheme}.
And studying those effects is the primary goal of this paper.

Mutations (Seiberg dualities) are a powerful instrument to investigate stability chambers outside the cyclic region in the phase diagram.
As a duality it establishes a one-to-one (yet implicitly defined) correspondence between fixed points for two different sets of data $(\mathfrak{Q},\zeta)$ and  $(\check{\mathfrak{Q}},\check{\zeta})$.
In general $\check{\mathfrak{Q}}\neq \mathfrak{Q}$.
In some cases it allows one to define at least a counting of fixed point for some quiver data $(\mathfrak{Q},W)$ for an out-of-cyclic values of $\zeta$ as fixed points on another quiver variety defined by data $(\check{\mathfrak{Q}},\check W)$ in a cyclic chamber $\check{\zeta}$ where the basis of BPS states is known to be given by molten crystals.

We will consider a specific model corresponding to the semi-Fock module of $\mathsf{Y}(\widehat{\fg\fl}_{1|1})$ \cite{Galakhov:2023mak,Nishinaka:2013pua}.
The corresponding quiver data\footnote{By the quiver data here and in what follows we imply the following collection of data: a quiver, a superpotential, equivariant  weights (flavor charges) of quiver morphisms.
Equivariant weights (flavor charges) of the fields reflect weights of a toric action on CY${}_3$. The torus scales all the local coordinates so that the top volume form remains invariant, so the equivariant torus is 2d.
We label its coordinates as $\epsilon_{1,2}$. 
In sec.\ref{sec:Yang_rem} the same $\epsilon_{1,2}$ play the role of deformation parameters of affine Yangian $\mathsf{Y}(\widehat{\fg\fl}_{1|1})$. 
CY${}_3$ in question is a resolution of a conic singular surface $xy = zw$ in $\IC^4$. 
We could extend the toric action on $\IC^4$ as $(x,y,z,w)\mapsto(e^{\epsilon_1+\epsilon_2}x,e^{-\epsilon_1-\epsilon_2}y,e^{\epsilon_1-\epsilon_2}z,e^{-\epsilon_1+\epsilon_2}w)$.
} are the following:
\begin{equation}\label{quiver}
	\begin{array}{c}
		\begin{tikzpicture}
			\draw[postaction={decorate},
			decoration={markings, mark= at position 0.55 with {\arrow{stealth}}, mark= at position 0.65 with {\arrow{stealth}}}] (0,0) to[out=15, in=165] node[pos=0.5, above] {$\scriptstyle A_{1,2}$} (2,0);
			\draw[postaction={decorate},
			decoration={markings, mark= at position 0.55 with {\arrow{stealth}}, mark= at position 0.65 with {\arrow{stealth}}}] (2,0) to[out=195,in=345] node[pos=0.5, below] {$\scriptstyle  B_{1,2}$} (0,0);
			\draw[postaction={decorate},
			decoration={markings, mark= at position 0.55 with {\arrow{stealth}}}] (1,-2) -- (0,0) node[pos=0.5, below left] {$\scriptstyle R$};
			\draw[postaction={decorate},
			decoration={markings, mark= at position 0.55 with {\arrow{stealth}}}] (2,0) -- (1,-2) node[pos=0.5, below right] {$\scriptstyle S$};
			\draw[fill=white] (0,0) circle (0.1);
			\draw[fill=gray] (2,0) circle (0.1);
			\begin{scope}[shift={(1,-2)}]
				\draw[fill=\myblue] (-0.1,-0.1) -- (-0.1,0.1) -- (0.1,0.1) -- (0.1,-0.1) -- cycle;
			\end{scope}
			\node[above left] at (-0.1,0.1) {$\scriptstyle \zeta^{\circ},\;d^{\circ}$};
			\node[above right] at (2.1,0.1) {$\scriptstyle \zeta^{\bullet},\;d^{\bullet}$};
		\end{tikzpicture}
	\end{array}\quad \begin{array}{c}
		W=\Tr\left(A_1B_1A_2B_2-A_1B_2A_2B_1+A_2RS\right)\,,\\
		\\
		\begin{array}{c|c|c|c|c|c|c}
			\mbox{Fields}&A_1 & A_2 & B_1 & B_2 & R & S\\
			\hline
			\mbox{Weights}&\epsilon_1 & -\epsilon_1 & \epsilon_2 & -\epsilon_2 & 0 &\epsilon_1
		\end{array}
	\end{array}
\end{equation}
We choose this particular model for the following reasons:
\begin{enumerate}
	\item For this type of problems it is expected that no MS wall will be intersected if $\zeta$-parameters are simply re-scaled.
	Therefore if the $\zeta$-space is 1d there are no other MS walls except a single one at $\zeta=0$.
	In our model the $\zeta$-space is 2d that is optimal for the phase portrait to be non-trivial and not too involved at the same moment.
	\item The phase diagram is described entirely for a cousin quiver having no framing $S$-map \cite{Chuang:2009crq,Nagao:2010kx}.
	In that situation the BPS spectrum could be constructed as a mutation dual to the cyclic spectrum of a quiver with modified framing explicitly.
	\item We hope that the property of having fixed points labeled by 2d diagrams of the semi-Fock module (defined by the current framing) in the cyclic chamber will be preserved by the wall-crossing, therefore we will acquire simpler 2d structures counting fixed points outside the cyclic chamber, compared to 3d pyramid partitions associated with a Macmahon-like module.
\end{enumerate}

\begin{figure}[ht!]
	\begin{center}
		\scalebox{0.9}{\begin{tikzpicture}[scale=1.4]
			%%%%%%%%%%%%%%%%%%%%%%%%%%%%%%%%%%%%%%
			\draw[fill = black!60!red, ultra thin] (0,0) -- (2,0) -- (2,2) -- (0,2) -- (0,0);
			\draw[fill = black!50!red, ultra thin] (0,0) -- (0,2.) to node[pos=0.5,above] {1} (-1.,2.) -- (0,0);
			\draw[fill = black!40!red, ultra thin] (0,0) -- (-1.,2.) to node[pos=0.5,above] {\small 2} (-1.33333,2.) -- (0,0);
			\draw[fill = black!30!red, ultra thin] (0,0) -- (-1.33333,2.) to node[pos=0.5,above] {\footnotesize 3} (-1.5,2.) -- (0,0);
			\draw[fill = black!20!red, ultra thin] (0,0) -- (-1.5,2.) to node[pos=0.5,above] {\scriptsize 4} (-1.6,2.) -- (0,0);
			\draw[fill = black!10!red, ultra thin] (0,0) -- (-1.6,2.) to node[pos=0.5,above] {\tiny 5} (-1.66667,2.) -- (0,0);
			\draw[fill = red, ultra thin] (0,0) -- (-1.66667,2.) -- (-1.71429,2.) -- (0,0);
			\draw[fill = white!10!red, ultra thin] (0,0) -- (-1.71429,2.) -- (-1.75,2.) -- (0,0);
			\draw[fill = white!20!red, ultra thin] (0,0) -- (-1.75,2.) -- (-1.77778,2.) -- (0,0);
			\draw[fill = white!30!red, ultra thin] (0,0) -- (-1.77778,2.) -- (-1.8,2.) -- (0,0);
			\draw[fill = white!40!red, ultra thin] (0,0) -- (-1.8,2.) -- (-1.81818,2.) -- (0,0);
			\node[above] at (-1.81818,2.) {\dots};
			\node[white] at (1,1) {$\begin{array}{c}
					\mbox{$R$-Crystal}\\
					\mbox{(Cyclic)}
				\end{array}$};
			%%%%%%%%%%%%%%%%%%%%%%%%%%%%%%%%%%%%%%
			\begin{scope}[rotate = -90, yscale=-1]
				\draw[fill = blue!10!white, ultra thin] (0,0) -- (2,0) -- (2,2) -- (0,2) -- (0,0);
				\draw[fill = blue!20!white, ultra thin] (0,0) -- (0,2.) to node[pos=0.5, above, rotate = 90] {1} (-1.,2.) -- (0,0);
				\draw[fill = blue!30!white, ultra thin] (0,0) -- (-1.,2.) to node[pos=0.5, above, rotate = 90] {\small 2} (-1.33333,2.) -- (0,0);
				\draw[fill = blue!40!white, ultra thin] (0,0) -- (-1.33333,2.) to node[pos=0.5, above, rotate = 90] {\footnotesize 3} (-1.5,2.) -- (0,0);
				\draw[fill = blue!50!white, ultra thin] (0,0) -- (-1.5,2.) to node[pos=0.5, above, rotate = 90] {\scriptsize 4} (-1.6,2.) -- (0,0);
				\draw[fill = blue!60!white, ultra thin] (0,0) -- (-1.6,2.) to node[pos=0.5, above, rotate = 90] {\tiny 5} (-1.66667,2.) -- (0,0);
				\draw[fill = blue!70!white, ultra thin] (0,0) -- (-1.66667,2.) -- (-1.71429,2.) -- (0,0);
				\draw[fill = blue!80!white, ultra thin] (0,0) -- (-1.71429,2.) -- (-1.75,2.) -- (0,0);
				\draw[fill = blue!90!white, ultra thin] (0,0) -- (-1.75,2.) -- (-1.77778,2.) -- (0,0);
				\draw[fill = blue, ultra thin] (0,0) -- (-1.77778,2.) -- (-1.8,2.) -- (0,0);
				\draw[fill = black!10!blue, ultra thin] (0,0) -- (-1.8,2.) -- (-1.81818,2.) -- (0,0);
				\node[above, rotate = 90] at (-1.9,2.) {\dots};
				\node at (1,1) {$\begin{array}{c}
						\mbox{$S$-Crystal}\\
						\mbox{(Cyclic)}
					\end{array}$};
			\end{scope}
			%%%%%%%%%%%%%%%%%%%%%%%%%%%%%%%%%%%%%%
			\begin{scope}[xscale=-1, rotate=90]
				\draw[fill = black!50!violet, ultra thin] (0,0) -- (0,2.) -- (-1.,2.) -- (0,0);
				\draw[fill = black!40!violet, ultra thin] (0,0) -- (-1.,2.) -- (-1.33333,2.) -- (0,0);
				\draw[fill = black!30!violet, ultra thin] (0,0) -- (-1.33333,2.) -- (-1.5,2.) -- (0,0);
				\draw[fill = black!20!violet, ultra thin] (0,0) -- (-1.5,2.) -- (-1.6,2.) -- (0,0);
				\draw[fill = black!10!violet, ultra thin] (0,0) -- (-1.6,2.) -- (-1.66667,2.) -- (0,0);
				\draw[fill = violet, ultra thin] (0,0) -- (-1.66667,2.) -- (-1.71429,2.) -- (0,0);
				\draw[fill = white!10!violet, ultra thin] (0,0) -- (-1.71429,2.) -- (-1.75,2.) -- (0,0);
				\draw[fill = white!20!violet, ultra thin] (0,0) -- (-1.75,2.) -- (-1.77778,2.) -- (0,0);
				\draw[fill = white!30!violet, ultra thin] (0,0) -- (-1.77778,2.) -- (-1.8,2.) -- (0,0);
				\draw[fill = white!40!violet, ultra thin] (0,0) -- (-1.8,2.) -- (-1.81818,2.) -- (0,0);
			\end{scope}
			%%%%%%%%%%%%%%%%%%%%%%%%%%%%%%%%%%%%%%
			\begin{scope}[scale=-1]
				\draw[fill = black!80!green, ultra thin] (0,0) -- (0,2.) -- (-1.,2.) -- (0,0);
				\draw[fill = black!70!green, ultra thin] (0,0) -- (-1.,2.) -- (-1.33333,2.) -- (0,0);
				\draw[fill = black!60!green, ultra thin] (0,0) -- (-1.33333,2.) -- (-1.5,2.) -- (0,0);
				\draw[fill = black!50!green, ultra thin] (0,0) -- (-1.5,2.) -- (-1.6,2.) -- (0,0);
				\draw[fill = black!40!green, ultra thin] (0,0) -- (-1.6,2.) -- (-1.66667,2.) -- (0,0);
				\draw[fill = black!30!green, ultra thin] (0,0) -- (-1.66667,2.) -- (-1.71429,2.) -- (0,0);
				\draw[fill = black!20!green, ultra thin] (0,0) -- (-1.71429,2.) -- (-1.75,2.) -- (0,0);
				\draw[fill = black!10!green, ultra thin] (0,0) -- (-1.75,2.) -- (-1.77778,2.) -- (0,0);
				\draw[fill = green, ultra thin] (0,0) -- (-1.77778,2.) -- (-1.8,2.) -- (0,0);
				\draw[fill = white!10!green, ultra thin] (0,0) -- (-1.8,2.) -- (-1.81818,2.) -- (0,0);
			\end{scope}
			%%%%%%%%%%%%%%%%%%%%%%%%%%%%%%%%%%%%%%
			\draw[dashed] (-2,2) -- (2,-2);
			\draw[-stealth] (-2.2,0) -- (2.2,0);
			\draw[-stealth] (0,-2.2) -- (0,2.2);
			\node[right] at (2.2,0) {$\zeta^{\circ}$};
			\node[above] at (0,2.2) {$\zeta^{\bullet}$};
			%%%%%%%%%%%%%%%%%%%%%%%%%%%%%%%%%%%%%%
			\draw (0,2) to[out=120,in=0] (-0.3,2.3) -- (-2,2.3) to[out=180,in=315] (-2.4,2.4);
			\begin{scope}[yscale=-1,rotate=90]
				\draw (0,2) to[out=120,in=0] (-0.3,2.3) -- (-2,2.3) to[out=180,in=315] (-2.4,2.4);
			\end{scope}
			\node[above left] at (-2.4,2.4) {Simple glasses};
			\begin{scope}[scale=-1]
				\draw (0,2) to[out=120,in=0] (-0.3,2.3) -- (-2,2.3) to[out=180,in=315] (-2.4,2.4);
				\begin{scope}[yscale=-1,rotate=90]
					\draw (0,2) to[out=120,in=0] (-0.3,2.3) -- (-2,2.3) to[out=180,in=315] (-2.4,2.4);
				\end{scope}
				\node[below right] at (-2.4,2.4) {Crooked glasses};
			\end{scope}
			%%%%%%%%%%%%%%%%%%%%%%%%%%%%%%%%%%%%%%
			\node[above, black!50!red] at (1,2) {$R$-dominated};
			\node[rotate=90, above, black!50!blue] at (-2,-1) {$S$-dominated};
		\end{tikzpicture}}
		\caption{Phase diagram.}\label{fig:new_phases}
	\end{center}
\end{figure}
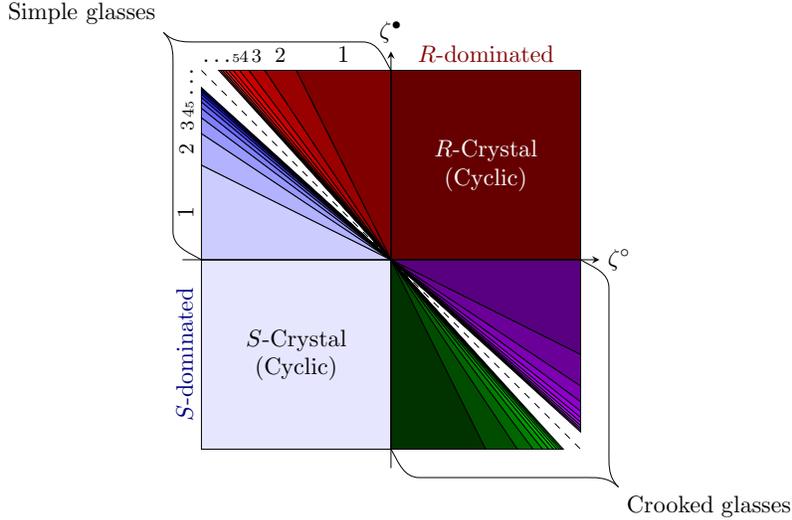

Eventually, we arrive to the following phase portrait in stability parameters $(\zeta^{\circ},\zeta^{\bullet})$ depicted in fig.\ref{fig:new_phases}.

The whole picture is divided in two halves by a dashed line $\zeta^{\circ}+\zeta^{\bullet}=0$.
This is a reflection of a $\IZ_2$-symmetry of the problem inverting directions of all the arrows and permuting quiver nodes.
Therefore we expect that each fixed point above the dashed line has a twin below the line where quiver maps are transposed.

Whole upper right ($\zeta^{\circ} > 0$, $\zeta^{\bullet} > 0$) and lower left ($\zeta^{\circ} < 0$, $\zeta^{\bullet} < 0$) quadrants are cyclic phases.
Both quiver representation vector spaces are cyclic in this case -- all the vectors (resp. co-vectors) are generated from a single vector $R$ (resp. co-vector $S$) by monomial words in $A_{1,2}$, $B_{1,2}$.
BPS spectra in these chambers where naturally identified with super-Young diagrams in \cite{Galakhov:2023mak}.

The top left quadrant ($\zeta^{\circ} < 0$, $\zeta^{\bullet} > 0$) is split in two parts related by the $\IZ_2$-symmetry by the dashed  line $\zeta^{\circ}+\zeta^{\bullet}=0$.
Furthermore each of those parts contains infinitely many chambers enumerated by $n\in \IN$ ($(n+1)\zeta^{\circ}+n\zeta^{\bullet}>0$, $n\zeta^{\circ}+(n-1)\zeta^{\bullet}<0$).
All they are mutant (Seiberg dual) to the cyclic chamber, therefore fixed points in those may be enumerated by super-Young diagrams.
However actual quiver representations in these fixed points are \emph{not cyclic}.
There are vectors that can not be represented as $M\cdot R$ (or $\left(S\cdot M\right)^{\dagger}$) where $M$ is a monomial in fields $A_{1,2}$, $B_{1,2}$.
However the property $R\neq 0$, $S=0$ (resp. $R=0$, $S\neq 0$) is preserved above (resp. below) the dashed line.
For this reason we call corresponding branches $R$-dominant (resp. $S$-dominant).
Since the absent cyclicity property was tightly related to the crystal melting rule we propose to call these new phases  \emph{simple glass}-$n$ phases.

Finally, it turns that in the bottom right quadrant ($\zeta^{\circ} > 0$, $\zeta^{\bullet} < 0$) $R$-dominant and $S$-dominant branches overlap.
Under these circumstances the mutation dualities can not be applied straightforwardly.
And we are not aware of the general phase structure of this quadrant.
For this reason we propose to call these phases \emph{crooked glasses}.
Nevertheless we manage to identify a dual theory for a branch of solutions for a crooked glass sub-phase close to the boundary ($\zeta^{\bullet}<0$, $\zeta^{\circ}+2\zeta^{\bullet}>0$).
In this case the fixed point counting is given by \emph{skew} super-Young diagrams where tiles fill in a quadrant with a non-trivial boundary.

In all the cases where we were able to identify the fixed point ordering with the (skew) super-Young diagrams we were able to calculate matrices ${\bf E}_{\lambda,\lambda+a}$, ${\bf F}_{\lambda,\lambda-a}$ on actual fixed points as equivariant Duistermaat-Heckman integrals over quiver moduli spaces.
In all these cases we confirm that fixed points with associated matrices ${\bf E}_{\lambda,\lambda+a}$, ${\bf F}_{\lambda,\lambda-a}$ form a representation of the (shifted) affine super-Yangian $\mathsf{Y}(\widehat{\fg\fl}_{1|1})$.
By this observation we find a confirmation example for a hypothesis proposed in \cite{Galakhov:2021xum} that under crossing an MS wall algebra $\fA$ transforms via \emph{automorphisms}.
This makes $\fA$  a perspective non-trivial invariant of the wall-crossing.

%%%%%%%%%%%%%%%%%%%%%%%%%%%%%%%%%%%%%%%%%%%%%%%%%%%%%%%%%%%%%%%%%%%
%%%%%%%%%%%%%%%%%%%%%%%%%%%%%%%%%%%%%%%%%%%%%%%%%%%%%%%%%%%%%%%%%%%
%%%%%%%%%%%%%%%%%%%%%%%%%%%%%%%%%%%%%%%%%%%%%%%%%%%%%%%%%%%%%%%%%%%
%%%%%%%%%%%%%%%%%%%%%%%%%%%%%%%%%%%%%%%%%%%%%%%%%%%%%%%%%%%%%%%%%%%

\section{Reminder on affine Yangian $\mathsf{Y}(\widehat{\fg\fl}_{1|1})$}\label{sec:Yang_rem}
\subsection{Affine Yangian $\mathsf{Y}(\widehat{\fg\fl}_{1|1})$ }

\begin{figure}[ht!]
	\begin{center}
		\begin{tikzpicture}
			\draw (0,0) to[out=30,in=150] (2,0) (0,0) to[out=-30,in=-150] (2,0);
			\draw[fill=white] (0,0) circle (0.1);
			\draw[fill=gray] (2,0) circle (0.1);
			\node[above] at (0,0.1) {$1$};
			\node[above] at (2,0.1) {$2$};
		\end{tikzpicture}
		\caption{Dynkin diagram for $\widehat{\fg\fl}_{1|1}$.}\label{fig:gl_1_1_Dynkin}
	\end{center}
\end{figure}

A Dynkin diagram for affine superalgebra $\widehat{\fg\fl}_{1|1}$ (also classified as a Coxeter diagram $\tilde A_1$) is depicted in fig.\ref{fig:gl_1_1_Dynkin} and represents a circular necklace with two odd nodes.
We call these nodes a white node and a black node and denote respectively as $\circ$ and $\bullet$.
If we would like to make a statement for both nodes we denote this situation as $\circ/\bullet$.

The affine Yangian algebra $\mathsf{Y}_{\epsilon_1,\epsilon_2}(\widehat{\fg\fl}_{1|1})$ consists of series of Chevalley raising, lowering and Cartan generators $e_n^{\circ/\bullet}$, $f_n^{\circ/\bullet}$, $\psi_k^{\circ/\bullet}$, $n\in\IZ_{\geq 0}$, $k\in\IZ$ and depends on two deformation parameters $\epsilon_{1,2}$.
These generators satisfy a set of relations:
\begingroup
\renewcommand*{\arraystretch}{1.5}
\begin{equation}\label{Yang_modes}
	\begin{array}{l}
		\left\{e^{\circ}_n,e^{\circ}_k\right\}=\left\{f^{\circ}_n,f^{\circ}_k\right\}=\left[\psi^{\circ}_n,e^{\circ}_k\right]=\left[\psi^{\circ}_n,f^{\circ}_k\right]=\left\{e^{\bullet}_n,e^{\bullet}_k\right\}=\left\{f^{\bullet}_n,f^{\bullet}_k\right\}=\left[\psi^{\bullet}_n,e^{\bullet}_k\right]=\left[\psi^{\bullet}_n,f^{\bullet}_k\right]=0\,,\\
		\left\{e_{n+2}^{\circ},e_{k}^{\bullet}\right\}-2\left\{e_{n+1}^{\circ},e_{k+1}^{\bullet}\right\}+\left\{e_{n}^{\circ},e_{k+2}^{\bullet}\right\}-\frac{\epsilon_1^2+\epsilon_2^2}{2}\left\{e_{n}^{\circ},e_{k}^{\bullet}\right\}+\frac{\epsilon_1^2-\epsilon_2^2}{2}\left[e_{n}^{\circ},e_{k}^{\bullet}\right]=0\,,\\
		\left\{f_{n+2}^{\circ},f_{k}^{\bullet}\right\}-2\left\{f_{n+1}^{\circ},f_{k+1}^{\bullet}\right\}+\left\{f_{n}^{\circ},f_{k+2}^{\bullet}\right\}-\frac{\epsilon_1^2+\epsilon_2^2}{2}\left\{f_{n}^{\circ},f_{k}^{\bullet}\right\}-\frac{\epsilon_1^2-\epsilon_2^2}{2}\left[f_{n}^{\circ},f_{k}^{\bullet}\right]=0\,,\\
		\left[\psi_{n+2}^{\circ},e_{k}^{\bullet}\right]-2\left[\psi_{n+1}^{\circ},e_{k+1}^{\bullet}\right]+\left[\psi_{n}^{\circ},e_{k+2}^{\bullet}\right]-\frac{\epsilon_1^2+\epsilon_2^2}{2}\left[\psi_{n}^{\circ},e_{k}^{\bullet}\right]+\frac{\epsilon_1^2-\epsilon_2^2}{2}\left\{\psi_{n}^{\circ},e_{k}^{\bullet}\right\}=0\,,\\
		\left[\psi_{n+2}^{\circ},f_{k}^{\bullet}\right]-2\left[\psi_{n+1}^{\circ},f_{k+1}^{\bullet}\right]+\left[\psi_{n}^{\circ},f_{k+2}^{\bullet}\right]-\frac{\epsilon_1^2+\epsilon_2^2}{2}\left[\psi_{n}^{\circ},f_{k}^{\bullet}\right]-\frac{\epsilon_1^2-\epsilon_2^2}{2}\left\{\psi_{n}^{\circ},f_{k}^{\bullet}\right\}=0\,,\\
		\left[\psi_{n+2}^{\bullet},e^{\circ}_{k}\right]-2\left[\psi_{n+1}^{\bullet},e_{k+1}^{\circ}\right]+\left[\psi_{n}^{\bullet},e_{k+2}^{\circ}\right]-\frac{\epsilon_1^2+\epsilon_2^2}{2}\left[\psi_{n}^{\bullet},e_{k}^{\circ}\right]-\frac{\epsilon_1^2-\epsilon_2^2}{2}\left\{\psi_{n}^{\bullet},e_{k}^{\circ}\right\}=0\,,\\
		\left[\psi_{n+2}^{\bullet},f_{k}^{\circ}\right]-2\left[\psi_{n+1}^{\bullet},f_{k+1}^{\circ}\right]+\left[\psi_{n}^{\bullet},f_{k+2}^{\circ}\right]-\frac{\epsilon_1^2+\epsilon_2^2}{2}\left[\psi_{n}^{\bullet},f_{k}^{\circ}\right]+\frac{\epsilon_1^2-\epsilon_2^2}{2}\left\{\psi_{n}^{\bullet},f_{k}^{\circ}\right\}=0\,,\\
		\left\{e^{\circ}_n,f^{\circ}_k\right\}=-\psi^{\circ}_{n+k},\quad \left\{e^{\bullet}_n,f^{\bullet}_k\right\}=-\psi^{\bullet}_{n+k},\quad \left\{e^{\circ}_n,f^{\bullet}_k\right\}=\left\{e^{\bullet}_n,f^{\circ}_k\right\}=0\,,\\
		\left[\psi_n^{\circ/\bullet},\psi_k^{\circ/\bullet}\right]=0\,.
	\end{array}
\end{equation}

Canonical  Serre relations \cite{BM}:
\begin{equation}\label{Serre}
	\begin{split}
		&\mathop{\rm Sym}\lm_{i,j}\mathop{\rm Sym}\lm_{k,l}\left\{e_i^{\circ},\left[e_k^{\bullet},\left\{e_j^{\circ},e_m^{\bullet}\right\}\right]\right\}=0\,,\\
		&\mathop{\rm Sym}\lm_{i,j}\mathop{\rm Sym}\lm_{k,l}\left\{f_i^{\circ},\left[f_k^{\bullet},\left\{f_j^{\circ},f_m^{\bullet}\right\}\right]\right\}=0\,.
	\end{split}
\end{equation}
\endgroup
in the case of $\mathsf{Y}(\widehat{\fg\fl}_{1|1})$ are actually redundant:  
they follow from the same-color anti-commutation relations in the first line of \eqref{Yang_modes}\footnote{We would like to thank Alexey Litvinov for pointing out this detail to us.}.

%%%%%%%%%%%%%%%%%%%%%%%%%%%%%%%%%%%%%%%%%%%%%%%%%%%%%%%%%%%%
%%%%%%%%%%%%%%%%%%%%%%%%%%%%%%%%%%%%%%%%%%%%%%%%%%%%%%%%%%%%
%%%%%%%%%%%%%%%%%%%%%%%%%%%%%%%%%%%%%%%%%%%%%%%%%%%%%%%%%%%%
%%%%%%%%%%%%%%%%%%%%%%%%%%%%%%%%%%%%%%%%%%%%%%%%%%%%%%%%%%%%

\subsection{Semi-Fock module}

Here we would like to exploit a notion of super-Young diagrams (super-partitions) introduced in \cite{Galakhov:2023mak}.
Therefore we recall the definition here.
We call by a super-partition $\lambda$ of a semi-integer number $u\in\IZ_{\geq 0}/2$ a sequence of semi-integer numbers:
\begin{equation}\label{parti}
	\lambda_1\geq \lambda_2\geq \lambda_3\geq\ldots\geq 0\,,
\end{equation}
such that $\sum\lm_{i}\lambda_i=u$ and if $\lambda_i$ is not integer then inequalities in sequence \eqref{parti} are strict: $\lambda_{i-1}>\lambda_i>\lambda_{i+1}$.

Super-partitions may be depicted with the help of super-Young diagrams.
The diagram is constructed as a filling of the bottom right plane quadrant with tiles, so that each next tile is supported on the top and left by a previous tile or a wall.
To the halves we assign triangular half-tiles.
The height of the tile column correspond to a number $\lambda_i$, for example:
\begin{equation}\label{diag_example}
	\left\{4,\frac{7}{2},2,2,\frac{3}{2},1\right\}=\begin{array}{c}
		\begin{tikzpicture}[scale=0.3]
			\draw[-stealth] (-0.5,0.5) -- (6.5,0.5);
			\draw[-stealth] (-0.5,0.5) -- (-0.5,-4.5);
			\node[right] at (6.5,0.5) {$\scriptstyle x$};
			\node[below] at (-0.5,-4.5) {$\scriptstyle y$};
			\foreach \i/\j in {0/-4, 0/-3, 0/-2, 0/-1, 0/0, 1/-3, 1/-2, 1/-1, 1/0, 2/-2, 2/-1, 2/0, 3/-2, 3/-1, 3/0, 4/-1, 4/0, 5/-1, 5/0}
			{
				\draw (\i,\j) -- (\i+1,\j);
			}
			\foreach \i/\j in {0/-3, 0/-2, 0/-1, 0/0, 1/-3, 1/-2, 1/-1, 1/0, 2/-2, 2/-1, 2/0, 3/-1, 3/0, 4/-1, 4/0, 5/0, 6/0}
			{
				\draw (\i,\j) -- (\i,\j-1);
			}
			\foreach \i/\j in {2/-3, 5/-1}
			{
				\draw (\i,\j) -- (\i-1,\j-1);
			}
		\end{tikzpicture}
	\end{array}\,.
\end{equation}

We would like to assure the reader that the concept of the super-partitions (super-Young diagrams) is very similar to the concept of ordinary partitions (ordinary Young diagrams). 
Let us enumerate all such diagrams at first few levels:
\begin{equation}\label{super-partitions}
	\begin{aligned}
		&  \begin{array}{|c|}\hline\mbox{Lvl.1:}\\ \hline \begin{array}{c}
				\begin{tikzpicture}[scale=0.3]
					\foreach \x/\y/\z/\w in {0/0/1/0, 0/0/0/-1, 0/-1/1/0}
					{
						\draw[thick] (\x,\y) -- (\z,\w);
					}
				\end{tikzpicture}
			\end{array} \\ \hline \theta_{\frac{1}{2}} \\ \hline \end{array}\quad
		\begin{array}{|c|}
			\hline
			\mbox{Lvl.2:}\\
			\hline
			\begin{array}{c}
				\begin{tikzpicture}[scale=0.3]
					\foreach \x/\y/\z/\w in {0/0/1/0, 0/0/0/-1, 0/-1/1/-1, 1/0/1/-1}
					{
						\draw[thick] (\x,\y) -- (\z,\w);
					}
				\end{tikzpicture}
			\end{array}\\
			\hline
			p_1\\
			\hline
		\end{array}\quad
		\begin{array}{|c|c|}
			\hline
			\multicolumn{2}{|c|}{\mbox{Lvl.3:}}\\
			\hline
			\begin{array}{c}
				\begin{tikzpicture}[scale=0.3]
					\foreach \x/\y/\z/\w in {0/0/1/0, 0/0/0/-1, 0/-1/1/-1, 1/0/1/-1, 0/-1/0/-2, 0/-2/1/-1}
					{
						\draw[thick] (\x,\y) -- (\z,\w);
					}
				\end{tikzpicture}
			\end{array}&\begin{array}{c}
				\begin{tikzpicture}[scale=0.3]
					\foreach \x/\y/\z/\w in {0/0/1/0, 0/0/0/-1, 0/-1/1/-1, 1/0/1/-1, 1/0/2/0, 1/-1/2/0}
					{
						\draw[thick] (\x,\y) -- (\z,\w);
					}
				\end{tikzpicture}
			\end{array}\\
			\hline
			\theta_{\frac{3}{2}} & p_1\theta_{\frac{1}{2}}\\
			\hline
		\end{array}\quad \begin{array}{|c|c|c|}
			\hline
			\multicolumn{3}{|c|}{\mbox{Lvl.4:}}\\
			\hline
			\begin{array}{c}
				\begin{tikzpicture}[scale=0.3]
					\foreach \x/\y/\z/\w in {0/0/1/0, 0/0/0/-1, 0/-1/1/-1, 1/0/1/-1, 0/-1/0/-2, 0/-2/1/-1, 1/0/2/0, 1/-1/2/0}
					{
						\draw[thick] (\x,\y) -- (\z,\w);
					}
				\end{tikzpicture}
			\end{array} & \begin{array}{c}
				\begin{tikzpicture}[scale=0.3]
					\foreach \x/\y/\z/\w in {0/0/1/0, 0/0/0/-1, 0/-1/1/-1, 1/0/1/-1, 0/-1/0/-2, 0/-2/1/-2, 1/-1/1/-2}
					{
						\draw[thick] (\x,\y) -- (\z,\w);
					}
				\end{tikzpicture}
			\end{array} & \begin{array}{c}
				\begin{tikzpicture}[scale=0.3]
					\foreach \x/\y/\z/\w in {0/0/1/0, 0/0/0/-1, 0/-1/1/-1, 1/0/1/-1, 1/0/2/0, 1/-1/2/-1, 2/0/2/-1}
					{
						\draw[thick] (\x,\y) -- (\z,\w);
					}
				\end{tikzpicture}
			\end{array}\\
			\hline
			%%%%%%%%%%%%%%%%%%%%%%%%%%%%%%%%%%%%%%%%%%%%%%%%
			\theta_{\frac{3}{2}}\theta_{\frac{1}{2}} & p_2 & p_1^2\\
			\hline
		\end{array}\\
		& \begin{array}{|c|c|c|c|}
			\hline
			\multicolumn{4}{|c|}{\mbox{Lvl.5:}}\\
			\hline
			\begin{array}{c}
				\begin{tikzpicture}[scale=0.3]
					\foreach \x/\y/\z/\w in {0/0/1/0, 0/0/0/-1, 0/-1/1/-1, 1/0/1/-1, 0/-1/0/-2, 0/-2/1/-2, 1/-1/1/-2, 1/0/2/0, 1/-1/2/0}
					{
						\draw[thick] (\x,\y) -- (\z,\w);
					}
				\end{tikzpicture}
			\end{array} & \begin{array}{c}
				\begin{tikzpicture}[scale=0.3]
					\foreach \x/\y/\z/\w in {0/0/1/0, 0/0/0/-1, 0/-1/1/-1, 1/0/1/-1, 1/0/2/0, 1/-1/2/-1, 2/0/2/-1, 0/-1/0/-2, 0/-2/1/-1}
					{
						\draw[thick] (\x,\y) -- (\z,\w);
					}
				\end{tikzpicture}
			\end{array} & \begin{array}{c}
				\begin{tikzpicture}[scale=0.3]
					\foreach \x/\y/\z/\w in {0/0/1/0, 0/0/0/-1, 0/-1/1/-1, 1/0/1/-1, 0/-1/0/-2, 0/-2/1/-2, 1/-1/1/-2, 0/-2/0/-3, 0/-3/1/-2}
					{
						\draw[thick] (\x,\y) -- (\z,\w);
					}
				\end{tikzpicture}
			\end{array} & \begin{array}{c}
				\begin{tikzpicture}[scale=0.3]
					\foreach \x/\y/\z/\w in {0/0/1/0, 0/0/0/-1, 0/-1/1/-1, 1/0/1/-1, 1/0/2/0, 1/-1/2/-1, 2/0/2/-1, 2/0/3/0, 2/-1/3/0}
					{
						\draw[thick] (\x,\y) -- (\z,\w);
					}
				\end{tikzpicture}
			\end{array}\\
			\hline
			%%%%%%%%%%%%%%%%%%%%%%%%%%%%%%%%%%%%%%%%%%%%%%%%%%%%%%%%
			p_2\theta_{\frac{1}{2}} & \theta_{\frac{3}{2}}p_1 & \theta_{\frac{5}{2}} & p_1^2\theta_{\frac{1}{2}}\\
			\hline
		\end{array} \quad
		\begin{array}{|c|c|c|c|c|}
			\hline
			\multicolumn{5}{|c|}{\mbox{Lvl.6:}}\\
			\hline
			\begin{array}{c}
				\begin{tikzpicture}[scale=0.3]
					\foreach \x/\y/\z/\w in {0/0/1/0, 0/0/0/-1, 0/-1/1/-1, 1/0/1/-1, 0/-1/0/-2, 0/-2/1/-2, 1/-1/1/-2, 1/0/2/0, 1/-1/2/0, 0/-2/0/-3, 0/-3/1/-2}
					{
						\draw[thick] (\x,\y) -- (\z,\w);
					}
				\end{tikzpicture}
			\end{array} & \begin{array}{c}
				\begin{tikzpicture}[scale=0.3]
					\foreach \x/\y/\z/\w in {0/0/1/0, 0/0/0/-1, 0/-1/1/-1, 1/0/1/-1, 0/-1/0/-2, 0/-2/1/-2, 1/-1/1/-2, 1/0/2/0, 1/-1/2/-1, 2/0/2/-1}
					{
						\draw[thick] (\x,\y) -- (\z,\w);
					}
				\end{tikzpicture}
			\end{array} & \begin{array}{c}
				\begin{tikzpicture}[scale=0.3]
					\foreach \x/\y/\z/\w in {0/0/1/0, 0/0/0/-1, 0/-1/1/-1, 1/0/1/-1, 1/0/2/0, 1/-1/2/-1, 2/0/2/-1, 0/-1/0/-2, 0/-2/1/-1, 2/0/3/0, 2/-1/3/0}
					{
						\draw[thick] (\x,\y) -- (\z,\w);
					}
				\end{tikzpicture}
			\end{array} & \begin{array}{c}
				\begin{tikzpicture}[scale=0.3]
					\foreach \x/\y/\z/\w in {0/0/1/0, 0/0/0/-1, 0/-1/1/-1, 1/0/1/-1, 0/-1/0/-2, 0/-2/1/-2, 1/-1/1/-2, 0/-2/0/-3, 0/-3/1/-3, 1/-2/1/-3}
					{
						\draw[thick] (\x,\y) -- (\z,\w);
					}
				\end{tikzpicture}
			\end{array} & \begin{array}{c}
				\begin{tikzpicture}[scale=0.3]
					\foreach \x/\y/\z/\w in {0/0/1/0, 0/0/0/-1, 0/-1/1/-1, 1/0/1/-1, 1/0/2/0, 1/-1/2/-1, 2/0/2/-1, 2/0/3/0, 2/-1/3/-1, 3/0/3/-1}
					{
						\draw[thick] (\x,\y) -- (\z,\w);
					}
				\end{tikzpicture}
			\end{array}\\
			\hline
			%%%%%%%%%%%%%%%%%%%%%%%%%%%%%%%%%%%%%%%%%%%%%%%%%%%%%%%%%%%%%%%%%%
			\theta_{\frac{5}{2}}\theta_{\frac{1}{2}}& p_2p_1 & \theta_{\frac{3}{2}}p_1\theta_{\frac{1}{2}} & p_3 & p_1^3\\
			\hline
		\end{array}
	\end{aligned}
\end{equation}

A generating function for the number of such diagrams, where function $c(\lambda)$ denotes the number of complete tiles and $h(\lambda)$ denotes the number of half-tiles reads:
\begin{equation}
	\sum\lm_{\lambda}q^{c(\lambda)}t^{h(\lambda)}:=1+t+q+2qt+\left(2q^2+qt^2\right)+4q^2 t + \left(3q^3 + 2q^2 t^2\right) +\ldots=\prod\lm_{k=1}^{\infty}\frac{1+q^{k-1}t^k}{1-q^k}\,.
\end{equation}
This expression corresponds to an index of a gas of non-interacting Bose and Fermi particles (see e.g. \cite{Manschot:2010qz}).
This allows us to identify super-partitions with monomials of a Verma module for the mixed super-Heisenberg algebra of free bosonic fields $p_{\pm k}$, $k\in \IN$ and free fermionic fields $\theta_{\pm\left(k-\frac{1}{2}\right)}$, $k\in\IN$ (see \eqref{super-partitions} and \cite[sec.3]{Galakhov:2023mak}).
In this fashion super-partitions are similar to ordinary partitions enumerating monomials in solely bosonic free fields.

Another similarity between diagrammatic notations of ordinary and super-partitions is that both may be used as a primitive depiction of atomic structure plots we will introduce in sec.\ref{sec:atomic_plot} for $\IC^3$ ($\mathsf{Y}(\widehat{\fg\fl}_{1})$ algebra) and for a conifold ($\mathsf{Y}(\widehat{\fg\fl}_{1|1})$ algebra) respectively.
In this fashion an analog of super-Young diagrams was introduced in \cite{Noshita:2021dgj}, however compared to our notations those authors introduced an additional division of a complete tile in two oppositely colored half-tiles.

We define a semi-Fock representation in the following way.
Vectors of the module $|\lambda\rangle$ are in one-to-one correspondence with super-Young diagrams $\lambda$.
The action of the generating functions for $\mathsf{Y}(\widehat{\fg\fl}_{1|1})$ operators has the following form:
\begin{equation}\label{rep}
	\begin{aligned}
		\sum\lm_{n=0}^{\infty}\frac{e_n^{\circ}}{z^{n+1}}|\lambda\rangle&=\sum\lm_{\shtile\in\lambda^+}\frac{{\bf E}_{\lambda,\lambda+\shtile}}{z-\omega_{\shtile}}|\lambda+\nhtile\rangle,\quad
		&\sum\lm_{n=0}^{\infty}\frac{e_n^{\bullet}}{z^{n+1}}|\lambda\rangle&=\sum\lm_{\shhtile\in\lambda^+}\frac{{\bf E}_{\lambda,\lambda+\shhtile}}{z-\omega_{\shhtile}}|\lambda+\nhhtile\rangle\,,\\
		%%%%%%%%%%%%%%%%%%%%%%%%%%%%%%%%%%%%%
		\sum\lm_{n=0}^{\infty}\frac{f_n^{\circ}}{z^{n+1}}|\lambda\rangle&=\sum\lm_{\shtile\in\lambda^-}\frac{{\bf F}_{\lambda,\lambda-\shtile}}{z-\omega_{\shtile}}|\lambda-\nhtile\rangle,\quad
		&\sum\lm_{n=0}^{\infty}\frac{f_n^{\bullet}}{z^{n+1}}|\lambda\rangle&=\sum\lm_{\shhtile\in\lambda^-}\frac{{\bf F}_{\lambda,\lambda-\shhtile}}{z-\omega_{\shhtile}}|\lambda-\nhhtile\rangle\,,\\
		%%%%%%%%%%%%%%%%%%%%%%%%%%%%%%%%%%%%%
		\sum\lm_{n=-\infty}^{\infty}\frac{\psi_n^{\circ}}{z^{n+1}}|\lambda\rangle&=\Psi_{\lambda}^{\circ}(z)|\lambda\rangle,\quad
		&\sum\lm_{n=-\infty}^{\infty}\frac{\psi_n^{\bullet}}{z^{n+1}}|\lambda\rangle&=\Psi_{\lambda}^{\bullet}(z)|\lambda\rangle\,.\\
	\end{aligned}
\end{equation}
Here $\lambda^+$($\lambda^-$) are elements, upper or lower half-tiles, that can be added to (subtracted from) super-partition $\lambda$, so that a new diagram denoted as $\lambda\pm\nhtile/\nhhtile$ is again a diagram of a super-partition.
Spectral parameters for tiles are defined in the following way:
\begin{equation}
	\begin{aligned}
		&\omega_{\shtile}=\epsilon_1\left(x_{\shtile}+y_{\shtile}\right)+\epsilon_2\left(x_{\shtile}-y_{\shtile}\right)\,,\\
		&\omega_{\shhtile}=\epsilon_1\left(x_{\shhtile}+y_{\shhtile}+1\right)+\epsilon_2\left(x_{\shhtile}-y_{\shhtile}\right)\,,
	\end{aligned}
\end{equation}
where by tile coordinates $x_{\shtile/\shhtile}$, $y_{\shtile/\shhtile}$ we imply coordinates of respective complete tile centers (see \eqref{diag_example}).

Matrices \eqref{rep} form a representation \footnote{
We should note that one has to include certain sign shifts to fix the statistics of generators.
See \cite[sec.3.5]{Galakhov:2020vyb} and \cite[sec.3.2]{Galakhov:2021vbo} for details.
} of the affine Yangian $\mathsf{Y}(\widehat{\fg\fl}_{1|1})$ if matrices ${\bf E}$, ${\bf F}$ and eigenvalues $\Psi$ satisfy \emph{hysteresis} relations:
%\begin{tcolorbox}
\begin{equation}\label{hysteresis}
	\begin{aligned}
		&{\bf E}_{\lambda+a,\lambda+a+b}{\bf F}_{\lambda+a+b,\lambda+b}={\bf F}_{\lambda+a,\lambda}{\bf E}_{\lambda,\lambda+b}\,,\\
		&\frac{{\bf E}_{\lambda,\lambda+a}{\bf E}_{\lambda+a,\lambda+a+b}}{{\bf E}_{\lambda,\lambda+b}{\bf E}_{\lambda+b,\lambda+a+b}}\varphi_{a,b}(\omega_a-\omega_b)=1\,,\\
		&\frac{{\bf F}_{\lambda+a+b,\lambda+a}{\bf F}_{\lambda+a,\lambda}}{{\bf F}_{\lambda+a+b,\lambda+b}{\bf F}_{\lambda+b,\lambda}}\varphi_{a,b}(\omega_a-\omega_b)=1\,,\\
		& {\bf E}_{\lambda,\lambda+a}{\bf F}_{\lambda+a,\lambda}=\mathop{\rm res}\lm_{z=\omega_a}\Psi^{a}_{\lambda}(z)\,,
	\end{aligned}
\end{equation}
%\end{tcolorbox}
where $a$, $b$ are either an upper tile $\nhtile$ (a white stone $\nhtile\to\circ$ in subscripts) or a lower tile $\nhhtile$ (a black stone $\nhhtile\to\bullet$ in subscripts).
Functions $\varphi$ and eigenvalues $\Psi$ are defined in the following way:\footnote{Let us stress that functions $\varphi$ are compatible with \eqref{Yang_modes} and may be restored canonically form the quiver theory via procedure \eqref{scheme}.
The result is a surprisingly simple expression in terms of quiver morphism weights (see e.g. \cite[eq.(4.11)]{Li:2020rij}).}
\begin{equation}\label{varphi}
	\begin{aligned}
	&\varphi_{\circ,\circ}(z)=-1,\quad \varphi_{\bullet,\bullet}(z)=-1,\quad \varphi_{\circ,\bullet}(z)=\frac{z^2-\epsilon_1^2}{z^2-\epsilon_2^2},\quad \varphi_{\bullet,\circ}(z)=\frac{z^2-\epsilon_2^2}{z^2-\epsilon_1^2}\,,\\
	&\Psi_{\lambda}^a(z)=\psi_\varnothing^a(z)\,\prod\lm_{b\in\lambda}\varphi_{a,b}(z-\omega_b)\,,
	\end{aligned}
\end{equation}
where
\begin{equation}\label{psi_rep}
	\psi_\varnothing^{\circ}(z)=\frac{c_{\circ}}{z},\quad \psi_\varnothing^{\bullet}(z)=c_{\bullet}\,(z+\epsilon_1)\,,
\end{equation}
and $c_{\circ/\bullet}$ are some generic constants.

%%%%%%%%%%%%%%%%%%%%%%%%%%%%%%%%%%%%%%%%%%%%%%%%%%%%%%%%%%%%%%%%%%%
%%%%%%%%%%%%%%%%%%%%%%%%%%%%%%%%%%%%%%%%%%%%%%%%%%%%%%%%%%%%%%%%%%%
%%%%%%%%%%%%%%%%%%%%%%%%%%%%%%%%%%%%%%%%%%%%%%%%%%%%%%%%%%%%%%%%%%%
%%%%%%%%%%%%%%%%%%%%%%%%%%%%%%%%%%%%%%%%%%%%%%%%%%%%%%%%%%%%%%%%%%%

\subsection{Representations of Yangian algebras}

The action of generators $e$ and $f$ increases and decreases the sum of dimensions $D:=d^\circ + d^\bullet$ by one.
Enumeration of states at every given $D$ depends on a representation and is an open problem.

A natural first attempt would be to build a kind of Verma module formed by the action of a sequence of raising generators $e$
on some initial state.
For some values of $\zeta$ this is straightforward and leads to representations, known as {\it molten crystals},
where the states can be labeled by a kind of Young tableaux or, perhaps, by their modification like {\it super} or 3d.
Additional {\it hysteresis properties} then allow to reduce this system to Young diagrams.
The main problem for this kind of representations is to explicitly describe matrices ${\bf E}$ and ${\bf F}$
acting between adjacent Young diagrams, which glue or erase one additional box.

A techincally convenient way to do this is to extract the diagrams from a rather different problem --
enumeration of the BPS states, i.e. the fixed points of the superpotential on the moduli space of ADHM solutions
associated with the quiver in question.
As a bonus, the matrices ${\bf E}$ and ${\bf F}$ are provided by a kind of Duistermaat-Heckmann integrals reducing the whole calculation to a calculation of Euler classes of tangent spaces to the fixed points we denote as ``Eul''.

However, the variety of fixed points appear sensitive to the values of parameters $\zeta$ --
and only in one \emph{cyclic} sector, when all $\zeta$ are of the same sign, their enumeration reduces to Young-like diagrams naturally.
In other sectors the variety is more complicated, and the idea of the Verma module is essentially modified.
%The new states with a different $D$ can be formed not only by the action of sequences of the raising operators $e$
%from initial one,
%but also can be a kind of its {\it pre}images.
Under this circumstances raising operators $e$ will not solely add an atom and raise $D$ by 1, rather they will add or even remove some peculiar \emph{clusters} of atoms raising or lowering $D$ accordingly.

We explain what this means by the examples in sec.\ref{sec:lot_of_examples} below.
The natural object for BPS states are not Young tableaux and diagrams but rather the more general {\it atomic structures},
to be introduced in the next section \ref{sec:atomic_plot}.
At the present level of understanding it is not guaranteed that ${\bf E}, {\bf F}$,
calculated from the BPS algebras, satisfy the commutation relations of the right Yangian --
but so far it was true in all the examples.
Still every time this should be checked  explicitly.

To conclude this section let us note that eventually for some non-cyclic phases we will be able to restore an enumeration of the module vectors by Young-like diagrams.
To achieve this goal we establish one-to-one correspondences -- dualities -- between fixed points in the theory in question in a non-cyclic phase and another (potentially completely different) theory in the cyclic phase.
Unfortunately, these manipulations are not canonical, we are not aware whether this approach is applicable in a generic situation.
And the equivalence between elements of the atomic structure plots and Young-like diagrams becomes rather obscure.

Nevertheless, these manipulations allow us to define matrices $\bf E$ and $\bf F$ outside the cyclic phase and confirm that the Verma module is a representation of a quiver Yangian by checking hysteresis relations \eqref{hysteresis}.
Moreover this procedure allows us to extract a quadruplet of functions $\varphi_{\circ,\circ}(z)$, $\varphi_{\bullet,\circ}(z)$, $\varphi_{\circ,\bullet}(z)$, $\varphi_{\bullet,\bullet}(z)$ for the Verma module in the non-cyclic phase.
Comparing those expressions with \eqref{varphi} and finding an agreement we conclude that the Verma module is a representation of the same affine super-Yangian $\mathsf{Y}(\widehat{\fg\fl}_{1|1})$.
And extracted vacuum charge functions $\psi_{\varnothing}^{\circ}(z)$, $\psi_{\varnothing}^{\bullet}(z)$ tell us what representation we acquire and how to mimic it as a Verma module in a cyclic phase of \emph{another} quiver theory with probably modified framing (see \cite{Galakhov:2021xum} for details).

\section{BPS vacua without disguise}\label{sec:atomic_plot}

\subsection{Vacuum equations}

In this section we level by level construct the fixed points in the space of quiver \eqref{quiver} representations.
The initial data contains:
\begin{itemize}
	\item Two non-negative integer numbers $d^{\circ}, d^{\bullet}$ that encode the corresponding quiver dimensions.
	\item Set of matrices corresponding to quiver arrows:
	\begin{align}
		\begin{aligned}
			A_{1}, A_2 \in \text{Mat}_{d^{\bullet} \times d^{\circ}}(\IC), \hspace{7mm}
			B_{1}, B_2 \in \text{Mat}_{d^{\circ} \times d^{\bullet}}(\IC), \hspace{7mm}
			R \in \text{Mat}_{d^{\bullet} \times 1}(\IC), \hspace{7mm}
			S \in \text{Mat}_{1 \times d^{\circ}}(\IC)\,.
		\end{aligned}
	\end{align}
	\item Stability parameters (real numbers) $\zeta^{\circ}$ and $\zeta^{\bullet}$.
	\item We impose the following three sets of equations on matrices $A_1, A_2, B_1, B_2, R, S$ with fixed values of $d^{\circ}, d^{\bullet}, \zeta^{\circ}, \zeta^{\bullet}$:
	\begin{enumerate}
		\item D-term equations:
		\begin{align}\label{D-term}
			\text{D-term}: \left\{ \hspace{2mm}
			\begin{aligned}
				- A_1^{\dagger} A_1 - A_2^{\dagger} A_2 + B_1 B_1^{\dagger} + B_2 B_2^{\dagger} + R R^{\dagger} = \zeta^{\circ} \cdot 1_{d^{\circ} \times d^{\circ}} \\
				A_1 A_1^{\dagger} + A_2 A_2^{\dagger} - B_1^{\dagger} B_1 - B_2^{\dagger} B_2 -  S^{\dagger} S  = \zeta^{\bullet} \cdot 1_{d^{\bullet} \times d^{\bullet}}
			\end{aligned}
			\right.
		\end{align}
		\item F-term equations that are derivatives of the superpotential:
		\begin{equation}
			W = \Tr \Big[ A_1 B_1 A_2 B_2 - A_1 B_2 A_2 B_1 + A_2 R S \Big]
		\end{equation}
		\begin{align}\label{F-term}
			\text{F-term}: \left\{ \hspace{2mm}
			\begin{aligned}
				\frac{\partial W}{\partial A_1} &= B_1 A_2 B_2 - B_2 A_2 B_1 = 0 \\
				\frac{\partial W}{\partial A_2} &= B_2 A_1 B_1 - B_1 A_1 B_2 + R S = 0 \\
				\frac{\partial W}{\partial B_1} &= A_2 B_2 A_1 - A_1 B_2 A_2 = 0 \\
				\frac{\partial W}{\partial B_2} &= A_1 B_1 A_2 - A_2 B_1 A_1 = 0 \\
                    \frac{\partial W}{\partial R} &= S A_2 \\
                    \frac{\partial W}{\partial S} &= A_2 R
			\end{aligned}
			\right.
		\end{align}
	\end{enumerate}
	The superpotential $W$ and all the D-term and F-term equations are invariant with respect to gauge transformations (unitary rotation of the basis in quiver nodes):
	\begin{equation}
		g^{\circ} \in U\left( d^{\circ} \right), \hspace{10mm} g^{\bullet} \in U \left(d^{\bullet} \right)
	\end{equation}
	\begin{equation}
		A_i \ \to \ g^{\bullet} A_i \left(g^{\circ} \right)^{\dagger}, \hspace{10mm}
		B_i \ \to \ g^{\circ} B_i \left(g^{\bullet} \right)^{\dagger}, \hspace{10mm}
		R \ \to \ g^{\circ} R, \hspace{10mm}
		S \ \to \ S \left(g^{\bullet} \right)^{\dagger}
	\end{equation}
	\item We define the following action of torus $U(1) \times U(1) = \exp\left( i \epsilon_1 \right) \times \exp \left( i \epsilon_2 \right) $ according to \eqref{quiver} on the space of solutions of D-term and F-term equations:
	\begin{align}
		\begin{aligned}
			A_1 \ \to \ \exp\left( i \epsilon_1 \right) \cdot A_1, &\hspace{10mm}
			A_2 \ \to \ \exp\left( - i \epsilon_1 \right) \cdot A_2, \\
			B_1 \ \to \ \exp\left( i \epsilon_2 \right) \cdot B_1, &\hspace{10mm}
			B_2 \ \to \ \exp\left( - i \epsilon_2 \right) \cdot B_2, \\
			R \ \to \  R, &\hspace{10mm}
			S \ \to \ \exp\left( i \epsilon_1 \right) \cdot S.
		\end{aligned}
	\end{align}
	We are interested in the {\it fixed points} of $U(1) \times U(1)$ action on the space of solutions of D-term and F-term equations. Due to the gauge freedom the fixed point condition is equivalent to the requirement that $U(1) \times U(1)$ action goes along the gauge orbit:
	\begin{align}
		\begin{aligned}
			g^{\bullet} A_1 \left(g^{\circ} \right)^{\dagger} = \exp\left( i \epsilon_1 \right) \cdot A_1, &\hspace{10mm}
			g^{\bullet} A_2 \left(g^{\circ} \right)^{\dagger} = \exp\left( - i \epsilon_1 \right) \cdot A_2, \\
			g^{\circ} B_1 \left(g^{\bullet} \right)^{\dagger} = \exp\left( i \epsilon_2 \right) \cdot B_1, &\hspace{10mm}
			g^{\circ} B_2 \left(g^{\bullet} \right)^{\dagger} = \exp\left( - i \epsilon_2 \right) \cdot B_2, \\
			g^{\circ} R = R, &\hspace{10mm}
			S \left(g^{\bullet} \right)^{\dagger} = \exp\left( i \epsilon_1 \right) \cdot S.
		\end{aligned}
	\end{align}
	where $g^{\circ}$ and $g^{\bullet}$ are proper unitary matrices. We reformulate these conditions in an infinitesimal form:
	\begin{align}
		\label{fixed point AB}
		\text{Fixed point cond.}: \left\{ \hspace{2mm}
		\begin{aligned}
			&\Phi^{\bullet} A_1 - A_1 \Phi^{\circ} = \epsilon_1 \cdot A_1 \\
			&\Phi^{\bullet} A_2 - A_2 \Phi^{\circ} = -\epsilon_1 \cdot A_2 \\
			&\Phi^{\circ} B_1 - B_1 \Phi^{\bullet} = \epsilon_2 \cdot B_1 \\
			&\Phi^{\circ} B_2 - B_2 \Phi^{\bullet} = -\epsilon_2 \cdot B_2 \\
			&\Phi^{\circ}\, R  = 0 \\
			&S \, \Phi^{\bullet} = \epsilon_1 \cdot S
		\end{aligned}
		\right.
	\end{align}
	where $g^{\circ} = \exp\left( i \Phi^{\circ} \right), g^{\bullet} = \exp\left( i \Phi^{\bullet} \right)$ with Hermitian matrices $\Phi^{\circ}, \Phi^{\bullet}$. Using gauge transformations one can always choose a basis where these Hermitian matrices are diagonal. 
	Finally, we have additional $d^{\circ} + d^{\bullet}$ variables from matrices $\Phi^{\circ}, \Phi^{\bullet}$ that should be found from the resulting system:
	\begin{equation}
		\label{resulting system}
		\text{D-term} + \text{F-term} + \text{Fixed point cond.}
	\end{equation}
\end{itemize}
In the subsequent sections we explicitly solve \eqref{resulting system} in various setups.
For each solution of the system we draw an atomic structure plot and indicate a region of the phase space $ (\zeta^{\circ}, \zeta^{\bullet}) = \mathbb{R}^2$ where the solution is valid.

Here we apply the most basic solution methods, however it would be interesting to learn if our calculations may be simplified and extended to larger families if more advanced method of counting fixed point are involved.
One of such methods \cite{Beaujard:2021fsk,Bao:2024noj} is to count admissible atomic structure pictures as labeling Jeffrey-Kirwan residues \cite{JK}.
As other approaches one might count enumerating attractor flow trees \cite{Andriyash:2010qv,Mozgovoy:2020has,Denef:2007vg,Denef:2000ar,Manschot:2010xp} or enumerating special Lagrangian cycles on a curve mirror dual to a Calabi-Yau threefold (see \cite{Banerjee:2024smk,Klemm:1996bj,Eager:2016yxd,Banerjee:2019apt} for reviews of this huge topic)

\subsection{Atomic structure plots}

The best known way to enumerate fixed points is with the help of (super)Young tableaux,
which actually reduce to (super)Young diagrams due to the {\it hysteresis property} of the representation \eqref{hysteresis}.
This is a natural representation metamorphosis of another schematic fixed point enumeration valid in a cyclic chamber (when either all the stability parameters are positive or negative) called molten crystals \cite{Okounkov:2003sp,Iqbal:2003ds,Ooguri:2009ijd,Chuang:2009crq}.
Details of this identification for the conifold in question could be found in \cite{Galakhov:2023mak}.

However the molten crystal structure is not valid in all the phase chambers in general.
For this reason we adopt a language of plot depictions for fixed points that seems to work in all the situations we encounter in this paper.
A depiction consists of oriented edges connecting colored dots -- ``atoms'' -- and generalizes the molten crystal pictures in a natural way.
Therefore we propose to call these pictures \emph{atomic structure plots}.

Since in the generic phase the quiver representation loses its cyclic property the atomic structure plot loses  certain regularity.
We might encounter pictures similar to long polymer-like strings of atoms attached to a crystal body, we propose to call these pictures \emph{simple glasses}.
On the other hand in other phases crystals might acquire in addition a version of a defect -- more than a single atom is presented at a single dot position -- we propose to call such types of pictures \emph{crooked glasses}.

\bigskip

\begin{comment}
In this subsection we will describe a graphical language to restrict the variety of solutions to a system \eqref{resulting system} of vacuum equations we will exploit in what follows extensively.
We call this kind of plots \emph{atomic structure plots}.
\end{comment}

\begin{figure}[ht!]
\begin{center}
	\begin{tikzpicture}[scale=0.4]
		\draw[thin,-stealth] (-5,0) -- (10,0);
		\draw[thin,-stealth] (0,-3) -- (0,3);
		\node[right] at (10,0) {$\scriptstyle \epsilon_1$};
		\node[above] at (0,3) {$\scriptstyle \epsilon_2$};
		%%%%%%%%%%%%%%%%%%%%%%%%%%%%%%%%%%%%%%%%%%
		\foreach \x/\y/\z/\w in {0/0/1/0, 1/-1/2/-1, 1/0/1/-1, 1/0/1/1, 1/1/2/1, 2/-2/3/-2, 2/-1/2/-2, 2/-1/2/0, 2/0/3/0, 2/1/2/0, 2/1/2/2, 2/2/3/2, 3/-3/4/-3, 3/-2/3/-3, 3/-2/3/-1, 3/-1/4/-1, 3/0/3/-1, 3/0/3/1, 3/1/4/1, 3/2/3/1, 3/2/3/3, 3/3/4/3, 4/-4/5/-4, 4/-3/4/-4, 4/-3/4/-2, 4/-2/5/-2, 4/-1/4/-2, 5/1/4/1, 5/1/6/1, 7/1/6/1, 7/1/8/1, 9/1/8/1, 9/1/10/1, 11/1/10/1, 5/3/4/3, 5/3/6/3, 7/3/6/3, 7/3/8/3, 9/3/8/3, 9/3/10/3, 11/3/10/3, 6/-4/5/-4, 6/-4/7/-4, 8/-4/7/-4, 8/-4/9/-4, 10/-4/9/-4, 10/-4/11/-4, 12/-4/11/-4, 6/-2/5/-2, 6/-2/7/-2, 8/-2/7/-2, 8/-2/9/-2, 10/-2/9/-2, 10/-2/11/-2, 12/-2/11/-2}
		{
			\draw[thick, postaction={decorate},decoration={markings,
				mark= at position 0.7 with {\arrow{stealth}}}] (\x,\y) -- (\z,\w);
		}
		\foreach \x/\y in {0/0, 1/-1, 1/1, 2/-2, 2/0, 2/2, 3/-3, 3/-1, 3/1, 3/3, 4/-4, 4/-2, 5/1, 7/1, 9/1, 11/1, 5/3, 7/3, 9/3, 11/3, 6/-4, 8/-4, 10/-4, 12/-4, 6/-2, 8/-2, 10/-2, 12/-2}
		{
			\draw[fill=white] (\x,\y) circle (0.2);
		}
		\foreach \x/\y in {1/0, 2/-1, 2/1, 3/-2, 3/0, 3/2, 4/-3, 4/-1, 4/1, 6/1, 8/1, 10/1, 4/3, 6/3, 8/3, 10/3, 5/-4, 7/-4, 9/-4, 11/-4, 5/-2, 7/-2, 9/-2, 11/-2}
		{
			\draw[fill=gray] (\x,\y) circle (0.2);
		}
		%%%%%%%%%%%%%%%%%%%%%%%%%%%%%%%%%%%%%%%%
		\begin{scope}[shift={(-1,0)}]
		\begin{scope}[xscale=-1]
			\foreach \x/\y/\z/\w in {0/0/1/0, 1/-1/2/-1, 1/0/1/-1, 1/0/1/1, 1/1/2/1, 2/-2/3/-2, 2/-1/2/-2, 2/-1/2/0, 2/0/3/0, 2/1/2/0, 2/1/2/2, 2/2/3/2, 3/0/3/1, 3/1/4/1, 3/2/3/1, 3/2/3/3}
			{
				\draw[thick, postaction={decorate},decoration={markings,
					mark= at position 0.7 with {\arrow{stealth}}}] (\z,\w) -- (\x,\y);
			}
			\foreach \x/\y in {0/0, 1/-1, 1/1, 2/-2, 2/0, 2/2, 3/1, 3/3}
			{
				\draw[fill=gray] (\x,\y) circle (0.2);
			}
			\foreach \x/\y in {1/0, 2/-1, 2/1, 3/-2, 3/0, 3/2, 4/1}
			{
				\draw[fill=white] (\x,\y) circle (0.2);
			}
		\end{scope}
		\end{scope}
		%%%%%%%%%%%%%%%%%%%%%%%%%%%%%%%%%%%%%%%%%%%%%%
		\draw[burgundy] (0,0) circle (0.4);
		\draw[\myblue] (-1,0) circle (0.4);
		\begin{scope}[shift={(-7,-3)}]
			\draw[-stealth] (0,0) -- (1,0) node[right] {$\scriptstyle A_1$};
			\draw[-stealth] (0,0) -- (-1,0) node[left] {$\scriptstyle A_2$};
			\draw[-stealth] (0,0) -- (0,1) node[above] {$\scriptstyle B_1$};
			\draw[-stealth] (0,0) -- (0,-1) node[below] {$\scriptstyle B_2$};
		\end{scope}
	\end{tikzpicture}
	\caption{\footnotesize Example of the atomic structure plot.} \label{fig:ASD}
\end{center}
\end{figure}
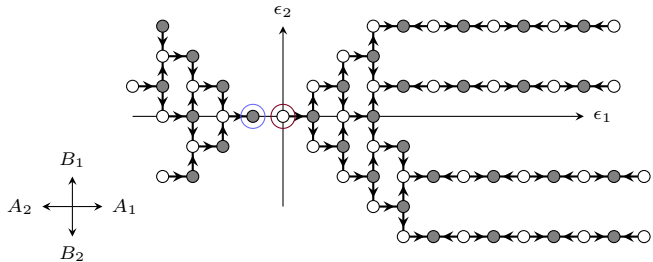

A plot
in this language is represented by a mesh of black\footnote{In print we mark them with the gray color, so that the resulting picture is less contrasting.} and white dots (``atoms'') connected by directed arrows (see e.g. fig.\ref{fig:ASD}).
Also we distinguish few atoms by marking them with red and blue circles.
The plot is drawn in a plane and all the atoms have definite integral coordinates $(x,y)$.
All the arrow edges are of length 1, and each one is co-directed with one of four vectors:
\begin{equation}
	A_1=\left(\begin{array}{cc}
		1 & 0
	\end{array}\right),\quad A_2=\left(\begin{array}{cc}
		-1 & 0
	\end{array}\right),\quad B_1=\left(\begin{array}{cc}
		0 & 1
	\end{array}\right),\quad B_2=\left(\begin{array}{cc}
		0 & -1
	\end{array}\right)\,.
\end{equation}
Note that these vectors coincide in the $\epsilon_{1,2}$-plane with weights \eqref{quiver} of corresponding fields.

Such a plot restricts partially the values of fields $A_{1,2}$, $B_{1,2}$, $R$, $S$ and $\Phi^{\circ/\bullet}$ according to the following rules:
\begin{enumerate}
	\item The white and black atoms represent graphically vectors of vector spaces $\IC^{d^{\circ}}$ and $\IC^{d^{\bullet}}$ corresponding to white $\circ$ and black $\bullet$  nodes of quiver \eqref{quiver} respectively.
	Therefore we arrive to the first rule:
	\begin{equation}\label{dimensions}
		d^{\circ}=\#\{\mbox{white atoms}\},\quad d^{\bullet}=\#\{\mbox{black atoms}\}\,.
	\end{equation}
	\item We enumerate white and black atoms in any order.
	An ordering change is a permutation element $\sigma\in S_d\subset U(d)$ of the gauge group associated with one of quiver nodes and is a symmetry of the problem in question.
	Fields $\Phi^{\circ/\bullet}$ acquire diagonal expectation values:
	\begin{equation}
		\Phi^{\circ/\bullet}={\rm diag}(\omega_1^{\circ/\bullet},\ldots, \omega_{d^{\circ/\bullet}}^{\circ/\bullet})\,,
	\end{equation}
	where eigenvalues $\omega_a^{\circ/\bullet}$ are defined by the coordinates $(x_a^{\circ/\bullet},y_a^{\circ/\bullet})$ of the $a^{\rm th}$ white/black atom in the $\epsilon_{1,2}$-plane:
	\begin{equation}
		\omega_a^{\circ/\bullet}=\epsilon_1x_a^{\circ/\bullet}+\epsilon_2y_a^{\circ/\bullet}\,.
	\end{equation}
	\item Directed edges of the graph are co-direct with one of four quiver morphism arrow.
	Let us call it $M$.
	An arrow $\begin{array}{c}
		\begin{tikzpicture}[scale=0.4]
			\draw[postaction={decorate},decoration={markings,
				mark= at position 0.6 with {\arrow{stealth}}}] (0,0) -- (1.5,0) node[pos=0.5,above] {$\scriptstyle M$};
			\draw[fill=white] (0,0) circle (0.2);
			\draw[fill=gray] (1.5,0) circle (0.2);
			\node[left] at (0,0) {$a$};
			\node[right] at (1.5,0) {$b$};
		\end{tikzpicture}
	\end{array}$ in the graph connecting the $a^{\rm th}$ and $b^{\rm th}$ atoms indicate that the $ba$ matrix element $m_{ba}$ of this morphism is non-zero:
	\begin{equation}
		M={\color{white!80!black}\left(\begin{array}{cccccc}
				* & * & * & * & * & *\\
				* & * & {\color{black} m_{ba}} & * & * & *\\
				* & * & * & * & * & *\\
				* & * & * & * & * & *\\
			\end{array}\right)}\,.
	\end{equation}
	All the other matrix elements of the four quiver morphisms not depicted in the graph by a directed edge are strictly zeroes.
	\item Similarly morphism $R$ (resp. $S$) maps the framing line $\IC$ only to (resp. from) the $a^{\rm th}$ vector of $\IC^{d^{\circ}}$ (resp. $\IC^{d^{\bullet}}$), so that the matrix has a non-zero element only at the $a^{\rm th}$ position:
	\begin{equation}
		R=\left(\begin{array}{c}
			\ldots\\
			0\\
			{\bf\color{burgundy} r}_a\\
			0\\
			\ldots
		\end{array}\right),\quad S=\left(\begin{array}{ccccc}
		\ldots&
		0&
		{\bf\color{\myblue} s}_a&
		0&
		\ldots
		\end{array}\right)\,.
	\end{equation}
	We mark the atom representing the corresponding vector by a {\color{burgundy} red} (resp. {\color{\myblue}blue}) circle.
	
	When we consider dual quiver \eqref{dual_quiver_2} the map form the framing node to a gauge node is called $\check{S}$.
	In this case in the atomic structure plot we denote by the {\color{burgundy} red} circle the  image of $\check{S}$.
	\item In cases when the coordinate frame orientation is obscure we add a quiver morphism rose picture to the plot.
\end{enumerate}

We should note that the strength of these diagrammatic rules is that field values constructed from an atomic structure plot resolve constraints \eqref{fixed point AB} \emph{automatically}.
However remaining equations \eqref{D-term} and \eqref{F-term} should be solved further to restrict matrix elements of the quiver morphisms.

As an illustration consider the following plot where we have enumerated the atoms in some random way:
\begin{equation}
\begin{aligned}
	&\begin{array}{c}
		\begin{tikzpicture}
			\draw[thick, postaction={decorate},decoration={markings,
				mark= at position 0.7 with {\arrow{stealth}}}] (0,0) -- (1,0) node[pos=0.5, above] {$\scriptstyle a_{1,34}$};
			\draw[thick, postaction={decorate},decoration={markings,
				mark= at position 0.7 with {\arrow{stealth}}}] (1,0) -- (1,1) node[pos=0.5, right] {$\scriptstyle b_{1,23}$};
			\draw[thick, postaction={decorate},decoration={markings,
				mark= at position 0.7 with {\arrow{stealth}}}] (1,0) -- (1,-1) node[pos=0.5, right] {$\scriptstyle b_{2,33}$};
			\draw[thick, postaction={decorate},decoration={markings,
				mark= at position 0.7 with {\arrow{stealth}}}] (1,1) -- (2,1) node[pos=0.5, above] {$\scriptstyle a_{1,34}$};
			\draw[thick, postaction={decorate},decoration={markings,
				mark= at position 0.7 with {\arrow{stealth}}}] (-2,0) -- (-1,0) node[pos=0.5, above] {$\scriptstyle a_{1,21}$};
			%%%%%%%%%%%%%%%%%%%%%%%%%%%%%%%%%%%%%%%%%%%%%%%%%%%%
			\draw[fill=white] (0,0) circle (0.15);
			\node at (0,0) {$\scriptstyle 4$};
			\draw[fill=gray] (1,0) circle (0.15);
			\node[white] at (1,0) {$\scriptstyle 3$};
			\draw[fill=white] (1,1) circle (0.15);
			\node at (1,1) {$\scriptstyle 2$};
			\draw[fill=white] (1,-1) circle (0.15);
			\node at (1,-1) {$\scriptstyle 3$};
			\draw[fill=gray] (2,1) circle (0.15);
			\node[white] at (2,1) {$\scriptstyle 1$};
			\draw[fill=white] (-2,0) circle (0.15);
			\node at (-2,0) {$\scriptstyle 1$};
			\draw[fill=gray] (-1,0) circle (0.15);
			\node[white] at (-1,0) {$\scriptstyle 2$};
			%%%%%%%%%%%%%%%%%%%%%%%%%%%%%%%%%%%%%%%%%%%%%
			\begin{scope}[shift={(-4,0)}]
				\begin{scope}[scale=0.4]
				\draw[-stealth] (0,0) -- (1,0) node[right] {$\scriptstyle A_1$};
				\draw[-stealth] (0,0) -- (-1,0) node[left] {$\scriptstyle A_2$};
				\draw[-stealth] (0,0) -- (0,1) node[above] {$\scriptstyle B_1$};
				\draw[-stealth] (0,0) -- (0,-1) node[below] {$\scriptstyle B_2$};
				\end{scope}
			\end{scope}
			%%%%%%%%%%%%%%%%%%%%%%%%%%%%%%%%
			\draw[thick, burgundy] (0,0) circle (0.25);
			\draw[thick, \myblue] (-1,0) circle (0.25);
		\end{tikzpicture}
	\end{array}\\
	& A_1=\left(\begin{array}{cccc}
		0& a_{1,12} & 0 & 0 \\
		a_{1,21} & 0 & 0 & 0\\
		0 & 0 & 0 & a_{1,34}\\
	\end{array}\right),\quad
	A_2=\left(\begin{array}{cccc}
		0& 0 & 0 & 0 \\
		0 & 0 & 0 & 0\\
		0 & 0 & 0 & 0\\
	\end{array}\right)\,,\\
	&B_1=\left(\begin{array}{ccc}
		0 & 0 & 0\\
		0 & 0 & b_{1,23}\\
		0 & 0 & 0\\
		0 & 0 & 0\\
	\end{array}\right),\quad
	B_2=\left(\begin{array}{ccc}
		0 & 0 & 0\\
		0 & 0 & 0\\
		0 & 0 & b_{2,33}\\
		0 & 0 & 0\\
	\end{array}\right),\quad
	R=\left(\begin{array}{c}
		0\\ 0\\ 0\\ r_4
	\end{array}\right),\quad S=\left(\begin{array}{ccc}
	0 & s_2 & 0 \\
	\end{array}\right)\,.
\end{aligned}
\end{equation}

\subsection{From crystals to glasses}

The atomic structure plots may be thought of as a modification of a graphical language of molten crystals \cite{Ooguri:2009ijd,Yamazaki:2010fz,Nishinaka:2013mba,Li:2020rij,Galakhov:2020vyb} given by oriented paths in a periodic quiver lattice on a torus to describe glasses as well.
For this we loosen a constraint that all the paths between atoms along oriented edges should start (or terminate) on some root atom.
This situation should not be confused with the case when there are multiple root atoms \cite{Galakhov:2021xum} representing seeds for different molten crystals growing from them.
Unlike molten crystals in the cyclic phase glasses describe quiver representations where some representation vectors can not be constructed as a holomorphic function of quiver fields\footnote{Let us remind here that maps $R$ and $S$ connecting to the framing node represent a distinct vector and a co-vector in the quiver representation.}:
\begin{equation}
	\begin{array}{c}
		\begin{tikzpicture}
			\draw[thick, postaction={decorate},decoration={markings,
				mark= at position 0.7 with {\arrow{stealth}}}] (0,0) -- (1,0) node[pos=0.5, below] {$\scriptstyle A_1$};
			\draw[thick, postaction={decorate},decoration={markings,
				mark= at position 0.7 with {\arrow{stealth}}}] (1,0) -- (1,1) node[pos=0.5, left] {$\scriptstyle B_1$};
			\draw[thick, postaction={decorate},decoration={markings,
				mark= at position 0.7 with {\arrow{stealth}}}] (1,0) -- (1,-1) node[pos=0.5, right] {$\scriptstyle B_2$};
			\draw[thick, postaction={decorate},decoration={markings,
				mark= at position 0.7 with {\arrow{stealth}}}] (1,1) -- (2,1) node[pos=0.5, below] {$\scriptstyle A_1$};
			\draw[thick, postaction={decorate},decoration={markings,
				mark= at position 0.7 with {\arrow{stealth}}}] (3,1) -- (2,1) node[pos=0.5, below] {$\scriptstyle A_2$};
			%%%%%%%%%%%%%%%%%%%%%%%%%%%%%%%%%%%%%%%%%%%%%%%%%%%%
			\draw[fill=white] (0,0) circle (0.1);
			\draw[fill=gray] (1,0) circle (0.1);
			\draw[fill=white] (1,1) circle (0.1);
			\draw[fill=white] (1,-1) circle (0.1);
			\draw[fill=gray] (2,1) circle (0.1);
			\draw[fill=white] (3,1) circle (0.1);
			%%%%%%%%%%%%%%%%%%%%%%%%%%%%%%%%%%%%%%%%%%%%%
			\draw[thick, burgundy] (0,0) circle (0.2);
			%%%%%%%%%%%%%%%%%%%%%%%%%%%%%%%%%%%%%%%%%%%%%%
			\node[left] at (-0.2,0) {$\scriptstyle v_1$};
			\node[right] at (1.1,0) {$\scriptstyle v_2$};
			\node[right] at (1.1,-1) {$\scriptstyle v_3$};
			\node[left] at (0.9,1) {$\scriptstyle v_4$};
			\node[below] at (2,0.9) {$\scriptstyle v_5$};
			\node[right] at (3.1,1) {$\scriptstyle v_6$};
		\end{tikzpicture}
	\end{array},\quad \begin{array}{c}
		v_1=R,\quad v_2=A_1R,\quad v_3=B_2A_1R,\quad v_4=B_1A_1R\,, \\
		\\
		v_5=A_1B_1A_1R=A_2v_6,\quad v_6\not\in\IC[A_1,A_2,B_1,B_2,R,S]\,.
	\end{array}
\end{equation}
Yet as an unoriented graph an atomic structure plot remains connected, therefore one might try to construct these new paths in glasses by reversing directions of some edges.
An oriented edge $\phi$ connecting atoms $a$ and $b$ in this language represents a matrix element of quiver morphism $\phi$ between quiver representation vectors $|b\rangle\sim\phi|a\rangle$, and the arrow direction indicates which atom-vector belongs to the image of $\phi$, which one belongs to the pre-image.
We could have swapped those spaces by conjugating $\phi\to\phi^{\dagger}$ and simultaneously reversing the oriented edge direction, so that $|a\rangle\sim\phi^{\dagger}|b\rangle$:
\begin{equation}
	\left(\begin{array}{c}
		\begin{tikzpicture}
			\draw[thick, postaction={decorate},decoration={markings,
				mark= at position 0.7 with {\arrow{stealth}}}] (0,0) -- (1,0) node[pos=0.5, above] {$\scriptstyle \phi$};
			\draw[fill=white] (0,0) circle (0.1);
			\draw[fill=gray] (1,0) circle (0.1);
			\node[left] at (-0.1,0) {$\scriptstyle a$};
			\node[right] at (1.1,0) {$\scriptstyle b$};
		\end{tikzpicture}
	\end{array}\right)^{\dagger}=\begin{array}{c}
		\begin{tikzpicture}
			\draw[thick, postaction={decorate},decoration={markings,
				mark= at position 0.7 with {\arrow{stealth}}}] (1,0) -- (0,0) node[pos=0.5, above] {$\scriptstyle \phi^{\dagger}$};
			\draw[fill=white] (0,0) circle (0.1);
			\draw[fill=gray] (1,0) circle (0.1);
			\node[left] at (-0.1,0) {$\scriptstyle a$};
			\node[right] at (1.1,0) {$\scriptstyle b$};
		\end{tikzpicture}
	\end{array}\,.
\end{equation}
However this manipulation would make some quiver representation vectors non-holomorphic functions of quiver fields and break the complex structure of the quiver representation as an algebraic variety.

\section{Examples}\label{sec:lot_of_examples}

Now we are ready for exhaustive consideration of simple examples, at small values of parameters
$d^{\circ}$ and $d^{\bullet}$.

\subsection{$d^{\circ} = 1, d^{\bullet} = 1$}\label{sec:d_1=1,d_2=1}
We consider the following ansatz:
\begin{equation}
    A_{1} = a_1, \hspace{5mm} A_{2} = a_2, \hspace{5mm} B_1 = b_1, \hspace{5mm} B_2 = b_2, \hspace{5mm} R = r, \hspace{5mm} S = s
\end{equation}
\begin{equation}
    \Phi^{\circ} = \omega_1, \hspace{5mm} \Phi^{\bullet} = \omega_2
\end{equation}
The system \eqref{resulting system} has the following form:
\begin{align}
\left\{ \hspace{2mm}
    \begin{aligned}
        & -|a_1|^2-|a_2|^2 + |b_1|^2 + |b_2|^2 + |r|^2 = \zeta^{\circ} \\
        & |a_1|^2+|a_2|^2 - |b_1|^2 - |b_2|^2 - |s|^2 = \zeta^{\bullet} \\
        & (\omega_2 - \omega_1 - \epsilon_1) a_1 = 0 \\
        & (\omega_2 - \omega_1 + \epsilon_1) a_2 = 0 \\
        & (\omega_2 - \omega_1 + \epsilon_2) b_1 = 0 \\
        & (\omega_2 - \omega_1 - \epsilon_2) b_2 = 0 \\
        & r \, s = 0 \\
        & s \, a_2 = 0 \\
        & r \, a_2 = 0 \\
        & \omega_1 \, r = 0 \\
        & (\omega_2 + \epsilon_1) s = 0
    \end{aligned}
\right.
\end{align}

\subsubsection{Case $s = 0, r \not = 0 $}

Immediately we have $ \omega_1 = 0, a_2 = 0$. The remaining system
    \begin{table}[h!]
    \centering
    \begin{tabular}{|c|c|c|c|}
    \hline
        $\omega_1$ & $0$ & $0$ & $0$ \\
    \hline
        $\omega_2$ & $\epsilon_1$ & $-\epsilon_2$ & $\epsilon_2$ \\
        \hline
        \rotatebox[origin=c]{90}{Atomic struct.}
        &
        $\begin{array}{c}
        \begin{tikzpicture}[scale=0.4]
			\draw[thick, postaction={decorate},decoration={markings,
				mark= at position 0.7 with {\arrow{stealth}}}] (0,0) -- (1,0);
                \draw[-stealth,thin] (-1.8,0) -- (1.8,0);
                \draw[-stealth,thin] (0,-1.8) -- (0,1.8);
                \node[right] at (1.8,0) {$\scriptstyle \epsilon_1$};
			\node[above] at (0,1.8) {$\scriptstyle \epsilon_2$};
			\draw[fill=white] (0,0) circle (0.2);
			\draw[fill=gray] (1,0) circle (0.2);
			\draw[burgundy] (0,0) circle (0.4);
	\end{tikzpicture}\end{array}$ &
		$\begin{array}{c}
        \begin{tikzpicture}[scale=0.4]
			\draw[thick, postaction={decorate},decoration={markings,
				mark= at position 0.7 with {\arrow{stealth}}}] (0,-1) -- (0,0);
                \draw[-stealth,thin] (-1.8,0) -- (1.8,0);
                \draw[-stealth,thin] (0,-1.8) -- (0,1.8);
                \node[right] at (1.8,0) {$\scriptstyle \epsilon_1$};
			\node[above] at (0,1.8) {$\scriptstyle \epsilon_2$};
			\draw[fill=white] (0,0) circle (0.2);
			\draw[fill=gray] (0,-1) circle (0.2);
			\draw[burgundy] (0,0) circle (0.4);
	\end{tikzpicture}\end{array}$ &
		$\begin{array}{c}
        \begin{tikzpicture}[scale=0.4]
			\draw[thick, postaction={decorate},decoration={markings,
				mark= at position 0.7 with {\arrow{stealth}}}] (0,1) -- (0,0);
                \draw[-stealth,thin] (-1.8,0) -- (1.8,0);
                \draw[-stealth,thin] (0,-1.8) -- (0,1.8);
                \node[right] at (1.8,0) {$\scriptstyle \epsilon_1$};
			\node[above] at (0,1.8) {$\scriptstyle \epsilon_2$};
			\draw[fill=white] (0,0) circle (0.2);
			\draw[fill=gray] (0,1) circle (0.2);
			\draw[burgundy] (0,0) circle (0.4);
	\end{tikzpicture}\end{array}$\\
        \hline
        \rotatebox[origin=c]{90}{Solution}
        &
        $\begin{aligned}
            &|a_1|^2 = \zeta^{\bullet} \\
            &b_1 = 0 \\
            &b_2 = 0 \\
            &|r|^2 = \zeta^{\circ} + \zeta^{\bullet}
        \end{aligned}$ &
        $\begin{aligned}
            &a_1 = 0 \\
            &|b_1|^2 = - \zeta^{\bullet} \\
            &b_2 = 0 \\
            &|r|^2 = \zeta^{\circ} + \zeta^{\bullet}
        \end{aligned}$ &
        $\begin{aligned}
            &a_1 = 0 \\
            &b_1 = 0 \\
            &|b_2|^2 = - \zeta^{\bullet} \\
            &|r|^2 = \zeta^{\circ} + \zeta^{\bullet}
        \end{aligned}$
        \\
        \hline
        \rotatebox[origin=c]{90}{Consrt.}
        &
        $\left\{ \begin{aligned}
             \zeta^{\bullet} &> 0 \\
             \zeta^{\circ} + \zeta^{\bullet} &> 0
        \end{aligned}
        \right. $ &
        $\left\{ \begin{aligned}
             \zeta^{\bullet} &< 0 \\
             \zeta^{\circ} + \zeta^{\bullet} &> 0
        \end{aligned}
        \right. $ &
        $\left\{ \begin{aligned}
             \zeta^{\bullet} &< 0 \\
             \zeta^{\circ} + \zeta^{\bullet} &> 0
        \end{aligned}
        \right. $ \\
        \hline
        \rotatebox[origin=c]{90}{Phase region}
        &
        $\begin{array}{c}
        \begin{tikzpicture}[scale=0.4]
        	\draw[fill=\graphcol, \graphcol] (0,0) -- (-2,2) -- (2,2) -- (2,0) -- cycle;
			\draw[-stealth] (-2.5,0) -- (2.5,0);
			\draw[-stealth] (0,-2.5) -- (0,2.5);
			\node[right] at (2.5,0) {$\scriptstyle \zeta^{\circ}$};
			\node[above] at (0,2.5) {$\scriptstyle \zeta^{\bullet}$};
        \end{tikzpicture} \end{array}$
        & $\begin{array}{c}\begin{tikzpicture}[scale=0.4]
        	\draw[fill=\graphcol, \graphcol] (0,0) -- (2,0) -- (2,-2) -- cycle;
			\draw[-stealth] (-2.5,0) -- (2.5,0);
			\draw[-stealth] (0,-2.5) -- (0,2.5);
			\node[right] at (2.5,0) {$\scriptstyle \zeta^{\circ}$};
			\node[above] at (0,2.5) {$\scriptstyle \zeta^{\bullet}$};
        \end{tikzpicture}\end{array}$
        & $\begin{array}{c}\begin{tikzpicture}[scale=0.4]
        	\draw[fill=\graphcol, \graphcol] (0,0) -- (2,0) -- (2,-2) -- cycle;
			\draw[-stealth] (-2.5,0) -- (2.5,0);
			\draw[-stealth] (0,-2.5) -- (0,2.5);
			\node[right] at (2.5,0) {$\scriptstyle \zeta^{\circ}$};
			\node[above] at (0,2.5) {$\scriptstyle \zeta^{\bullet}$};
        \end{tikzpicture}\end{array}$ \\
        \hline
    \end{tabular}
    \caption{Three solutions of case $s=0, r\not=0$.}
    \label{tab:solution s=0 d,d = 1,1}
\end{table}
    \begin{align}
    \left\{ \hspace{2mm}
    \begin{aligned}
        & -|a_1|^2 + |b_1|^2 + |b_2|^2 + |r|^2 = \zeta^{\circ} \\
        & |a_1|^2 - |b_1|^2 - |b_2|^2  = \zeta^{\bullet} \\
        & (\omega_2 - \epsilon_1) a_1 = 0 \\
        & (\omega_2 + \epsilon_2) b_1 = 0 \\
        & (\omega_2 - \epsilon_2) b_2 = 0
    \end{aligned}
\right.
    \end{align}
    has three solutions sketched in tab.\ref{tab:solution s=0 d,d = 1,1}.
\subsubsection{Case $s \not=0, r = 0$}
Immediately we have $a_2 = 0, \omega_2 = -\epsilon_1$. The remaining system:
    \begin{align}
\left\{ \hspace{2mm}
    \begin{aligned}
        & -|a_1|^2 + |b_1|^2 + |b_2|^2 = \zeta^{\circ} \\
        & |a_1|^2- |b_1|^2 - |b_2|^2 - |s|^2 = \zeta^{\bullet} \\
        & (\omega_1 + 2\epsilon_1) a_1 = 0 \\
        & (\omega_1 - \epsilon_2 + \epsilon_1) b_1 = 0 \\
        & (\omega_1 + \epsilon_2 + \epsilon_1) b_2 = 0
    \end{aligned}
\right.
\end{align}

\begin{table}[h!]
    \centering
    \begin{tabular}{|c|c|c|c|}
    \hline
        $\omega_1$ & $-2\epsilon_1$ & $\epsilon_2 - \epsilon_1$ & $-\epsilon_2-\epsilon_1$ \\
    \hline
        $\omega_2$ & $-\epsilon_1$ & $-\epsilon_1$ & $-\epsilon_1$ \\
        \hline
        \rotatebox[origin=c]{90}{Atomic struct.}
        &
        $\begin{array}{c}
        \begin{tikzpicture}[scale=0.4]
			\draw[thick, postaction={decorate},decoration={markings,
				mark= at position 0.7 with {\arrow{stealth}}}] (-2,0) -- (-1,0);
                \draw[-stealth,thin] (-2.8,0) -- (0.8,0);
                \draw[-stealth,thin] (0,-1.8) -- (0,1.8);
                \node[right] at (0.8,0) {$\scriptstyle \epsilon_1$};
			\node[above] at (0,1.8) {$\scriptstyle \epsilon_2$};
			\draw[fill=gray] (-1,0) circle (0.2);
			\draw[fill=white] (-2,0) circle (0.2);
			\draw[\myblue] (-1,0) circle (0.4);
	\end{tikzpicture} \end{array}$&
	$\begin{array}{c}
		\begin{tikzpicture}[scale=0.4]
			\draw[thick, postaction={decorate},decoration={markings,
				mark= at position 0.7 with {\arrow{stealth}}}] (-1,0) -- (-1,1);
			\draw[-stealth,thin] (-2.8,0) -- (0.8,0);
			\draw[-stealth,thin] (0,-1.8) -- (0,1.8);
			\node[right] at (0.8,0) {$\scriptstyle \epsilon_1$};
			\node[above] at (0,1.8) {$\scriptstyle \epsilon_2$};
			\draw[fill=white] (-1,1) circle (0.2);
			\draw[fill=gray] (-1,0) circle (0.2);
			\draw[\myblue] (-1,0) circle (0.4);
	\end{tikzpicture} \end{array}$
	&
	 $\begin{array}{c}
	 	\begin{tikzpicture}[scale=0.4]
	 		\draw[thick, postaction={decorate},decoration={markings,
	 			mark= at position 0.7 with {\arrow{stealth}}}] (-1,0) -- (-1,-1);
	 		\draw[-stealth,thin] (-2.8,0) -- (0.8,0);
	 		\draw[-stealth,thin] (0,-1.8) -- (0,1.8);
	 		\node[right] at (0.8,0) {$\scriptstyle \epsilon_1$};
	 		\node[above] at (0,1.8) {$\scriptstyle \epsilon_2$};
	 		\draw[fill=white] (-1,-1) circle (0.2);
	 		\draw[fill=gray] (-1,0) circle (0.2);
	 		\draw[\myblue] (-1,0) circle (0.4);
	 \end{tikzpicture}\end{array}$
		\\
        \hline
        \rotatebox[origin=c]{90}{Solution}
        &
        $\begin{aligned}
            &|a_1|^2 = -\zeta^{\circ} \\
            &b_1 = 0 \\
            &b_2 = 0 \\
            &|s|^2 = - \zeta^{\circ} - \zeta^{\bullet}
        \end{aligned}$
        &
        $\begin{aligned}
            &a_1 = 0 \\
            &|b_1|^2 = \zeta^{\circ} \\
            &b_2 = 0 \\
            &|s|^2 = -\zeta^{\circ} - \zeta^{\bullet}
        \end{aligned}$
        &
        $\begin{aligned}
            &a_1 = 0 \\
            &b_1 = 0 \\
            &|b_2|^2 = \zeta^{\circ} \\
            &|s|^2 = - \zeta^{\circ} - \zeta^{\bullet}
        \end{aligned}$
        \\
        \hline
        \rotatebox[origin=c]{90}{Constr.}
        &
        $\left\{ \begin{aligned}
             \zeta^{\circ} & < 0 \\
             \zeta^{\circ} + \zeta^{\bullet} & <  0
        \end{aligned}
        \right. $
        &
        $\left\{ \begin{aligned}
             \zeta^{\circ} &> 0 \\
             \zeta^{\circ} + \zeta^{\bullet} &< 0
        \end{aligned}
        \right. $
        &
        $\left\{ \begin{aligned}
             \zeta^{\circ} &> 0 \\
             \zeta^{\circ} + \zeta^{\bullet} &< 0
        \end{aligned}
        \right. $ \\
        \hline
        \rotatebox[origin=c]{90}{Phase region}
        &
        $\begin{array}{c}\begin{tikzpicture}[scale=0.4]
        	\draw[fill=\graphcol, \graphcol] (0,0) -- (-2,2) -- (-2,-2) -- (0,-2) -- cycle;
			\draw[-stealth] (-2.5,0) -- (2.5,0);
			\draw[-stealth] (0,-2.5) -- (0,2.5);
			\node[right] at (2.5,0) {$\scriptstyle \zeta^{\circ}$};
			\node[above] at (0,2.5) {$\scriptstyle \zeta^{\bullet}$};
        \end{tikzpicture} \end{array}$
        &
        $\begin{array}{c}\begin{tikzpicture}[scale=0.4]
        	\draw[fill=\graphcol, \graphcol] (0,0) -- (0,-2) -- (2,-2) -- cycle;
			\draw[-stealth] (-2.5,0) -- (2.5,0);
			\draw[-stealth] (0,-2.5) -- (0,2.5);
			\node[right] at (2.5,0) {$\scriptstyle \zeta^{\circ}$};
			\node[above] at (0,2.5) {$\scriptstyle \zeta^{\bullet}$};
        \end{tikzpicture}\end{array}$
        &
        $\begin{array}{c}\begin{tikzpicture}[scale=0.4]
        	\draw[fill=\graphcol, \graphcol] (0,0) -- (0,-2) -- (2,-2) -- cycle;
			\draw[-stealth] (-2.5,0) -- (2.5,0);
			\draw[-stealth] (0,-2.5) -- (0,2.5);
			\node[right] at (2.5,0) {$\scriptstyle \zeta^{\circ}$};
			\node[above] at (0,2.5) {$\scriptstyle \zeta^{\bullet}$};
        \end{tikzpicture} \end{array}$\\
        \hline
    \end{tabular}
    \caption{Three solutions of case $r=0, s\not=0$.}
    \label{tab:solution r=0 d,d = 1,1}
\end{table}
has three solutions sketched in tab.\ref{tab:solution r=0 d,d = 1,1}.
Comparing with solutions from tab.\ref{tab:solution s=0 d,d = 1,1} we observe they are related by the $\IZ_2$-symmetry.

\subsection{$d^{\circ} = 2, d^{\bullet} = 1$}
We consider the following ansatz:
\begin{equation}
    A_{1} = \begin{pmatrix}
        a_1 & a_2
    \end{pmatrix}, \hspace{5mm}
    A_{2} = \begin{pmatrix}
        a_3 & a_4
    \end{pmatrix}, \hspace{5mm}
    B_1 = \begin{pmatrix}
        b_1 \\
        b_2
    \end{pmatrix}, \hspace{5mm}
    B_2 = \begin{pmatrix}
        b_3 \\
        b_4
    \end{pmatrix}, \hspace{5mm}
    R = \begin{pmatrix}
        r_1 \\
        r_2
    \end{pmatrix}, \hspace{5mm}
    S = \begin{pmatrix}
        s
    \end{pmatrix}
\end{equation}
\begin{equation}
    \Phi^{\circ} = \begin{pmatrix}
        \omega_1 & 0 \\
        0 & \omega_2
    \end{pmatrix}, \hspace{5mm}
    \Phi^{\bullet} = \begin{pmatrix}
        \omega_2
    \end{pmatrix}
\end{equation}

\begin{table}[h!]
    \centering
    \scalebox{0.8}{\begin{tabular}{|c|c|c|c|c|c|c|}
    \hline
        $\omega_1$
        & $2\epsilon_1$
        & $\epsilon_1 + \epsilon_2$
        & $\epsilon_1 - \epsilon_2$
        & $2 \epsilon_2$
        & $-2 \epsilon_2$
        & $-\epsilon_1 +  \epsilon_2$
        \\
    \hline
        $\omega_2$
        & $0$
        & $0$
        & $0$
        & $0$
        & $0$
        & $-\epsilon_1 - \epsilon_2$
        \\
        \hline
        $\omega_3$
        & $\epsilon_1$
        & $\epsilon_1$
        & $\epsilon_1$
        & $\epsilon_2$
        & $-\epsilon_2$
        & $-\epsilon_1$
        \\
        \hline
        \rotatebox[origin=c]{90}{Atomic struct.}
        &
        $\begin{array}{c}
             \begin{tikzpicture}[scale=0.4]
			\draw[thick, postaction={decorate},decoration={markings,
			mark= at position 0.7 with {\arrow{stealth}}}] (1,0) -- (2,0);
		\draw[thick, postaction={decorate},decoration={markings,
			mark= at position 0.7 with {\arrow{stealth}}}] (0,0) -- (1,0);
                \draw[-stealth] (-1.8,0) -- (2.8,0);
                \draw[-stealth] (0,-2.8) -- (0,2.8);
                \node[right] at (2.8,0) {$\scriptstyle \epsilon_1$};
			\node[above] at (0,2.8) {$\scriptstyle \epsilon_2$};
			\draw[fill=gray] (1,0) circle (0.2);
			\draw[fill=white] (0,0) circle (0.2);
                \draw[fill=white] (2,0) circle (0.2);
                \draw[burgundy] (0,0) circle (0.5);
	\end{tikzpicture}
        \end{array}$
        &
        $\begin{array}{c}
             \begin{tikzpicture}[scale=0.4]
			\draw[thick, postaction={decorate},decoration={markings,
			mark= at position 0.7 with {\arrow{stealth}}}] (1,0) -- (1,1);
		\draw[thick, postaction={decorate},decoration={markings,
			mark= at position 0.7 with {\arrow{stealth}}}] (0,0) -- (1,0);
                \draw[-stealth] (-1.8,0) -- (1.8,0);
                \draw[-stealth] (0,-2.8) -- (0,2.8);
                \node[right] at (1.8,0) {$\scriptstyle \epsilon_1$};
			\node[above] at (0,2.8) {$\scriptstyle \epsilon_2$};
			\draw[fill=gray] (1,0) circle (0.2);
			\draw[fill=white] (0,0) circle (0.2);
                \draw[fill=white] (1,1) circle (0.2);
                \draw[burgundy] (0,0) circle (0.5);
	    \end{tikzpicture}
        \end{array}$
        &
        $\begin{array}{c}
             \begin{tikzpicture}[scale=0.4]
			\draw[thick, postaction={decorate},decoration={markings,
			mark= at position 0.7 with {\arrow{stealth}}}] (1,0) -- (1,-1);
		\draw[thick, postaction={decorate},decoration={markings,
			mark= at position 0.7 with {\arrow{stealth}}}] (0,0) -- (1,0);
                \draw[-stealth] (-1.8,0) -- (1.8,0);
                \draw[-stealth] (0,-2.8) -- (0,2.8);
                \node[right] at (1.8,0) {$\scriptstyle \epsilon_1$};
			\node[above] at (0,2.8) {$\scriptstyle \epsilon_2$};
			\draw[fill=gray] (1,0) circle (0.2);
			\draw[fill=white] (0,0) circle (0.2);
                \draw[fill=white] (1,-1) circle (0.2);
                \draw[burgundy] (0,0) circle (0.5);
	\end{tikzpicture}
        \end{array}$
        &
        $\begin{array}{c}
	\begin{tikzpicture}[scale=0.4]
		\draw[thick, postaction={decorate},decoration={markings,
			mark= at position 0.7 with {\arrow{stealth}}}] (0,1) -- (0,0);
		\draw[thick, postaction={decorate},decoration={markings,
			mark= at position 0.7 with {\arrow{stealth}}}] (0,1) -- (0,2);
		\draw[-stealth,thin] (-1.8,0) -- (1.8,0);
		\draw[-stealth,thin] (0,-2.8) -- (0,2.8);
		\node[right] at (1.8,0) {$\scriptstyle \epsilon_1$};
		\node[above] at (0,2.8) {$\scriptstyle \epsilon_2$};
		\draw[fill=gray] (0,1) circle (0.2);
		\draw[fill=white] (0,0) circle (0.2);
		\draw[fill=white] (0,2) circle (0.2);
		\draw[burgundy] (0,0) circle (0.5);
	\end{tikzpicture}\end{array}$
        &
        $\begin{array}{c}
	\begin{tikzpicture}[scale=0.4]
		\draw[thick, postaction={decorate},decoration={markings,
			mark= at position 0.7 with {\arrow{stealth}}}] (0,-1) -- (0,0);
		\draw[thick, postaction={decorate},decoration={markings,
			mark= at position 0.7 with {\arrow{stealth}}}] (0,-1) -- (0,-2);
		\draw[-stealth,thin] (-1.8,0) -- (1.8,0);
		\draw[-stealth,thin] (0,-2.8) -- (0,2.8);
		\node[right] at (1.8,0) {$\scriptstyle \epsilon_1$};
		\node[above] at (0,2.8) {$\scriptstyle \epsilon_2$};
		\draw[fill=gray] (0,-1) circle (0.2);
		\draw[fill=white] (0,0) circle (0.2);
		\draw[fill=white] (0,-2) circle (0.2);
		\draw[burgundy] (0,0) circle (0.5);
	\end{tikzpicture}\end{array}$
        &
        $\begin{array}{c}
	\begin{tikzpicture}[scale=0.4]
		\draw[thick, postaction={decorate},decoration={markings,
			mark= at position 0.7 with {\arrow{stealth}}}] (-1,0) -- (-1,-1);
		\draw[thick, postaction={decorate},decoration={markings,
			mark= at position 0.7 with {\arrow{stealth}}}] (-1,0) -- (-1,1);
		\draw[-stealth,thin] (-1.8,0) -- (1.8,0);
		\draw[-stealth,thin] (0,-2.8) -- (0,2.8);
		\node[right] at (1.8,0) {$\scriptstyle \epsilon_1$};
		\node[above] at (0,2.8) {$\scriptstyle \epsilon_2$};
		\draw[fill=gray] (-1,0) circle (0.2);
		\draw[fill=white] (-1,1) circle (0.2);
		\draw[fill=white] (-1,-1) circle (0.2);
		\draw[\myblue] (-1,0) circle (0.5);
	\end{tikzpicture}\end{array}$
        \\
        \hline
        \rotatebox[origin=c]{90}{Solution}
        &
        $\begin{aligned}
            &a_1 = a_4 = b_4 = 0 \\
            &b_1 = b_2 = b_3 = 0 \\
            &r_1 = s = 0\\
            &|a_3|^2 = -\zeta^{\circ} \\
            &|a_2|^2 =  \zeta^{\circ} + \zeta^{\bullet}\\
            &|r_2|^2 = 2 \zeta^{\circ} + \zeta^{\bullet}
        \end{aligned}$
        &
        $\begin{aligned}
            &a_1 = a_3 = a_4 = 0 \\
            &b_2 = b_3 = b_4 = 0 \\
            &r_1 = s = 0\\
            &|b_1|^2 = \zeta^{\circ} \\
            &|a_2|^2 =  \zeta^{\circ} + \zeta^{\bullet}\\
            &|r_2|^2 = 2 \zeta^{\circ} + \zeta^{\bullet}
        \end{aligned}$
        &
        $\begin{aligned}
            &a_1 = a_3 = a_4 = 0 \\
            &b_1 = b_2 = b_4 = 0 \\
            &r_1 = s = 0\\
            &|b_3|^2 = \zeta^{\circ} \\
            &|a_2|^2 =  \zeta^{\circ} + \zeta^{\bullet}\\
            &|r_2|^2 = 2 \zeta^{\circ} + \zeta^{\bullet}
        \end{aligned}$
        &
        $\begin{aligned}
            &a_1 = a_4 = a_3 = 0 \\
            &b_1 = b_4 = a_4 = 0 \\
            &r_2 = s = 0\\
            &|b_2|^2 = \zeta^{\circ} \\
            &|b_3|^2 =  -\zeta^{\circ} - \zeta^{\bullet}\\
            &|r_1|^2 = 2 \zeta^{\circ} + \zeta^{\bullet}
        \end{aligned}$
        &
        $\begin{aligned}
            &a_1 = a_4 = a_3 = 0 \\
            &b_1 = b_4 = a_4 = 0 \\
            &r_2 = s = 0\\
            &|b_2|^2 = \zeta^{\circ} \\
            &|b_3|^2 =  -\zeta^{\circ} - \zeta^{\bullet}\\
            &|r_1|^2 = 2 \zeta^{\circ} + \zeta^{\bullet}
        \end{aligned}$
        &
        $\begin{aligned}
            &a_1 = a_4 = a_3 = 0 \\
            &b_2 = b_3 = a_4 = 0 \\
            &r_1 = r_2 = 0\\
            &|b_1|^2 = \zeta^{\circ} \\
            &|b_4|^2 = \zeta^{\circ} \\
            &|s|^2 = -2 \zeta^{\circ} - \zeta^{\bullet}
        \end{aligned}$
        \\
        \hline
        \rotatebox[origin=c]{90}{Consrt.}
        &
        $\left\{ \begin{aligned}
             \zeta^{\circ} & < 0 \\
             \zeta^{\circ} + \zeta^{\bullet} & >  0 \\
             2\zeta^{\circ} + \zeta^{\bullet} & >  0
        \end{aligned}
        \right. $
        &
        $\left\{ \begin{aligned}
             \zeta^{\circ} & > 0 \\
             \zeta^{\circ} + \zeta^{\bullet} & >  0 \\
             2\zeta^{\circ} + \zeta^{\bullet} & >  0
        \end{aligned}
        \right. $
        &
        $\left\{ \begin{aligned}
             \zeta^{\circ} & > 0 \\
             \zeta^{\circ} + \zeta^{\bullet} & >  0 \\
             2\zeta^{\circ} + \zeta^{\bullet} & >  0
        \end{aligned}
        \right. $
        &
        $\left\{ \begin{aligned}
             \zeta^{\circ} & > 0 \\
             \zeta^{\circ} + \zeta^{\bullet} & <  0 \\
             2\zeta^{\circ} + \zeta^{\bullet} & >  0
        \end{aligned}
        \right. $
        &
        $\left\{ \begin{aligned}
             \zeta^{\circ} & > 0 \\
             \zeta^{\circ} + \zeta^{\bullet} & <  0 \\
             2\zeta^{\circ} + \zeta^{\bullet} & >  0
        \end{aligned}
        \right. $
        &
        $\left\{ \begin{aligned}
             \zeta^{\circ} & > 0 \\
             2\zeta^{\circ} + \zeta^{\bullet} & <  0
        \end{aligned}
        \right. $
        \\
        \hline
        \rotatebox[origin=c]{90}{Phase region}
        &
        $\begin{array}{c}
             \begin{tikzpicture}[scale=0.4]
            \draw[fill=\graphcol, \graphcol] (0,0) -- (-1,2) -- (0,2) -- cycle;
			\draw[-stealth] (-2.5,0) -- (2.5,0);
			\draw[-stealth] (0,-2.5) -- (0,2.5);
			\node[right] at (2.5,0) {$\scriptstyle \zeta^{\circ}$};
			\node[above] at (0,2.5) {$\scriptstyle \zeta^{\bullet}$};
            \end{tikzpicture}
        \end{array}$
        &
        $\begin{array}{c}
             \begin{tikzpicture}[scale=0.4]
            \draw[fill=\graphcol, \graphcol] (0,0) -- (0,2) -- (2,2) -- (2,-2) -- cycle;
			\draw[-stealth] (-2.5,0) -- (2.5,0);
			\draw[-stealth] (0,-2.5) -- (0,2.5);
			\node[right] at (2.5,0) {$\scriptstyle \zeta^{\circ}$};
			\node[above] at (0,2.5) {$\scriptstyle \zeta^{\bullet}$};
        \end{tikzpicture}
        \end{array}$
        &
        $\begin{array}{c}
             \begin{tikzpicture}[scale=0.4]
            \draw[fill=\graphcol, \graphcol] (0,0) -- (0,2) -- (2,2) -- (2,-2) -- cycle;
			\draw[-stealth] (-2.5,0) -- (2.5,0);
			\draw[-stealth] (0,-2.5) -- (0,2.5);
			\node[right] at (2.5,0) {$\scriptstyle \zeta^{\circ}$};
			\node[above] at (0,2.5) {$\scriptstyle \zeta^{\bullet}$};
        \end{tikzpicture}
        \end{array}$
        &
        $\begin{array}{c}
             \begin{tikzpicture}[scale=0.4]
            \draw[fill=\graphcol, \graphcol] (0,0) -- (2,-2) -- (1,-2) -- cycle;
			\draw[-stealth] (-2.5,0) -- (2.5,0);
			\draw[-stealth] (0,-2.5) -- (0,2.5);
			\node[right] at (2.5,0) {$\scriptstyle \zeta^{\circ}$};
			\node[above] at (0,2.5) {$\scriptstyle \zeta^{\bullet}$};
        \end{tikzpicture}
        \end{array}$
        &
        $\begin{array}{c}
             \begin{tikzpicture}[scale=0.4]
            \draw[fill=\graphcol, \graphcol] (0,0) -- (2,-2) -- (1,-2) -- cycle;
			\draw[-stealth] (-2.5,0) -- (2.5,0);
			\draw[-stealth] (0,-2.5) -- (0,2.5);
			\node[right] at (2.5,0) {$\scriptstyle \zeta^{\circ}$};
			\node[above] at (0,2.5) {$\scriptstyle \zeta^{\bullet}$};
        \end{tikzpicture}
        \end{array}$
        &
        $\begin{array}{c}
             \begin{tikzpicture}[scale=0.4]
            \draw[fill=\graphcol, \graphcol] (0,0) -- (1,-2) -- (0,-2) -- cycle;
			\draw[-stealth] (-2.5,0) -- (2.5,0);
			\draw[-stealth] (0,-2.5) -- (0,2.5);
			\node[right] at (2.5,0) {$\scriptstyle \zeta^{\circ}$};
			\node[above] at (0,2.5) {$\scriptstyle \zeta^{\bullet}$};
        \end{tikzpicture}
        \end{array}$
        \\
        \hline
    \end{tabular}}
    \caption{Solutions for $d^{\circ}=2, d^{\bullet}=1$.}
    \label{tab:solution s=0 d,d = 2,1}
\end{table}
Solutions are summarized in tab.\ref{tab:solution s=0 d,d = 2,1}.

\subsection{$d^{\circ} = 1, d^{\bullet} = 2$}

We consider the following ansatz:
\begin{equation}
    A_{1} = \begin{pmatrix}
        a_1 \\
        a_2
    \end{pmatrix}, \hspace{5mm}
    A_{2} = \begin{pmatrix}
        a_3 \\
        a_4
    \end{pmatrix}, \hspace{5mm}
    B_1 = \begin{pmatrix}
        b_1 & b_2
    \end{pmatrix}, \hspace{5mm}
    B_2 = \begin{pmatrix}
        b_3 & b_4
    \end{pmatrix}, \hspace{5mm}
    R = \begin{pmatrix}
        r_1
    \end{pmatrix}, \hspace{5mm}
    S = \begin{pmatrix}
        s_1 & s_2
    \end{pmatrix}
\end{equation}
\begin{equation}
    \Phi^{\circ} = \begin{pmatrix}
        \omega_1
    \end{pmatrix} , \hspace{5mm}
    \Phi^{\bullet} = \begin{pmatrix}
        \omega_2 & 0 \\
        0 & \omega_3
    \end{pmatrix}
\end{equation}

\begin{table}[h!]
    \centering
    %\rotatebox[origin=c]{-90}{
    \scalebox{0.8}{\begin{tabular}{|c|c|c|c|c|c|c|}
    \hline
        $\omega_1$
        & $-2\epsilon_1$
        & $-2\epsilon_1$
        & $-2\epsilon_1$
        & $\epsilon_2-\epsilon_1$
        & $-\epsilon_2-\epsilon_1$
        & $0$
        \\
    \hline
        $\omega_2$
        & $-3\epsilon_1$
        & $-2\epsilon_1 + \epsilon_2$
        & $-2\epsilon_1 - \epsilon_2$
        & $-\epsilon_1 +2 \epsilon_2$
        & $-\epsilon_1 -2 \epsilon_2$
        & $\epsilon_2$
        \\
        \hline
        $\omega_3$
        & $-\epsilon_1$
        & $-\epsilon_1$
        & $-\epsilon_1$
        & $-\epsilon_1$
        & $-\epsilon_1$
        & $-\epsilon_2$
        \\
        \hline
        \rotatebox[origin=c]{90}{Atomic struct.}
        &
        $\begin{array}{c}
        \begin{tikzpicture}[scale=0.4]
			\draw[thick, postaction={decorate},decoration={markings,
				mark= at position 0.7 with {\arrow{stealth}}}] (-2,0) -- (-1,0);
            \draw[thick, postaction={decorate},decoration={markings,
            	mark= at position 0.7 with {\arrow{stealth}}}] (-2,0) -- (-3,0);
            \draw[-stealth,thin] (-3.8,0) -- (1.8,0);
            \draw[-stealth,thin] (0,-2.8) -- (0,2.8);
            \node[right] at (1.8,0) {$\scriptstyle \epsilon_1$};
			\node[above] at (0,2.8) {$\scriptstyle \epsilon_2$};
			\draw[fill=gray] (-1,0) circle (0.2);
			\draw[fill=white] (-2,0) circle (0.2);
            \draw[fill=gray] (-3,0) circle (0.2);
            \draw[\myblue] (-1,0) circle (0.5);
	\end{tikzpicture} \end{array}$
        &
        $\begin{array}{c}
        \begin{tikzpicture}[scale=0.4]
			\draw[thick, postaction={decorate},decoration={markings,
				mark= at position 0.7 with {\arrow{stealth}}}] (-2,0) -- (-1,0);
            \draw[thick, postaction={decorate},decoration={markings,
            	mark= at position 0.7 with {\arrow{stealth}}}] (-2,1) -- (-2,0);
            \draw[-stealth,thin] (-2.8,0) -- (1.8,0);
            \draw[-stealth,thin] (0,-2.8) -- (0,2.8);
            \node[right] at (1.8,0) {$\scriptstyle \epsilon_1$};
			\node[above] at (0,2.8) {$\scriptstyle \epsilon_2$};
			\draw[fill=gray] (-1,0) circle (0.2);
			\draw[fill=white] (-2,0) circle (0.2);
            \draw[fill=gray] (-2,1) circle (0.2);
            \draw[\myblue] (-1,0) circle (0.5);
	\end{tikzpicture}\end{array}$
        &
        $\begin{array}{c}
        \begin{tikzpicture}[scale=0.4]
			\draw[thick, postaction={decorate},decoration={markings,
				mark= at position 0.7 with {\arrow{stealth}}}] (-2,0) -- (-1,0);
            \draw[thick, postaction={decorate},decoration={markings,
            	mark= at position 0.7 with {\arrow{stealth}}}] (-2,-1) -- (-2,0);
            \draw[-stealth,thin] (-2.8,0) -- (1.8,0);
            \draw[-stealth,thin] (0,-2.8) -- (0,2.8);
            \node[right] at (1.8,0) {$\scriptstyle \epsilon_1$};
			\node[above] at (0,2.8) {$\scriptstyle \epsilon_2$};
			\draw[fill=gray] (-1,0) circle (0.2);
			\draw[fill=white] (-2,0) circle (0.2);
            \draw[fill=gray] (-2,-1) circle (0.2);
            \draw[\myblue] (-1,0) circle (0.5);
	\end{tikzpicture}\end{array}$
	&
	$\begin{array}{c}
	\begin{tikzpicture}[scale=0.4]
		\draw[thick, postaction={decorate},decoration={markings,
			mark= at position 0.7 with {\arrow{stealth}}}] (-1,1) -- (-1,0);
		\draw[thick, postaction={decorate},decoration={markings,
			mark= at position 0.7 with {\arrow{stealth}}}] (-1,1) -- (-1,2);
		\draw[-stealth,thin] (-1.8,0) -- (1.8,0);
		\draw[-stealth,thin] (0,-2.8) -- (0,2.8);
		\node[right] at (1.8,0) {$\scriptstyle \epsilon_1$};
		\node[above] at (0,2.8) {$\scriptstyle \epsilon_2$};
		\draw[fill=gray] (-1,0) circle (0.2);
		\draw[fill=white] (-1,1) circle (0.2);
		\draw[fill=gray] (-1,2) circle (0.2);
		\draw[\myblue] (-1,0) circle (0.5);
	\end{tikzpicture} \end{array}$
	&
	$\begin{array}{c}
	\begin{tikzpicture}[scale=0.4]
		\draw[thick, postaction={decorate},decoration={markings,
			mark= at position 0.7 with {\arrow{stealth}}}] (-1,-1) -- (-1,0);
		\draw[thick, postaction={decorate},decoration={markings,
			mark= at position 0.7 with {\arrow{stealth}}}] (-1,-1) -- (-1,-2);
		\draw[-stealth,thin] (-1.8,0) -- (1.8,0);
		\draw[-stealth,thin] (0,-2.8) -- (0,2.8);
		\node[right] at (1.8,0) {$\scriptstyle \epsilon_1$};
		\node[above] at (0,2.8) {$\scriptstyle \epsilon_2$};
		\draw[fill=gray] (-1,0) circle (0.2);
		\draw[fill=white] (-1,-1) circle (0.2);
		\draw[fill=gray] (-1,-2) circle (0.2);
		\draw[\myblue] (-1,0) circle (0.5);
	\end{tikzpicture}\end{array}$
	&
	$\begin{array}{c}
	\begin{tikzpicture}[scale=0.4]
		\draw[thick, postaction={decorate},decoration={markings,
			mark= at position 0.7 with {\arrow{stealth}}}] (0,1) -- (0,0);
		\draw[thick, postaction={decorate},decoration={markings,
			mark= at position 0.7 with {\arrow{stealth}}}] (0,-1) -- (0,0);
		\draw[-stealth,thin] (-1.8,0) -- (1.8,0);
		\draw[-stealth,thin] (0,-2.8) -- (0,2.8);
		\node[right] at (1.8,0) {$\scriptstyle \epsilon_1$};
		\node[above] at (0,2.8) {$\scriptstyle \epsilon_2$};
		\draw[fill=gray] (0,1) circle (0.2);
		\draw[fill=white] (0,0) circle (0.2);
		\draw[fill=gray] (0,-1) circle (0.2);
		\draw[burgundy] (0,0) circle (0.5);
	\end{tikzpicture}\end{array}$
        \\
        \hline
        \rotatebox[origin=c]{90}{Solution}
        &
        $\begin{aligned}
            &a_1 = a_4 = b_4 = 0 \\
            &b_1 = b_2 = b_3 =  0 \\
            &s_1 = r = 0\\
            &|a_3|^2 = \zeta^{\bullet} \\
            &|a_2|^2 =  -\zeta^{\circ} - \zeta^{\bullet}\\
            &|s_2|^2 = -2 \zeta^{\bullet} - \zeta^{\circ}
        \end{aligned}$
        &
        $\begin{aligned}
            &a_1 = a_3 = a_4 = 0 \\
            &b_1 = b_2 = b_4 = 0 \\
            &r = s_1 = 0\\
            &|b_3|^2 = -\zeta^{\bullet} \\
            &|a_2|^2 =  -\zeta^{\circ} - \zeta^{\bullet}\\
            &|s_2|^2 = -2 \zeta^{\bullet} - \zeta^{\circ}
        \end{aligned}$
        &
        $\begin{aligned}
            &a_1 = a_3 = a_4 = 0 \\
            &b_2 = b_3 = b_4 = 0 \\
            &r = s_2 = 0\\
            &|b_1|^2 = -\zeta^{\bullet} \\
            &|a_2|^2 =  -\zeta^{\circ} - \zeta^{\bullet}\\
            &|s_2|^2 = - 2 \zeta^{\bullet} - \zeta^{\circ}
        \end{aligned}$
        &
        $\begin{aligned}
        	&a_1 = a_3 = a_4 = 0 \\
        	&b_1 = b_4 = a_2 = 0 \\
        	&r = s_1 = 0\\
        	&|b_3|^2 = -\zeta^{\bullet} \\
        	&|b_2|^2 =  \zeta^{\circ} + \zeta^{\bullet}\\
        	&|s_2|^2 = - 2 \zeta^{\bullet} - \zeta^{\circ}
        \end{aligned}$
        &
        $\begin{aligned}
        	&a_1 = a_3 = a_4 = 0 \\
        	&b_2 = b_3 = a_2 = 0 \\
        	&r = s_1 = 0\\
        	&|b_1|^2 = -\zeta^{\bullet} \\
        	&|b_4|^2 =  \zeta^{\circ} + \zeta^{\bullet}\\
        	&|s_2|^2 = - 2 \zeta^{\bullet} - \zeta^{\circ}
        \end{aligned}$
        &
        $\begin{aligned}
        	&a_1 = a_2 = a_3 = 0 \\
        	&b_1 = b_4 = a_4 =  0 \\
        	&s_1 = s_2 = 0\\
        	&|b_2|^2 = - \zeta^{\bullet} \\
        	&|b_3|^2 =  - \zeta^{\bullet}\\
        	&|r|^2 = 2 \zeta^{\bullet} + \zeta^{\circ}
        \end{aligned}$
        \\
        \hline
        \rotatebox[origin=c]{90}{Constr.}
        &
        $\left\{ \begin{aligned}
             \zeta^{\bullet} & >0 \\
             \zeta^{\circ} + \zeta^{\bullet} & <  0 \\
             2\zeta^{\bullet} + \zeta^{\circ} & <  0
        \end{aligned}
        \right. $
        &
        $\left\{ \begin{aligned}
             \zeta^{\bullet} & < 0 \\
             \zeta^{\circ} + \zeta^{\bullet} & <  0 \\
             2\zeta^{\circ} + \zeta^{\bullet} & <  0
        \end{aligned}
        \right. $
        &
        $\left\{ \begin{aligned}
             \zeta^{\bullet} & < 0 \\
             \zeta^{\circ} + \zeta^{\bullet} & <  0 \\
             2\zeta^{\circ} + \zeta^{\bullet} & <  0
        \end{aligned}
        \right. $
        &
        $\left\{ \begin{aligned}
        	\zeta^{\bullet} & < 0 \\
        	\zeta^{\circ} + \zeta^{\bullet} & >  0 \\
        	2\zeta^{\circ} + \zeta^{\bullet} & <  0
        \end{aligned}
        \right. $
        &
        $\left\{ \begin{aligned}
        	\zeta^{\bullet} & < 0 \\
        	\zeta^{\circ} + \zeta^{\bullet} & >  0 \\
        	2\zeta^{\circ} + \zeta^{\bullet} & <  0
        \end{aligned}
        \right. $
        &
        $\left\{ \begin{aligned}
        	\zeta^{\bullet} & < 0 \\
        	2\zeta^{\bullet} + \zeta^{\circ} & > 0
        \end{aligned}
        \right. $
        \\
        \hline
        \rotatebox[origin=c]{90}{Phase region}
        &
        $\begin{array}{c}
        \begin{tikzpicture}[scale=0.4]
        	\draw[fill=\graphcol, \graphcol] (0,0) -- (-2,1) -- (-2,0) -- cycle;
			\draw[-stealth] (-2.5,0) -- (2.5,0);
			\draw[-stealth] (0,-2.5) -- (0,2.5);
			\node[right] at (2.5,0) {$\scriptstyle \zeta^{\circ}$};
			\node[above] at (0,2.5) {$\scriptstyle \zeta^{\bullet}$};
        \end{tikzpicture}\end{array}$
        &
        $\begin{array}{c}
        \begin{tikzpicture}[scale=0.4]
        	\draw[fill=\graphcol, \graphcol] (0,0) -- (-2,0) -- (-2,-2) -- (2,-2) -- cycle;
			\draw[-stealth] (-2.5,0) -- (2.5,0);
			\draw[-stealth] (0,-2.5) -- (0,2.5);
			\node[right] at (2.5,0) {$\scriptstyle \zeta^{\circ}$};
			\node[above] at (0,2.5) {$\scriptstyle \zeta^{\bullet}$};
        \end{tikzpicture}\end{array}$
        &
        $\begin{array}{c}
        \begin{tikzpicture}[scale=0.4]
        	\draw[fill=\graphcol, \graphcol] (0,0) -- (-2,0) -- (-2,-2) -- (2,-2) -- cycle;
			\draw[-stealth] (-2.5,0) -- (2.5,0);
			\draw[-stealth] (0,-2.5) -- (0,2.5);
			\node[right] at (2.5,0) {$\scriptstyle \zeta^{\circ}$};
			\node[above] at (0,2.5) {$\scriptstyle \zeta^{\bullet}$};
        \end{tikzpicture}\end{array}$
        &
        $\begin{array}{c}
        \begin{tikzpicture}[scale=0.4]
        	\draw[fill=\graphcol, \graphcol] (0,0) -- (2,-1) -- (2,-2) -- cycle;
        	\draw[-stealth] (-2.5,0) -- (2.5,0);
        	\draw[-stealth] (0,-2.5) -- (0,2.5);
        	\node[right] at (2.5,0) {$\scriptstyle \zeta^{\circ}$};
        	\node[above] at (0,2.5) {$\scriptstyle \zeta^{\bullet}$};
        \end{tikzpicture}\end{array}$
        &
        $\begin{array}{c}
        \begin{tikzpicture}[scale=0.4]
        	\draw[fill=\graphcol, \graphcol] (0,0) -- (2,-1) -- (2,-2) -- cycle;
        	\draw[-stealth] (-2.5,0) -- (2.5,0);
        	\draw[-stealth] (0,-2.5) -- (0,2.5);
        	\node[right] at (2.5,0) {$\scriptstyle \zeta^{\circ}$};
        	\node[above] at (0,2.5) {$\scriptstyle \zeta^{\bullet}$};
        \end{tikzpicture}\end{array}$
        &
        $\begin{array}{c}
        \begin{tikzpicture}[scale=0.4]
        	\draw[fill=\graphcol, \graphcol] (0,0) -- (2,0) -- (2,-1) -- cycle;
        	\draw[-stealth] (-2.5,0) -- (2.5,0);
        	\draw[-stealth] (0,-2.5) -- (0,2.5);
        	\node[right] at (2.5,0) {$\scriptstyle \zeta^{\circ}$};
        	\node[above] at (0,2.5) {$\scriptstyle \zeta^{\bullet}$};
        \end{tikzpicture}\end{array}$
        \\
        \hline
    \end{tabular}}
    \caption{Solutions for $d^{\circ}=1, d^{\bullet}=2$.}
    \label{tab:solution s=0 d,d = 1,2}
\end{table}

Solutions are summarized in tab.\ref{tab:solution s=0 d,d = 1,2}.
Solutions described in this table are also related to solutions in tab.\ref{tab:solution s=0 d,d = 2,1} by the $\IZ_2$-symmetry.

\subsection{Polymer tail example} \label{sec:polymer_ex}

Consider a simple fixed point in a glass-$n$ phase given by the simplest super-Young diagram according to \eqref{diagram_rules}:
\begin{equation}
	\nhtile\quad \mapsto\quad
	\begin{array}{c}
		\begin{tikzpicture}[scale=0.8]
			\draw[postaction={decorate},
			decoration={markings, mark= at position 0.65 with {\arrow{stealth}}}] (0,0) -- (1,0) node[pos=0.5, above] {$\scriptstyle A_1$};
			\draw[postaction={decorate},
			decoration={markings, mark= at position 0.65 with {\arrow{stealth}}}] (2,0) -- (1,0) node[pos=0.5, above] {$\scriptstyle A_2$};
			\draw[dashed, postaction={decorate},
			decoration={markings, mark= at position 0.65 with {\arrow{stealth}}}] (2,0) -- (3,0) node[pos=0.5, above] {$\scriptstyle A_1$};
			\draw[postaction={decorate},
			decoration={markings, mark= at position 0.65 with {\arrow{stealth}}}] (4,0) -- (3,0) node[pos=0.5, above] {$\scriptstyle A_2$};
			\begin{scope}[shift={(1,0)}]
				\draw[\myblue, rounded corners] (-0.2,0.3) -- (-0.2,-0.3) -- (1.2,-0.3) -- (1.2,0.3) -- cycle;
			\end{scope}
			\begin{scope}[shift={(3,0)}]
				\draw[\myblue, rounded corners] (-0.2,0.3) -- (-0.2,-0.3) -- (1.2,-0.3) -- (1.2,0.3) -- cycle;
			\end{scope}
			\draw[\myblue, rounded corners] (0.8,0) -- (0.8,-0.5) -- (2.5,-0.5) -- (2.5,-0.7);
			\draw[\myblue, rounded corners] (4.2,0) -- (4.2,-0.5) -- (2.5,-0.5) -- (2.5,-0.7);
			\node[below] at (2.5,-0.7) {$n$ times};
			\draw[fill=white] (0,0) circle (0.1);
			\draw[fill=gray] (1,0) circle (0.1);
			\draw[fill=white] (2,0) circle (0.1);
			\draw[fill=gray] (3,0) circle (0.1);
			\draw[fill=white] (4,0) circle (0.1);
		\end{tikzpicture}
	\end{array}
\end{equation}
According to \eqref{diagram_rules} we find the following ansatz for the fixed point morphism values:
\begin{equation}
	\begin{aligned}
		R=\left(\begin{array}{c}
			\sqrt{u} \\ 0 \\ 0 \\ \vdots
		\end{array}\right)\in {\rm Mat}_{(n+1)\times 1}(\IC),\;A_1=\left(\begin{array}{ccccc}
			\sqrt{r_1} & 0 & 0 & 0 &\ldots\\
			0 & \sqrt{r_2} & 0 & 0 &\ldots\\
			0 & 0 & \sqrt{r_3} & 0 &\ldots\\
			\ldots
		\end{array}\right)\in {\rm Mat}_{n\times (n+1)}(\IC)\,,\\
		A_2=\left(\begin{array}{ccccc}
			0 &\sqrt{s_1} & 0 & 0 & \ldots\\
			0 &0 & \sqrt{s_2} & 0 & \ldots\\
			0 &0 & 0 & \sqrt{s_3} & \ldots\\
			\ldots
		\end{array}\right)\in {\rm Mat}_{n\times (n+1)}(\IC),\; B_{1,2}=0\in {\rm Mat}_{(n+1)\times n}(\IC),\; S=0\in {\rm Mat}_{n\times 1}(\IC)\,.
	\end{aligned}
\end{equation}
We assume that all the parameters $u$, $r_k$, $s_k$ have meanings of lengths, therefore they are all real and positive.

Substituting these ans\"atze in the vacuum equations:
\begin{equation}
	\begin{aligned}
		&R^{\dagger}R-A_1^{\dagger}A_1-A_2^{\dagger}A_2=\zeta_1\bbone_{(n+1)\times(n+1)}\,,\\\
		&A_1^{\dagger}A_1+A_2^{\dagger}A_2=\zeta_2\bbone_{n\times n}\,,
	\end{aligned}
\end{equation}
we find a solution for parameters $u$, $r_k$, $s_k$:
\begin{equation}
	\begin{aligned}
		& u=(n+1)\zeta_1+n\zeta_2\,,\\
		& r_k=(n+1-k)(\zeta_1+\zeta_2),\quad k=1,\ldots,n\,,\\
		& s_k=-(n+1-k)\zeta_1-(n-k)\zeta_2,\quad k=1,\ldots,n\,.
	\end{aligned}
\end{equation}
It is easy to deduce that all the parameters are positive only in the following chamber:
\begin{equation}
	\zeta_1<0,\quad \frac{n+1}{n}(-\zeta_1)<\zeta_2<\frac{n}{n-1}(-\zeta_1)\,,
\end{equation}
which corresponds to the simple glass-$n$ phase.

\subsection{Example with a cycle}\label{sec:ex_cycle}

Next we consider an example of an atomic structure plot with a cycle:
\begin{equation}\label{diag_w_cyc}
	\begin{array}{c}
		\begin{tikzpicture}
			\draw[-stealth] (0,0) -- (-0.3,-0.3) node[below left] {$\scriptstyle B_2$};
			\draw[-stealth] (0,0) -- (0.3,-0.3) node[below right] {$\scriptstyle A_1$};
			\draw[-stealth] (0,0) -- (0.3,0.3) node[above right] {$\scriptstyle B_1$};
			\draw[-stealth] (0,0) -- (-0.3,0.3) node[above left] {$\scriptstyle A_2$};
		\end{tikzpicture}
	\end{array}
	\quad
	\begin{array}{c}
		\begin{tikzpicture}[rotate=-45, scale=0.4]
			\foreach \x/\y/\z/\w in {0/0/1/0, 1/-1/2/-1, 1/0/1/-1, 1/0/1/1, 1/1/2/1, 2/-1/2/0, 2/1/2/0}
			{
				\draw[postaction={decorate},decoration={markings,
					mark= at position 0.6 with {\arrow{stealth}}}] (\x,\y) -- (\z,\w);
			}
			\foreach \x/\y in {0/0, 1/-1, 1/1, 2/0}
			{
				\draw[fill=white] (\x,\y) circle (0.2);
			}
			\foreach \x/\y in {1/0, 2/-1, 2/1}
			{
				\draw[fill=gray] (\x,\y) circle (0.2);
			}
			\draw[thick, burgundy] (0,0) circle (0.5);
		\end{tikzpicture}
	\end{array}\quad\mbox{\color{black!30!orange} a) }
	\begin{array}{c}
		\begin{tikzpicture}
			\begin{scope}[scale = 0.3]
				\foreach \x/\y/\z/\w in {0/0/1/0, 0/0/0/-1, 0/-1/1/-1, 1/0/1/-1, 0/-1/0/-2, 0/-2/1/-2, 1/-1/1/-2, 1/0/2/0, 1/-1/2/-1, 2/0/2/-1, 1/-2/2/-1}
				{
					\draw[thick] (\x,\y) -- (\z,\w);
				}
			\end{scope}
		\end{tikzpicture}
	\end{array}\quad\mbox{\color{\myblue} b) }
	\begin{array}{c}
		\begin{tikzpicture}
			\begin{scope}[scale = 0.3]
				\foreach \x/\y/\z/\w in {0/0/1/0, 0/0/0/-1, 0/-1/1/-1, 1/0/1/-1, 0/-1/0/-2, 0/-2/1/-2, 1/-1/1/-2, 1/0/2/0, 1/-1/2/-1, 2/0/2/-1, 1/-2/2/-2, 2/-1/2/-2, 0/-2/0/-3, 0/-3/1/-2, 2/0/3/0, 2/-1/3/0}
				{
					\draw[thick] (\x,\y) -- (\z,\w);
				}
				\draw[fill=burgundy] (0,0) -- (1,0) -- (0,-1) -- cycle;
			\end{scope}
		\end{tikzpicture}
	\end{array}\quad \begin{array}{c}
		\begin{tikzpicture}
			\draw[fill=\graphcol, \graphcol] (0,0) -- (1,0) -- (1,1) -- (0,1) -- cycle;
			\draw[fill=white!40!blue, white!40!blue] (0,0) -- (1,0) -- (1,-0.5) -- cycle;
			\node[right, black!30!orange] at (1, 0.5) {\small a)};
			\node[right, \myblue] at (1, -0.3) {\small b)};
			\draw[-stealth] (-1,0) -- (1.2,0);
			\draw[-stealth] (0,-1) -- (0,1.2);
			\node[above] at (0,1.2) {$\scriptstyle \zeta^{\bullet}$};
			\node[right] at (1.2,0) {$\scriptstyle \zeta^{\circ}$};
		\end{tikzpicture}
	\end{array}\,.
\end{equation}

According to this plot we construct the following ansatz for quiver morphisms:
\begin{equation}
	A_1=\left(
	\begin{array}{cccc}
		x_2 & 0 & 0 & 0 \\
		0 & x_3 & 0 & 0 \\
		0 & 0 & x_4 & 0 \\
	\end{array}
	\right), \; A_2=0, \; B_1=\left(
	\begin{array}{ccc}
		0 & 0 & 0 \\
		0 & 0 & 0 \\
		x_5 & 0 & 0 \\
		0 & x_6 & 0 \\
	\end{array}
	\right),\; B_2=\left(
	\begin{array}{ccc}
		0 & 0 & 0 \\
		x_7 & 0 & 0 \\
		0 & 0 & 0 \\
		0 & 0 & x_8 \\
	\end{array}
	\right),\; R=\left(
	\begin{array}{c}
		x_1 \\
		0 \\
		0 \\
		0 \\
	\end{array}
	\right),\; S=0\,.
\end{equation}

The D-term equations relate absolute values of variables $x_i$.
Let us denote their squares as $y_i=|x_i|^2$.
Then D-term equations read:
\begin{equation}\label{eqs_001}
	y_1-y_2 =\zeta^{\circ},\; y_7-y_3=\zeta^{\circ},\; y_5-y_4=\zeta^{\circ},\; y_6+y_8=\zeta^{\circ},\; y_2-y_5-y_7=\zeta^{\bullet},\; y_3-y_6=\zeta^{\bullet},\;  y_4-y_8=\zeta^{\bullet}\,.
\end{equation}

In addition due to a cycle in plot \eqref{diag_w_cyc} we have a relation $x_3 x_6 x_7=x_4 x_5 x_8$ that we promote to a square:
\begin{equation}\label{eqs_002}
	y_3 y_6 y_7-y_4 y_5 y_8=0\,.
\end{equation}

We can solve linear equations \eqref{eqs_001} with respect to, say, $y_3$ explicitly,
\begin{equation}
	\begin{aligned}
		&y_1= 4 \zeta^{\circ}+3 \zeta^{\bullet},\; y_2= 3 \left(\zeta^{\circ}+\zeta^{\bullet}\right),\; y_4= \zeta^{\circ}+2 \zeta^{\bullet}-y_3,\; y_5=2 \left(\zeta^{\circ}+\zeta^{\bullet}\right)-y_3\,,\\
		&y_6= y_3-\zeta^{\bullet},\; y_7= \zeta^{\circ}+y_3,\; y_8= \zeta^{\circ}+\zeta^{\bullet}-y_3\,.
	\end{aligned}
\end{equation}
Substituting these relations into \eqref{eqs_002} we obtain a cubic equation for $y_3$ that factorizes:
\begin{equation}
	\left(2 y_3-\zeta^{\circ}-2 \zeta^{\bullet}\right) \left(y_3^2-\left(\zeta^{\circ}+2 \zeta^{\bullet}\right) y_3+2 \left(\zeta^{\circ}+\zeta^{\bullet}\right)^2\right)=0\,.
\end{equation}

Therefore we acquire three solution branches:
\begin{equation}
	\begin{aligned}
		&y_1=4 \zeta^{\circ}+3 \zeta^{\bullet},\; y_2= 3 \left(\zeta^{\circ}+\zeta^{\bullet}\right),\; y_3= \delta +\frac{1}{2} \left(\zeta^{\circ}+2 \zeta^{\bullet}\right),\; y_4= -\delta +\frac{\zeta^{\circ}}{2}+\zeta^{\bullet}\,,\\
		&y_5= -\delta +\frac{3 \zeta^{\circ}}{2}+\zeta^{\bullet},\; y_6= \delta +\frac{\zeta^{\circ}}{2},\; y_7= \delta +\frac{3 \zeta^{\circ}}{2}+\zeta^{\bullet},\; y_8= \frac{1}{2} \left(\zeta^{\circ}-2 \delta \right)\,,
	\end{aligned}
\end{equation}
where either $\delta=0$, or
\begin{equation}\label{discriminant}
	\delta^2=-4\left(\frac{3-\sqrt{2}}{2}\zeta^{\circ}+\zeta^{\bullet}\right)\left(\frac{3+\sqrt{2}}{2}\zeta^{\circ}+\zeta^{\bullet}\right)\,.
\end{equation}

Let us note that in both cases $\delta=0$ or $\delta\neq 0$ solutions $y_{3,4}$, $y_{5,7}$ and $y_{6,8}$ come in paired as $u+\delta$, $u-\delta$ for some $u$.
Then in any case we have a restriction following from $y_i\geq 0$ that all such $u\geq 0$.
Then we arrive to the following restriction sector where a solution might exist:
\begin{equation}\label{sector}
	\left.\begin{array}{l}
		4\zeta^{\circ}+3\zeta^{\bullet}\geq 0\\
		\zeta^{\circ}+\zeta^{\bullet}\geq 0\\
		\zeta^{\circ}+2\zeta^{\bullet}\geq 0\\
		3\zeta^{\circ}+2\zeta^{\bullet}\geq 0\\
		\zeta^{\circ}\geq 0
	\end{array}\right\} \quad\Longrightarrow\quad \begin{array}{l}
		\zeta^{\circ}\geq 0\\
		\zeta^{\circ}+2\zeta^{\bullet}\geq 0
	\end{array}\,.
\end{equation}
Let us note that under these circumstances both brackets in \eqref{discriminant} are non-negative, therefore $\delta^2\leq 0$ has no solution.
It follows that the only one solution $\delta=0$ survives in the sector defined in \eqref{sector}.

Sector \eqref{sector} is depicted in \eqref{diag_w_cyc} and it clearly overlaps with two phases in fig.\ref{fig:new_phases}: the cyclic (crystal) phase ( a) in \eqref{diag_w_cyc}) and the phase of crooked glasses ( b) in \eqref{diag_w_cyc}).
Let us note moreover that the same atomic structure plot is encrypted with two different super-Young diagrams in these two phases (one of this super-Young diagrams is skew (see sec.\ref{sec:crooked})).

\subsection{Plot on a cover}\label{sec:double_black}
Let us consider the following plot drawn on a two-fold cover of the weight plane:
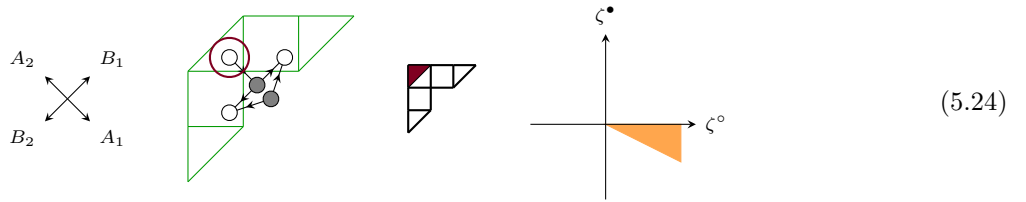
\begin{equation}\label{crooked_ex}
	\begin{array}{c}
		\begin{tikzpicture}
			\draw[-stealth] (0,0) -- (-0.3,-0.3) node[below left] {$\scriptstyle B_2$};
			\draw[-stealth] (0,0) -- (0.3,-0.3) node[below right] {$\scriptstyle A_1$};
			\draw[-stealth] (0,0) -- (0.3,0.3) node[above right] {$\scriptstyle B_1$};
			\draw[-stealth] (0,0) -- (-0.3,0.3) node[above left] {$\scriptstyle A_2$};
		\end{tikzpicture}
	\end{array}
	\quad
	\begin{array}{c}
		\begin{tikzpicture}[scale=1.3]
			\begin{scope}[shift = {(9,0)}]
				\begin{scope}[shift={(-0.141421,0.141421)}]
					\begin{scope}[scale=0.565685]
						\foreach \x/\y/\z/\w in {1/0/0/-1, 0/-1/1/-1, 1/0/1/-1, 0/-1/0/-2, 0/-2/1/-2, 1/-1/1/-2, 1/0/2/0, 1/-1/2/-1, 2/0/2/-1, 0/-2/0/-3, 0/-3/1/-2, 2/0/3/0, 2/-1/3/0}
						{
							\draw[thin, black!40!green] (\x,\y) -- (\z,\w);
						}
					\end{scope}
				\end{scope}
				\begin{scope}[rotate=-45, scale=0.4]
					\begin{scope}[shift={(1,0)}]
						\foreach \x/\y/\z/\w in {0/0/1/0, 1/0/1/-1, 1/0/1/1, 1.5/0/1/-1, 1.5/0/1/1}
						{
							\draw[postaction={decorate},decoration={markings,
								mark= at position 0.6 with {\arrow{stealth}}}] (\x,\y) -- (\z,\w);
						}
						\foreach \x/\y in {0/0, 1/-1, 1/1}
						{
							\draw[fill=white] (\x,\y) circle (0.2);
						}
						\foreach \x/\y in {1/0, 1.5/0}
						{
							\draw[fill=gray] (\x,\y) circle (0.2);
						}
						\draw[thick, burgundy] (0,0) circle (0.5);
					\end{scope}
				\end{scope}
			\end{scope}
		\end{tikzpicture}
	\end{array}\quad
	\begin{array}{c}
		\begin{tikzpicture}
			\begin{scope}[scale = 0.3]
				\foreach \x/\y/\z/\w in {0/0/1/0, 0/0/0/-1, 0/-1/1/-1, 1/0/1/-1, 0/-1/0/-2, 0/-2/1/-2, 1/-1/1/-2, 1/0/2/0, 1/-1/2/-1, 2/0/2/-1, 0/-2/0/-3, 0/-3/1/-2, 2/0/3/0, 2/-1/3/0}
				{
					\draw[thick] (\x,\y) -- (\z,\w);
				}
				\draw[fill=burgundy] (0,0) -- (1,0) -- (0,-1) -- cycle;
			\end{scope}
		\end{tikzpicture}
	\end{array}\quad
	\begin{array}{c}
		\begin{tikzpicture}
			\draw[fill=\graphcol, \graphcol] (0,0) -- (1,0) -- (1,-0.5) -- cycle;
			\draw[-stealth] (-1,0) -- (1.2,0);
			\draw[-stealth] (0,-1) -- (0,1.2);
			\node[above] at (0,1.2) {$\scriptstyle \zeta^{\bullet}$};
			\node[right] at (1.2,0) {$\scriptstyle \zeta^{\circ}$};
		\end{tikzpicture}
	\end{array}
\end{equation}
so that two black atoms coincide.

According to the plot we have the following matrix elements:
\begin{equation}
	A_1=\left(
	\begin{array}{ccc}
		x_2 & 0 & 0 \\
		0 & 0 & 0 \\
	\end{array}
	\right),\; A_2=0,\; B_1=\left(
	\begin{array}{cc}
		0 & 0 \\
		0 & 0 \\
		x_3 & x_4 \\
	\end{array}
	\right),\; B_2=\left(
	\begin{array}{cc}
		0 & 0 \\
		x_5 & x_6 \\
		0 & 0 \\
	\end{array}
	\right),\; R=\left(
	\begin{array}{c}
		x_1 \\
		0 \\
		0 \\
	\end{array}
	\right),\; S=0\,.
\end{equation}

The set of the vacuum equations impose the following constraints:
\begin{equation}\label{double_eqs}
	\begin{aligned}
		&|x_1|^2-|x_2|^2=\zeta^{\circ},\; |x_5|^2+|x_6|^2=\zeta^{\circ},\; |x_3|^2+|x_4|^2 =\zeta^{\circ}\,,\\
		&|x_2|^2-|x_3|^2-|x_5|^2=\zeta^{\bullet},\; -|x_4|^2-|x_6|^2=\zeta^{\bullet},\; -x_3 x_4-x_5 x_6=0 \,.
	\end{aligned}
\end{equation}
A single solution to these equations reads:
\begin{equation}
	|x_1|=\sqrt{3 \zeta^{\circ}+2 \zeta^{\bullet}},\; |x_2|= \sqrt{2 \zeta^{\circ}+2 \zeta^{\bullet}},\; |x_3|= \sqrt{ \zeta^{\circ}+\frac{\zeta^{\bullet}}{2}},\; |x_4|= \sqrt{-\frac{\zeta^{\bullet}}{2}},\; |x_5|= \sqrt{\zeta^{\circ}+\frac{\zeta^{\bullet}}{2}},\; |x_6|=\sqrt{-\frac{\zeta^{\bullet}}{2}}\,.
\end{equation}
Apparently, this solution exists in the following sector belonging to a set of crooked glass phases:
\begin{equation}
	\zeta^{\bullet}\leq 0,\quad 2\zeta^{\circ}+\zeta^{\bullet}\geq 0\,.
\end{equation}

Let us note that the last equation in \eqref{double_eqs} implies that one of four variables $x_{3,4,5,6}$ should have a minus sign phase to satisfy this equation.
This property is not modified by the $G_{\IC}$-flow we consider in sec.\ref{sec:integrals}.
Therefore as a point in the $G_{\IC}$-orbit we choose the following parameterization: $x_3=1$, $x_4=1$, $x_5=1$, $x_6=-1$.

\subsection{$S$-dominated solution in the upper left half-plane}
Consider the following atomic structure:
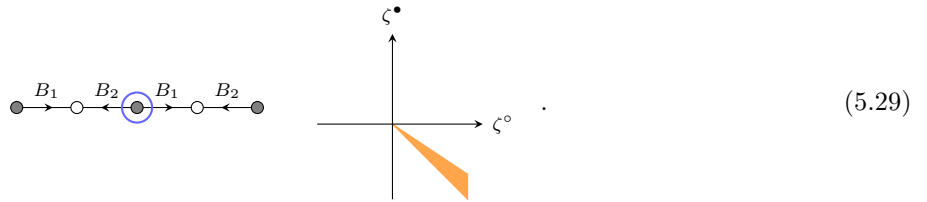
\begin{equation}\label{eqs_003}
	\begin{array}{c}
	\begin{tikzpicture}[scale=0.4]
		\draw[postaction={decorate},decoration={markings,
			mark= at position 0.6 with {\arrow{stealth}}}] (0,0) -- (2,0) node[pos=0.5,above] {$\scriptstyle B_1$};
		\draw[postaction={decorate},decoration={markings,
			mark= at position 0.6 with {\arrow{stealth}}}] (-4,0) -- (-2,0) node[pos=0.5,above] {$\scriptstyle B_1$};
		\draw[postaction={decorate},decoration={markings,
			mark= at position 0.6 with {\arrow{stealth}}}] (0,0) -- (-2,0) node[pos=0.5,above] {$\scriptstyle B_2$};
		\draw[postaction={decorate},decoration={markings,
			mark= at position 0.6 with {\arrow{stealth}}}] (4,0) -- (2,0) node[pos=0.5,above] {$\scriptstyle B_2$};
		\draw[fill=gray] (-4,0) circle (0.2) (0,0) circle (0.2) (4,0) circle (0.2);
		\draw[fill=white] (-2,0) circle (0.2) (2,0) circle (0.2);
		\draw[thick, \myblue] (0,0) circle (0.5);
	\end{tikzpicture}
	\end{array}\quad
\begin{array}{c}
\begin{tikzpicture}
	\draw[fill=\graphcol, \graphcol] (0,0) -- (1,-0.666667) -- (1,-1) -- cycle;
	\draw[-stealth] (-1,0) -- (1.2,0);
	\draw[-stealth] (0,-1) -- (0,1.2);
	\node[above] at (0,1.2) {$\scriptstyle \zeta^{\bullet}$};
	\node[right] at (1.2,0) {$\scriptstyle \zeta^{\circ}$};
\end{tikzpicture}
\end{array}\,.
\end{equation}

According to this plot we choose the following parameterization of the quiver morphisms:
\begin{equation}
	R=A_1=A_2=0,\; B_1=\left(\begin{array}{ccc}
		0 & 0 & b_2\\
		0 & b_1 & 0\\
	\end{array}\right),\; B_2=\left(\begin{array}{ccc}
	b_3 & 0 & 0\\
	0 & 0 & b_4\\
\end{array}\right),\; S=\left(\begin{array}{ccc}
0 & 0 & s\\
\end{array}\right)\,.
\end{equation}
Vacuum equations lead to the following relations for matrix elements:
\begin{equation}
	\begin{aligned}
		|b_2|^2+|b_3|^2=\zeta^{\circ},\; |b_1|^2+|b_4|^2=\zeta^{\circ},\; -|b_3|^2-|s|^2=\zeta^{\bullet},\; -|b_1|^2=\zeta^{\bullet},\; -|b_2|^2-|b_4|^2=\zeta^{\bullet}\,.
	\end{aligned}
\end{equation}
The solution reads:
\begin{equation}
	|b_1|=\sqrt{-\zeta^{\bullet}},\; |b_2|=\sqrt{ -\zeta^{\circ}-2 \zeta^{\bullet}},\; |b_3|=\sqrt{ 2 \left(\zeta^{\circ}+\zeta^{\bullet}\right)},\; |b_4|=\sqrt{ \zeta^{\circ}+\zeta^{\bullet}},\; |s|=\sqrt{ -2 \zeta^{\circ}-3 \zeta^{\bullet}}\,.
\end{equation}
This solution is valid in the following sector depicted in \eqref{eqs_003}:
\begin{equation}
	\zeta^{\circ}+\zeta^{\bullet}\geq 0,\quad 2\zeta^{\circ}+3\zeta^{\bullet}\leq 0\,.
\end{equation}
We should stress that this is an $S$-dominant solution in the phase portrait above the dashed line.

%%%%%%%%%%%%%%%%%%%%%%%%%%%%%%%%%%%%%%%%%%%%%%%%%%%%%%%%%%%%%%%%%%%
%%%%%%%%%%%%%%%%%%%%%%%%%%%%%%%%%%%%%%%%%%%%%%%%%%%%%%%%%%%%%%%%%%%
%%%%%%%%%%%%%%%%%%%%%%%%%%%%%%%%%%%%%%%%%%%%%%%%%%%%%%%%%%%%%%%%%%%
%%%%%%%%%%%%%%%%%%%%%%%%%%%%%%%%%%%%%%%%%%%%%%%%%%%%%%%%%%%%%%%%%%%

\section{Phases of $\widehat{\fg\fl}_{1|1}$ quiver theory}\label{sec:phases}

\subsection{Mutations (a.k.a. Seiberg dualities)} \label{sec:mutation}

Quiver mutation \cite{Alim:2011kw,Benini:2014mia} establishes a correspondence (a Seiberg duality \cite{Seiberg:1994pq,Intriligator:1995au}) between quiver representation moduli spaces of two pairs quiver-plus-superpotetnial $(\mathfrak{Q},W)$ and $(\check{\mathfrak{Q}},\check{W})$.
In particular, one is able to identify fixed points on those varieties.
Vectors of stability parameters and quiver dimensions are transformed accordingly $(\zeta_i,d_i)\leftrightarrow(\check{\zeta}_i,\check{d}_i)$.

Given $(\mathfrak{Q},W,\zeta_i,d_i)$ one constructs $(\check{\mathfrak{Q}},\check{W},\check{\zeta}_i,\check{d}_i)$ with the same nodes based on a particular choice of a selected node $a$ among them.
If the theory has some matter multiplets charged adjointly with respect to the gauge group associated with $a$ the procedure encounters certain difficulties
\cite{Kutasov:1995ss}.
However in the case when there is only the matter charged (anti-)fundamentally with respect to $a$ the dulization procedure is rather straightforward.

To dualize the quiver with respect to node $a$ for each pair of arrows $q\cdot p$ passing through node $a$ introduce a dual field $M_{pq}=qp$, substitute each product $qp$ in $W$ by $M_{pq}$, then we inverse arrows to $\check q$ and $\check p$ and, finally, add the corresponding term to acquire a dual superpotential:
\begin{equation}
	\check W=W(M_{pq})+\Tr M_{pq}{\check p}{\check q}\,.
\end{equation}
The dual dimensions are the same in all the nodes other than $a$.
And for $\check{d}_a$ we have:
\begin{equation}
	\check{d}_a=\mathop{\rm max}\lm_{b\in\{{\rm neighbors}\}}d_b-d_a\,.
\end{equation}
The stability parameters $\zeta$ are dualized in such a way that the net stability parameter (a slope in \cite{donaldson1983new}) $\sum_{a\in\fQ_0}\zeta_a d_a$  remains invariant.

We dualize quiver \eqref{quiver} with respect to the white node following \cite{Chuang:2009crq,Nagao:2010kx,Galakhov:2021xum,Nishinaka:2013pua}.
The resulting dimensions and stability parameters read:
\begin{equation}\label{duality_rel_1}
	({\check d}^{\circ},{\check d}^{\bullet})=(2d^{\bullet}-d^{\circ},d^{\bullet}),\quad ({\check \zeta}^{\circ},{\check \zeta}^{\bullet})=(-\zeta^{\circ},\zeta^{\bullet}+2\zeta^{\circ})\,.
\end{equation}

In our case we introduce dual fields:
\begin{equation}\label{du_morph_1}
	\begin{split}
		&H_{ij}=A_iB_j,\quad i,j=1,2\,,\\
		&M_i=A_iR,\quad i=1,2\,.
	\end{split}
\end{equation}

New quiver and superpotential read:
\begin{equation}\label{mut_1}
	\begin{split}
		\begin{array}{c}
			\begin{tikzpicture}
				\draw[postaction={decorate},
				decoration={markings, mark= at position 0.55 with {\arrow{stealth}}, mark= at position 0.65 with {\arrow{stealth}}}] (2,0) to[out=165, in=15] node[pos=0.5, above] {$\scriptstyle {\check A}_{1,2}$} (0,0);
				\draw[postaction={decorate},
				decoration={markings, mark= at position 0.55 with {\arrow{stealth}}, mark= at position 0.65 with {\arrow{stealth}}}] (0,0) to[out=345,in=195] node[pos=0.5, below] {$\scriptstyle  {\check B}_{1,2}$} (2,0);
				\draw[postaction={decorate},
				decoration={markings, mark= at position 0.55 with {\arrow{stealth}}}] (0,0) -- (1,-2) node[pos=0.5, below left] {$\scriptstyle {\check R}$};
				\draw[postaction={decorate},
				decoration={markings, mark= at position 0.55 with {\arrow{stealth}}}] (2,0) -- (1,-2) node[pos=0.5, above left] {$\scriptstyle S$};
				\draw[postaction={decorate},
				decoration={markings, mark= at position 0.55 with {\arrow{stealth}}, mark= at position 0.65 with {\arrow{stealth}}}] (1,-2) to[out=30,in=270] node[pos=0.5,below right] {$\scriptstyle M_{1,2}$} (2,0);
				\draw[postaction={decorate},
				decoration={markings, mark= at position 0.65 with {\arrow{stealth}}, mark= at position 0.7 with {\arrow{stealth}},mark= at position 0.75 with {\arrow{stealth}}, mark= at position 0.8 with {\arrow{stealth}}}] (2,0) to[out=330,in=270] (3,0) to[out=90,in=30] (2,0);
				\node[right] at (3,0) {$\scriptstyle H_{ij}$};
				\draw[fill=white] (0,0) circle (0.1);
				\draw[fill=gray] (2,0) circle (0.1);
				\begin{scope}[shift={(1,-2)}]
					\draw[fill=\myblue] (-0.1,-0.1) -- (-0.1,0.1) -- (0.1,0.1) -- (0.1,-0.1) -- cycle;
				\end{scope}
				\node[left] at (-0.1,0) {$\scriptstyle {\check\zeta}^{\circ},\;{\check d}^{\circ}$};
				\node[above] at (2,0.1) {$\scriptstyle {\check\zeta}^{\bullet},\;{\check d}^{\bullet}$};
			\end{tikzpicture}
		\end{array}\\
		{\check W} =\Tr\left[H_{11}H_{22}-H_{12}H_{21}+M_2S+\sum\lm_{i,j=1}^2H_{ij}{\check B}_j{\check A}_i+M_2S+\sum\lm_{i=1}^2M_i{\check R}{\check A}_i\right]\,.
	\end{split}
\end{equation}

Now we integrate over some fields:
\begin{equation}\label{du_morph_2}
	H_{11}=-{\check B}_2{\check A}_2,\quad H_{22}=-{\check B}_1{\check A}_1,\quad H_{12}={\check B}_1{\check A}_2,\quad H_{21}={\check B}_2{\check A}_1,\quad M_2=0\,.
\end{equation}
Also we redefine field $M_1={\check S}$.
The resulting quiver and superpotential read:
\begin{equation}\label{dual_quiver}
	\begin{array}{c}
		\begin{tikzpicture}
			\draw[postaction={decorate},
			decoration={markings, mark= at position 0.55 with {\arrow{stealth}}, mark= at position 0.65 with {\arrow{stealth}}}] (2,0) to[out=165, in=15] node[pos=0.5, above] {$\scriptstyle {\check A}_{1,2}$} (0,0);
			\draw[postaction={decorate},
			decoration={markings, mark= at position 0.55 with {\arrow{stealth}}, mark= at position 0.65 with {\arrow{stealth}}}] (0,0) to[out=345,in=195] node[pos=0.5, below] {$\scriptstyle  {\check B}_{1,2}$} (2,0);
			\draw[postaction={decorate},
			decoration={markings, mark= at position 0.55 with {\arrow{stealth}}}] (0,0) -- (1,-2) node[pos=0.5, below left] {$\scriptstyle {\check R}$};
			\draw[postaction={decorate},
			decoration={markings, mark= at position 0.55 with {\arrow{stealth}}}] (1,-2) -- (2,0) node[pos=0.5, below right] {$\scriptstyle {\check S}$};
			\draw[fill=white] (0,0) circle (0.1);
			\draw[fill=gray] (2,0) circle (0.1);
			\begin{scope}[shift={(1,-2)}]
				\draw[fill=\myblue] (-0.1,-0.1) -- (-0.1,0.1) -- (0.1,0.1) -- (0.1,-0.1) -- cycle;
			\end{scope}
			\node[above left] at (-0.1,0.1) {$\scriptstyle {\check\zeta}^{\circ},\;{\check d}^{\bullet}$};
			\node[above right] at (2.1,0.1) {$\scriptstyle {\check\zeta}^{\circ},\;{\check d}^{\bullet}$};
		\end{tikzpicture}
	\end{array}\quad \begin{array}{c}
		{\check W}=\Tr\left({\check B}_1{\check A}_2{\check B}_2{\check A}_1-{\check B}_1{\check A}_1{\check B}_2{\check A}_2+{\check A}_1{\check S}{\check R}\right)\,,\\
		\\
		\begin{array}{c|c|c|c|c|c|c}
			\mbox{Fields}&{\check A}_1 & {\check A}_2 & {\check B}_1 & {\check B}_2 & {\check R} & {\check S}\\
			\hline
			\mbox{Weights}&-\epsilon_1 & \epsilon_1 & -\epsilon_2 & \epsilon_2 & 0 &\epsilon_1
		\end{array}
	\end{array}
\end{equation}

We conclude that this quiver is Seiberg {\bf self-dual}.
Eventually, we could classify different duality chambers by a non-negative integer $n$.
Dimensions of stable BPS states in the corresponding phase could be also calculated in terms of dual crystal dimensions $(d^{\circ}_{(n)},d^{\bullet}_{(n)})$:
\begin{equation}
	\begin{aligned}
		\mbox{Glass-$n$ phase: }&\zeta^{\circ}<0,\quad -\frac{n+1}{n}\zeta^{\circ}<\zeta^{\bullet}<-\frac{n}{n-1}\zeta^{\circ} \,, \\
		&(d^{\circ},d^{\bullet})=\left\{\begin{array}{ll}
			\left((n+1)d_{(n)}^{\circ}-n d_{(n)}^{\bullet},nd_{(n)}^{\circ}-(n-1) d_{(n)}^{\bullet}\right), & n\mbox{ even}\,;\\
			\left((n+1)d_{(n)}^{\bullet}-n d_{(n)}^{\circ},nd_{(n)}^{\bullet}-(n-1) d_{(n)}^{\circ}\right), & n\mbox{ odd}\,.\\
		\end{array}\right.\,
	\end{aligned}
\end{equation}

\subsection{Mutant phases: simple glasses}\label{sec:simp_glasses}

Seiberg duality (mutation) implies that enumeration of fixed points in both phases is identical.
In other words, we can enumerate the fixed points in the crystal and non-crystal phase by the same set of super-Young diagrams.
As we will see in what follows fixed points are not described by molten crystals therefore we propose to call these phases \emph{glasses}.

However the interpretation in terms of atomic structures is different.
The difference in terms of atomic structures imposes a difference in the form of fixed point values of matrices $A_i$, $B_i$, $R$, $S$.
In practice, we will observe atoms that break the crystal melting rule.
Alternatively, we are not able to construct all the atoms as paths in the quiver path algebra.

In this section we will construct atomic structures for phases located in the top left quadrant of the phase diagram depicted in fig.\ref{fig:new_phases}.
We call these types of phases \emph{simple glasses} as the relation between atomic structures of the fixed points and super-Young diagrams is established explicitly.

We start with the glass-1 phase of theory \eqref{quiver} depicted in fig.\ref{fig:new_phases}:
\begin{equation}
	\zeta^{\circ}<0,\quad 2\zeta^{\circ}+\zeta^{\bullet}>0\,.
\end{equation}
This phase is dual to the cyclic phase of \eqref{dual_quiver}:
\begin{equation}
	\check{\zeta}^{\circ}>0,\quad	\check{\zeta}^{\bullet}>0\,.
\end{equation}

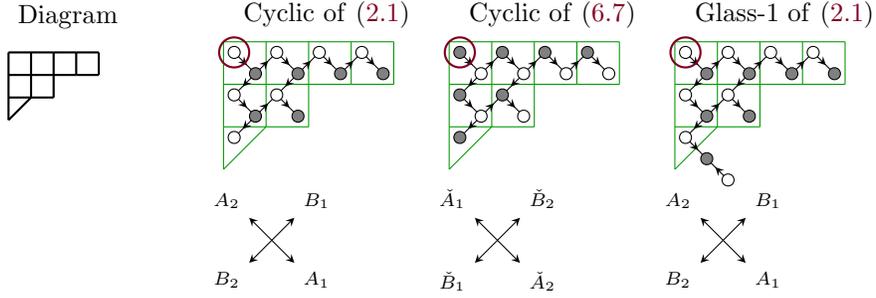
\begin{figure}[ht!]
	\begin{center}
		\begin{tikzpicture}
			\begin{scope}[scale=0.3]
				\foreach \x/\y/\z/\w in {0/0/1/0, 0/0/0/-1, 0/-1/1/-1, 1/0/1/-1, 0/-1/0/-2, 0/-2/1/-2, 1/-1/1/-2, 1/0/2/0, 1/-1/2/-1, 2/0/2/-1, 1/-2/2/-2, 2/-1/2/-2, 2/0/3/0, 2/-1/3/-1, 3/0/3/-1, 3/0/4/0, 3/-1/4/-1, 4/0/4/-1, 0/-2/0/-3, 0/-3/1/-2}
				{
					\draw[thick] (\x,\y) -- (\z,\w);
				}
			\end{scope}
			%%%%%%%%%%%%%%%%%%%%%
			\begin{scope}[shift = {(3,0)}]
				\begin{scope}[shift={(-0.141421,0.141421)}]
					\begin{scope}[scale=0.565685]
						\foreach \x/\y/\z/\w in {0/0/1/0, 0/0/0/-1, 0/-1/1/-1, 1/0/1/-1, 0/-1/0/-2, 0/-2/1/-2, 1/-1/1/-2, 1/0/2/0, 1/-1/2/-1, 2/0/2/-1, 1/-2/2/-2, 2/-1/2/-2, 2/0/3/0, 2/-1/3/-1, 3/0/3/-1, 3/0/4/0, 3/-1/4/-1, 4/0/4/-1, 0/-2/0/-3, 0/-3/1/-2}
						{
							\draw[thin, black!40!green] (\x,\y) -- (\z,\w);
						}
					\end{scope}
				\end{scope}
				\begin{scope}[rotate=-45, scale=0.4]
					\foreach \x/\y/\z/\w in {0/0/1/0, 1/-1/2/-1, 1/0/1/-1, 1/0/1/1, 1/1/2/1, 2/-1/2/-2, 2/-1/2/0, 2/0/3/0, 2/1/2/0, 2/1/2/2, 2/2/3/2, 3/2/3/3, 3/3/4/3}
					{
						\draw[postaction={decorate},decoration={markings,
							mark= at position 0.6 with {\arrow{stealth}}}] (\x,\y) -- (\z,\w);
					}
					\foreach \x/\y in {0/0, 1/-1, 1/1, 2/-2, 2/0, 2/2, 3/3}
					{
						\draw[fill=white] (\x,\y) circle (0.2);
					}
					\foreach \x/\y in {1/0, 2/-1, 2/1, 3/0, 3/2, 4/3}
					{
						\draw[fill=gray] (\x,\y) circle (0.2);
					}
					\draw[thick, burgundy] (0,0) circle (0.5);
				\end{scope}
			\end{scope}
			%%%%%%%%%%%%%%%%%%%%%
			\begin{scope}[shift = {(6,0)}]
				\begin{scope}[shift={(-0.141421,0.141421)}]
					\begin{scope}[scale=0.565685]
						\foreach \x/\y/\z/\w in {0/0/1/0, 0/0/0/-1, 0/-1/1/-1, 1/0/1/-1, 0/-1/0/-2, 0/-2/1/-2, 1/-1/1/-2, 1/0/2/0, 1/-1/2/-1, 2/0/2/-1, 1/-2/2/-2, 2/-1/2/-2, 2/0/3/0, 2/-1/3/-1, 3/0/3/-1, 3/0/4/0, 3/-1/4/-1, 4/0/4/-1, 0/-2/0/-3, 0/-3/1/-2}
						{
							\draw[thin, black!40!green] (\x,\y) -- (\z,\w);
						}
					\end{scope}
				\end{scope}
				\begin{scope}[rotate=-45, scale=0.4]
					\foreach \x/\y/\z/\w in {0/0/1/0, 1/-1/2/-1, 1/0/1/-1, 1/0/1/1, 1/1/2/1, 2/-1/2/-2, 2/-1/2/0, 2/0/3/0, 2/1/2/0, 2/1/2/2, 2/2/3/2, 3/2/3/3, 3/3/4/3}
					{
						\draw[postaction={decorate},decoration={markings,
							mark= at position 0.6 with {\arrow{stealth}}}] (\x,\y) -- (\z,\w);
					}
					\foreach \x/\y in {0/0, 1/-1, 1/1, 2/-2, 2/0, 2/2, 3/3}
					{
						\draw[fill=gray] (\x,\y) circle (0.2);
					}
					\foreach \x/\y in {1/0, 2/-1, 2/1, 3/0, 3/2, 4/3}
					{
						\draw[fill=white] (\x,\y) circle (0.2);
					}
					\draw[thick, burgundy] (0,0) circle (0.5);
				\end{scope}
			\end{scope}
			%%%%%%%%%%%%%%%%%%%%%
			\begin{scope}[shift = {(9,0)}]
				\begin{scope}[shift={(-0.141421,0.141421)}]
					\begin{scope}[scale=0.565685]
						\foreach \x/\y/\z/\w in {0/0/1/0, 0/0/0/-1, 0/-1/1/-1, 1/0/1/-1, 0/-1/0/-2, 0/-2/1/-2, 1/-1/1/-2, 1/0/2/0, 1/-1/2/-1, 2/0/2/-1, 1/-2/2/-2, 2/-1/2/-2, 2/0/3/0, 2/-1/3/-1, 3/0/3/-1, 3/0/4/0, 3/-1/4/-1, 4/0/4/-1, 0/-2/0/-3, 0/-3/1/-2}
						{
							\draw[thin, black!40!green] (\x,\y) -- (\z,\w);
						}
					\end{scope}
				\end{scope}
				\begin{scope}[rotate=-45, scale=0.4]
					\foreach \x/\y/\z/\w in {0/0/1/0, 1/-1/2/-1, 1/0/1/-1, 1/0/1/1, 1/1/2/1, 2/-2/3/-2, 2/-1/2/-2, 2/-1/2/0, 2/0/3/0, 2/1/2/0, 2/1/2/2, 2/2/3/2, 3/2/3/3, 3/3/4/3, 4/-2/3/-2}
					{
						\draw[postaction={decorate},decoration={markings,
							mark= at position 0.6 with {\arrow{stealth}}}] (\x,\y) -- (\z,\w);
					}
					\foreach \x/\y in {0/0, 1/-1, 1/1, 2/-2, 2/0, 2/2, 3/3, 4/-2}
					{
						\draw[fill=white] (\x,\y) circle (0.2);
					}
					\foreach \x/\y in {1/0, 2/-1, 2/1, 3/0, 3/2, 4/3, 3/-2}
					{
						\draw[fill=gray] (\x,\y) circle (0.2);
					}
					\draw[thick, burgundy] (0,0) circle (0.5);
				\end{scope}
			\end{scope}
			%%%%%%%%%%%%%%%%%%%%%%%%%%%%%%%%%%%%%%%%%%%%%%%%%
			\begin{scope}[shift = {(6.5,-2.5)}]
				\draw[-stealth] (0,0) -- (-0.3,-0.3) node[below left] {$\scriptstyle \check B_1$};
				\draw[-stealth] (0,0) -- (0.3,-0.3) node[below right] {$\scriptstyle \check A_2$};
				\draw[-stealth] (0,0) -- (0.3,0.3) node[above right] {$\scriptstyle \check B_2$};
				\draw[-stealth] (0,0) -- (-0.3,0.3) node[above left] {$\scriptstyle \check A_1$};
			\end{scope}
			\begin{scope}[shift = {(3.5,-2.5)}]
				\draw[-stealth] (0,0) -- (-0.3,-0.3) node[below left] {$\scriptstyle B_2$};
				\draw[-stealth] (0,0) -- (0.3,-0.3) node[below right] {$\scriptstyle A_1$};
				\draw[-stealth] (0,0) -- (0.3,0.3) node[above right] {$\scriptstyle B_1$};
				\draw[-stealth] (0,0) -- (-0.3,0.3) node[above left] {$\scriptstyle A_2$};
			\end{scope}
			\begin{scope}[shift = {(9.5,-2.5)}]
				\draw[-stealth] (0,0) -- (-0.3,-0.3) node[below left] {$\scriptstyle B_2$};
				\draw[-stealth] (0,0) -- (0.3,-0.3) node[below right] {$\scriptstyle A_1$};
				\draw[-stealth] (0,0) -- (0.3,0.3) node[above right] {$\scriptstyle B_1$};
				\draw[-stealth] (0,0) -- (-0.3,0.3) node[above left] {$\scriptstyle A_2$};
			\end{scope}
			%%%%%%%%%%%%%%%%%%%%%%%%%%%%%%%%%%%%%%%%%%%%%%%%%%%%%%%%%
			\node[above right] at (0,0.2) {Diagram};
			\node[above right] at (3,0.2) {Cyclic of \eqref{quiver}};
			\node[above right] at (6,0.2) {Cyclic of \eqref{dual_quiver}};
			\node[above right] at (9,0.2) {Glass-1 of \eqref{quiver}};
		\end{tikzpicture}
		\caption{Comparison of cyclic and glass phases}\label{fig:q_dq_qg}
	\end{center}
\end{figure}

In a way completely analogous to \cite{Galakhov:2023mak} we associate complete tiles to a pairs of white and black atoms in the cyclic phase of \eqref{quiver}.
For the dual quiver \eqref{dual_quiver} the roles of black and white quiver nodes are interchanged (see fig.\ref{fig:q_dq_qg}).
The duality relations between quiver morphisms \eqref{du_morph_1}, \eqref{du_morph_2} (where we neglect inessential signs in those relations):
\begin{equation}
	A_1B_1=\check{B}_2\check{A}_2,\quad A_2B_2=\check{B}_1\check{A}_1,\quad A_1B_2=\check{B}_1\check{A}_2,\quad A_2B_1=\check{B}_2\check{A}_1,\quad \check{S}=A_1R\,.
\end{equation}
and the relation between dimensions \eqref{duality_rel_1} indicate that under mutation the black atoms in the atomic structure plot remain intact whereas the white atoms in joints are reflected:
\begin{equation}
	\mbox{mutation:}\quad\left(\begin{array}{c}
		\begin{tikzpicture}[scale=0.7]
			\draw[postaction={decorate},decoration={markings,
				mark= at position 0.6 with {\arrow{stealth}}}] (0,0) -- (1,0) node[pos=0.5,above] {$\scriptstyle \check A_2$};
			\draw[postaction={decorate},decoration={markings,
				mark= at position 0.6 with {\arrow{stealth}}}] (1,0) -- (1,1) node[pos=0.5,right] {$\scriptstyle \check B_2$};
			\draw[fill=gray] (0,0) circle (0.1) (1,1) circle (0.1);
			\draw[fill=white] (1,0) circle (0.1);
		\end{tikzpicture}
	\end{array},\begin{array}{c}
		\begin{tikzpicture}[scale=0.7]
			\draw[postaction={decorate},decoration={markings,
				mark= at position 0.6 with {\arrow{stealth}}}] (0,0) -- (1,0) node[pos=0.5,below] {$\scriptstyle \check A_2$};
			\draw[postaction={decorate},decoration={markings,
				mark= at position 0.6 with {\arrow{stealth}}}] (1,0) -- (1,-1) node[pos=0.5,right] {$\scriptstyle \check B_1$};
			\draw[fill=gray] (0,0) circle (0.1) (1,-1) circle (0.1);
			\draw[fill=white] (1,0) circle (0.1);
		\end{tikzpicture}
	\end{array}\right)\;\mapsto\;\left(\begin{array}{c}
		\begin{tikzpicture}[scale=0.7]
			\draw[postaction={decorate},decoration={markings,
				mark= at position 0.6 with {\arrow{stealth}}}] (0,0) -- (0,1) node[pos=0.5,left] {$\scriptstyle B_1$};
			\draw[postaction={decorate},decoration={markings,
				mark= at position 0.6 with {\arrow{stealth}}}] (0,1) -- (1,1) node[pos=0.5,below] {$\scriptstyle A_1$};
			\draw[fill=gray] (0,0) circle (0.1) (1,1) circle (0.1);
			\draw[fill=white] (0,1) circle (0.1);
		\end{tikzpicture}
	\end{array},\begin{array}{c}
		\begin{tikzpicture}[scale=0.7]
			\draw[postaction={decorate},decoration={markings,
				mark= at position 0.6 with {\arrow{stealth}}}] (0,0) -- (0,-1) node[pos=0.5,left] {$\scriptstyle B_2$};
			\draw[postaction={decorate},decoration={markings,
				mark= at position 0.6 with {\arrow{stealth}}}] (0,-1) -- (1,-1) node[pos=0.5,above] {$\scriptstyle A_1$};
			\draw[fill=gray] (0,0) circle (0.1) (1,-1) circle (0.1);
			\draw[fill=white] (0,-1) circle (0.1);
		\end{tikzpicture}
	\end{array}\right)\,.
\end{equation}
According to these graphical relations to derive the atomic structure plot in the glass-1 phase it seems sufficient to simply construct the atomic structure in the dual picture of \eqref{dual_quiver} and then shift all the white atoms backwards leaving the black atoms at their original places.
However the mutation transformation turn out to be not that simple as it could be easily seen from the dimension relations we now analyze.

Let us consider a super-Young diagram $\lambda$.
Let us denote the number of complete tiles in $\lambda$ as $\kappa_{\stile}$ and the number of half-tiles as $\kappa_{\shtile}$.
In the dual theory in the cyclic phase each complete tile corresponds to a pair of a black and a white atoms, and a half-tile corresponds to a black atom:
\begin{equation}
	\check{d}^{\circ}=\kappa_{\stile},\quad \check{d}^{\bullet}=\kappa_{\stile}+\kappa_{\shtile}\,.
\end{equation}
For the dimensions we acquires the following values:
\begin{equation}
	d^{\circ}=\kappa_{\stile}+2\kappa_{\shtile},\quad 	d^{\bullet}=\kappa_{\stile}+\kappa_{\shtile}\,.
\end{equation}
Therefore as in the cyclic phase a complete tile contributes with a pair of a white and a black nodes, whereas a half-tile contributes with a pair of white nodes and a black node.

In the generic glass-$n$ phase the relation reads:
\begin{equation}
	d^{\circ}=\kappa_{\stile}+(n+1)\kappa_{\shtile},\quad 	d^{\bullet}=\kappa_{\stile}+n\,\kappa_{\shtile}\,.
\end{equation}
Therefore we conclude that the complete tile contributes with a pair, whereas a half-tile corresponds to a length-$n$ ``polymer'' tail we complete with edges of $A_1$ and $A_2$.
The simplest solution of such form is given in sec.\ref{sec:polymer_ex}.

\begin{figure}[ht!]
	\begin{center}
		\begin{tikzpicture}
			\begin{scope}[scale = 0.3]
				\foreach \x/\y/\z/\w in {0/0/1/0, 0/0/0/-1, 0/-1/1/-1, 1/0/1/-1, 0/-1/0/-2, 0/-2/1/-2, 1/-1/1/-2, 1/0/2/0, 1/-1/2/-1, 2/0/2/-1, 0/-2/0/-3, 0/-3/1/-3, 1/-2/1/-3, 1/-2/2/-2, 2/-1/2/-2, 2/0/3/0, 2/-1/3/-1, 3/0/3/-1, 0/-3/0/-4, 0/-4/1/-4, 1/-3/1/-4, 1/-3/2/-3, 2/-2/2/-3, 2/-2/3/-1, 3/0/4/0, 3/-1/4/0, 0/-4/0/-5, 0/-5/1/-4}
				{
					\draw[thick] (\x,\y) -- (\z,\w);
				}
			\end{scope}
			%%%%%%%%%%%%%%%%%%%%%
			\begin{scope}[shift = {(3,0)}]
				\begin{scope}[shift={(-0.141421,0.141421)}]
					\begin{scope}[scale=0.565685]
						\foreach \x/\y/\z/\w in {0/0/1/0, 0/0/0/-1, 0/-1/1/-1, 1/0/1/-1, 0/-1/0/-2, 0/-2/1/-2, 1/-1/1/-2, 1/0/2/0, 1/-1/2/-1, 2/0/2/-1, 0/-2/0/-3, 0/-3/1/-3, 1/-2/1/-3, 1/-2/2/-2, 2/-1/2/-2, 2/0/3/0, 2/-1/3/-1, 3/0/3/-1, 0/-3/0/-4, 0/-4/1/-4, 1/-3/1/-4, 1/-3/2/-3, 2/-2/2/-3, 2/-2/3/-1, 3/0/4/0, 3/-1/4/0, 0/-4/0/-5, 0/-5/1/-4}
						{
							\draw[thin, black!40!green] (\x,\y) -- (\z,\w);
						}
					\end{scope}
				\end{scope}
				\begin{scope}[rotate=-45, scale=0.4]
					\foreach \x/\y/\z/\w in {0/0/1/0, 1/-1/2/-1, 1/0/1/-1, 1/0/1/1, 1/1/2/1, 2/-2/3/-2, 2/-1/2/-2, 2/-1/2/0, 2/0/3/0, 2/1/2/0, 2/1/2/2, 2/2/3/2, 3/-3/4/-3, 3/-2/3/-3, 3/-2/3/-1, 3/-1/4/-1, 3/0/3/-1, 3/0/3/1, 3/2/3/1, 3/2/3/3, 4/-3/4/-4}
					{
						\draw[postaction={decorate},decoration={markings,
							mark= at position 0.6 with {\arrow{stealth}}}] (\x,\y) -- (\z,\w);
					}
					\foreach \x/\y in {0/0, 1/-1, 1/1, 2/-2, 2/0, 2/2, 3/-3, 3/-1, 3/1, 3/3, 4/-4}
					{
						\draw[fill=white] (\x,\y) circle (0.2);
					}
					\foreach \x/\y in {1/0, 2/-1, 2/1, 3/-2, 3/0, 3/2, 4/-3, 4/-1}
					{
						\draw[fill=gray] (\x,\y) circle (0.2);
					}
					\draw[thick, burgundy] (0,0) circle (0.5);
				\end{scope}
			\end{scope}
			%%%%%%%%%%%%%%%%%%%%%%%%%%%%%%
			\begin{scope}[shift = {(6,0)}]
				\begin{scope}[shift={(-0.141421,0.141421)}]
					\begin{scope}[scale=0.565685]
						\foreach \x/\y/\z/\w in {0/0/1/0, 0/0/0/-1, 0/-1/1/-1, 1/0/1/-1, 0/-1/0/-2, 0/-2/1/-2, 1/-1/1/-2, 1/0/2/0, 1/-1/2/-1, 2/0/2/-1, 0/-2/0/-3, 0/-3/1/-3, 1/-2/1/-3, 1/-2/2/-2, 2/-1/2/-2, 2/0/3/0, 2/-1/3/-1, 3/0/3/-1, 0/-3/0/-4, 0/-4/1/-4, 1/-3/1/-4, 1/-3/2/-3, 2/-2/2/-3, 2/-2/3/-1, 3/0/4/0, 3/-1/4/0, 0/-4/0/-5, 0/-5/1/-4}
						{
							\draw[thin, black!40!green] (\x,\y) -- (\z,\w);
						}
					\end{scope}
				\end{scope}
				\begin{scope}[rotate=-45, scale=0.4]
					\foreach \x/\y/\z/\w in {0/0/1/0, 1/-1/2/-1, 1/0/1/-1, 1/0/1/1, 1/1/2/1, 2/-2/3/-2, 2/-1/2/-2, 2/-1/2/0, 2/0/3/0, 2/1/2/0, 2/1/2/2, 2/2/3/2, 3/-3/4/-3, 3/-2/3/-3, 3/-2/3/-1, 3/-1/4/-1, 3/0/3/-1, 3/0/3/1, 3/1/4/1, 3/2/3/1, 3/2/3/3, 3/3/4/3, 4/-4/5/-4, 4/-3/4/-4, 5/1/4/1, 5/3/4/3, 6/-4/5/-4}
					{
						\draw[postaction={decorate},decoration={markings,
							mark= at position 0.6 with {\arrow{stealth}}}] (\x,\y) -- (\z,\w);
					}
					\foreach \x/\y in {0/0, 1/-1, 1/1, 2/-2, 2/0, 2/2, 3/-3, 3/-1, 3/1, 3/3, 4/-4, 5/1, 5/3, 6/-4}
					{
						\draw[fill=white] (\x,\y) circle (0.2);
					}
					\foreach \x/\y in {1/0, 2/-1, 2/1, 3/-2, 3/0, 3/2, 4/-3, 4/-1, 4/1, 4/3, 5/-4}
					{
						\draw[fill=gray] (\x,\y) circle (0.2);
					}
					\draw[thick, burgundy] (0,0) circle (0.5);
				\end{scope}
			\end{scope}
			%%%%%%%%%%%%%%%%%%%%%%%%%%%%%%
			\begin{scope}[shift = {(9,0)}]
				\begin{scope}[shift={(-0.141421,0.141421)}]
					\begin{scope}[scale=0.565685]
						\foreach \x/\y/\z/\w in {0/0/1/0, 0/0/0/-1, 0/-1/1/-1, 1/0/1/-1, 0/-1/0/-2, 0/-2/1/-2, 1/-1/1/-2, 1/0/2/0, 1/-1/2/-1, 2/0/2/-1, 0/-2/0/-3, 0/-3/1/-3, 1/-2/1/-3, 1/-2/2/-2, 2/-1/2/-2, 2/0/3/0, 2/-1/3/-1, 3/0/3/-1, 0/-3/0/-4, 0/-4/1/-4, 1/-3/1/-4, 1/-3/2/-3, 2/-2/2/-3, 2/-2/3/-1, 3/0/4/0, 3/-1/4/0, 0/-4/0/-5, 0/-5/1/-4}
						{
							\draw[thin, black!40!green] (\x,\y) -- (\z,\w);
						}
					\end{scope}
				\end{scope}
				\begin{scope}[rotate=-45, scale=0.4]
					\foreach \x/\y/\z/\w in {0/0/1/0, 1/-1/2/-1, 1/0/1/-1, 1/0/1/1, 1/1/2/1, 2/-2/3/-2, 2/-1/2/-2, 2/-1/2/0, 2/0/3/0, 2/1/2/0, 2/1/2/2, 2/2/3/2, 3/-3/4/-3, 3/-2/3/-3, 3/-2/3/-1, 3/-1/4/-1, 3/0/3/-1, 3/0/3/1, 3/1/4/1, 3/2/3/1, 3/2/3/3, 3/3/4/3, 4/-4/5/-4, 4/-3/4/-4, 5/1/4/1, 5/1/6/1, 7/1/6/1, 7/1/8/1, 9/1/8/1, 5/3/4/3, 5/3/6/3, 7/3/6/3, 7/3/8/3, 9/3/8/3, 6/-4/5/-4, 6/-4/7/-4, 8/-4/7/-4, 8/-4/9/-4, 10/-4/9/-4}
					{
						\draw[postaction={decorate},decoration={markings,
							mark= at position 0.6 with {\arrow{stealth}}}] (\x,\y) -- (\z,\w);
					}
					\foreach \x/\y in {0/0, 1/-1, 1/1, 2/-2, 2/0, 2/2, 3/-3, 3/-1, 3/1, 3/3, 4/-4, 5/1, 7/1, 9/1, 5/3, 7/3, 9/3, 6/-4, 8/-4, 10/-4}
					{
						\draw[fill=white] (\x,\y) circle (0.2);
					}
					\foreach \x/\y in {1/0, 2/-1, 2/1, 3/-2, 3/0, 3/2, 4/-3, 4/-1, 4/1, 6/1, 8/1, 4/3, 6/3, 8/3, 5/-4, 7/-4, 9/-4}
					{
						\draw[fill=gray] (\x,\y) circle (0.2);
					}
					\draw[thick, burgundy] (0,0) circle (0.5);
				\end{scope}
			\end{scope}
			\node[above right] at (0,0.2) {Diagram};
			\node[above right] at (3,0.2) {Cyclic};
			\node[above right] at (6,0.2) {Glass-1};
			\node[above right] at (9,0.2) {Glass-3};
			\begin{scope}[shift = {(1.5,-2.5)}]
				\draw[-stealth] (0,0) -- (-0.3,-0.3) node[below left] {$\scriptstyle B_2$};
				\draw[-stealth] (0,0) -- (0.3,-0.3) node[below right] {$\scriptstyle A_1$};
				\draw[-stealth] (0,0) -- (0.3,0.3) node[above right] {$\scriptstyle B_1$};
				\draw[-stealth] (0,0) -- (-0.3,0.3) node[above left] {$\scriptstyle A_2$};
			\end{scope}
		\end{tikzpicture}
		\caption{Super-partition and fixed points in various phases.}\label{fig:diag_examples}
	\end{center}
\end{figure}
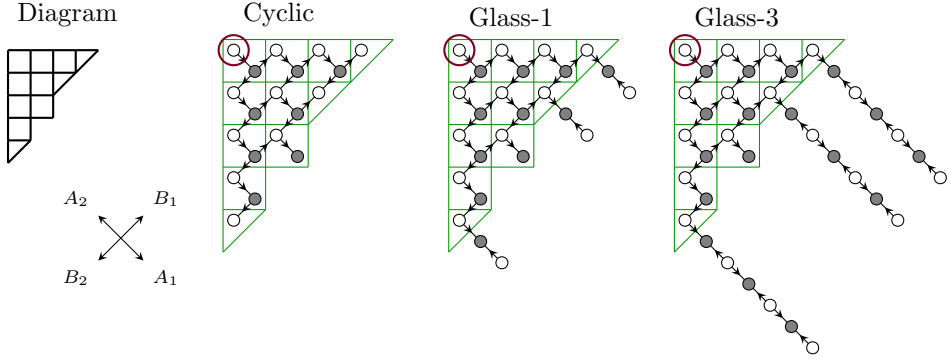

We summarize these rules of associating an atomic structure to a super-Young diagram in a glass-$n$ phase  in the following table (for an example of an application of these rules, see fig.\ref{fig:diag_examples}\footnote{To simplify reading and comparing processes for these diagrams here and in what follows we will put a super-Young diagram sample in the green color beneath an atomic structure plot if this is possible.}):
\begin{equation}\label{diagram_rules}
	\begin{array}{c}
		\begin{tikzpicture}
			\node{$\begingroup
				\renewcommand{\arraystretch}{1.5}
				\begin{array}{c|c|c|c}
					& \mbox{Diagram} & \mbox{Crystal} & \mbox{Glass-}n\\
					\hline
					\mbox{Complete tile} & \begin{array}{c}
						\begin{tikzpicture}[scale=0.2]
							\draw (0,0) -- (1,0) -- (1,-1) -- (0,-1) -- cycle;
						\end{tikzpicture}
					\end{array} & \begin{array}{c}
						\begin{tikzpicture}[rotate=-45, scale=0.7]
							\draw[postaction={decorate},
							decoration={markings, mark= at position 0.65 with {\arrow{stealth}}}] (0,0) -- (1,0) node[pos=0.5, above right] {$\scriptstyle A_1$};
							\draw[fill=white] (0,0) circle (0.1);
							\draw[fill=gray] (1,0) circle (0.1);
						\end{tikzpicture}
					\end{array}& \begin{array}{c}
						\begin{tikzpicture}[rotate=-45, scale=0.7]
							\draw[postaction={decorate},
							decoration={markings, mark= at position 0.65 with {\arrow{stealth}}}] (0,0) -- (1,0) node[pos=0.5, above right] {$\scriptstyle A_1$};
							\draw[fill=white] (0,0) circle (0.1);
							\draw[fill=gray] (1,0) circle (0.1);
						\end{tikzpicture}
					\end{array}\\
					\hline
					\mbox{Incomplete tile} & \begin{array}{c}
						\begin{tikzpicture}[scale=0.2]
							\draw (0,0) -- (1,0) -- (0,-1) -- cycle;
						\end{tikzpicture}
					\end{array} & \begin{array}{c}
						\begin{tikzpicture}[rotate=-45, scale=0.7]
							\draw[fill=white] (0,0) circle (0.1);
						\end{tikzpicture}
					\end{array} & \begin{array}{c}
						\begin{tikzpicture}[rotate=-45, scale=0.7]
							\draw[postaction={decorate},
							decoration={markings, mark= at position 0.65 with {\arrow{stealth}}}] (0,0) -- (1,0) node[pos=0.5, above right] {$\scriptstyle A_1$};
							\draw[postaction={decorate},
							decoration={markings, mark= at position 0.65 with {\arrow{stealth}}}] (2,0) -- (1,0) node[pos=0.5, above right] {$\scriptstyle A_2$};
							\draw[dashed, postaction={decorate},
							decoration={markings, mark= at position 0.65 with {\arrow{stealth}}}] (2,0) -- (3,0) node[pos=0.5, above right] {$\scriptstyle A_1$};
							\draw[postaction={decorate},
							decoration={markings, mark= at position 0.65 with {\arrow{stealth}}}] (4,0) -- (3,0) node[pos=0.5, above right] {$\scriptstyle A_2$};
							\begin{scope}[shift={(1,0)}]
								\draw[\myblue, rounded corners] (-0.2,0.3) -- (-0.2,-0.3) -- (1.2,-0.3) -- (1.2,0.3) -- cycle;
							\end{scope}
							\begin{scope}[shift={(3,0)}]
								\draw[\myblue, rounded corners] (-0.2,0.3) -- (-0.2,-0.3) -- (1.2,-0.3) -- (1.2,0.3) -- cycle;
							\end{scope}
							\draw[\myblue, rounded corners] (0.8,0) -- (0.8,-0.5) -- (2.5,-0.5) -- (2.5,-0.7);
							\draw[\myblue, rounded corners] (4.2,0) -- (4.2,-0.5) -- (2.5,-0.5) -- (2.5,-0.7);
							\node[below left] at (2.5,-0.7) {$n$ times};
							\draw[fill=white] (0,0) circle (0.1);
							\draw[fill=gray] (1,0) circle (0.1);
							\draw[fill=white] (2,0) circle (0.1);
							\draw[fill=gray] (3,0) circle (0.1);
							\draw[fill=white] (4,0) circle (0.1);
						\end{tikzpicture}
					\end{array}\\
					\hline
					\mbox{Vertical tile joint} & \begin{array}{c}
						\begin{tikzpicture}[scale=0.2]
							\draw (0,0) -- (1,0) -- (1,-1) -- (0,-1) -- cycle (1,-1) -- (1,-2) -- (0,-2) -- (0,-1);
						\end{tikzpicture}
					\end{array} &  \begin{array}{c}
						\begin{tikzpicture}[rotate=-45, scale=0.7]
							\draw[postaction={decorate},
							decoration={markings, mark= at position 0.65 with {\arrow{stealth}}}] (0,0) -- (1,0) node[pos=0.5, above right] {$\scriptstyle A_1$};
							\draw[postaction={decorate},
							decoration={markings, mark= at position 0.65 with {\arrow{stealth}}}] (1,-1) -- (2,-1) node[pos=0.5, above right] {$\scriptstyle A_1$};
							\draw[postaction={decorate},
							decoration={markings, mark= at position 0.65 with {\arrow{stealth}}}] (1,0) -- (1,-1) node[pos=0.5, above left] {$\scriptstyle B_2$};
							\draw[fill=white] (0,0) circle (0.1);
							\draw[fill=gray] (1,0) circle (0.1);
							\draw[fill=white] (1,-1) circle (0.1);
							\draw[fill=gray] (2,-1) circle (0.1);
							\begin{scope}[shift={(1,-0.5)}]
								\begin{scope}[xscale=0.5]
									\draw[\myblue] (0,0) circle (0.8);
								\end{scope}
							\end{scope}
						\end{tikzpicture}
					\end{array}& \begin{array}{c}
						\begin{tikzpicture}[rotate=-45, scale=0.7]
							\draw[postaction={decorate},
							decoration={markings, mark= at position 0.65 with {\arrow{stealth}}}] (0,0) -- (1,0) node[pos=0.5, above right] {$\scriptstyle A_1$};
							\draw[postaction={decorate},
							decoration={markings, mark= at position 0.65 with {\arrow{stealth}}}] (1,-1) -- (2,-1) node[pos=0.5, above right] {$\scriptstyle A_1$};
							\draw[postaction={decorate},
							decoration={markings, mark= at position 0.65 with {\arrow{stealth}}}] (1,0) -- (1,-1) node[pos=0.5, above left] {$\scriptstyle B_2$};
							\draw[fill=white] (0,0) circle (0.1);
							\draw[fill=gray] (1,0) circle (0.1);
							\draw[fill=white] (1,-1) circle (0.1);
							\draw[fill=gray] (2,-1) circle (0.1);
							\begin{scope}[shift={(1,-0.5)}]
								\begin{scope}[xscale=0.5]
									\draw[\myblue] (0,0) circle (0.8);
								\end{scope}
							\end{scope}
						\end{tikzpicture}
					\end{array}\\
					\hline
					\mbox{Horizontal tile joint} & \begin{array}{c}
						\begin{tikzpicture}[scale=0.2]
							\draw (0,0) -- (1,0) -- (1,-1) -- (0,-1) -- cycle (1,0) -- (2,0) -- (2,-1) -- (1,-1);
						\end{tikzpicture}
					\end{array} & \begin{array}{c}
						\begin{tikzpicture}[rotate=-45, scale=0.7, yscale=-1]
							\draw[postaction={decorate},
							decoration={markings, mark= at position 0.65 with {\arrow{stealth}}}] (0,0) -- (1,0) node[pos=0.5, below left] {$\scriptstyle A_1$};
							\draw[postaction={decorate},
							decoration={markings, mark= at position 0.65 with {\arrow{stealth}}}] (1,-1) -- (2,-1) node[pos=0.5, above right] {$\scriptstyle A_1$};
							\draw[postaction={decorate},
							decoration={markings, mark= at position 0.65 with {\arrow{stealth}}}] (1,0) -- (1,-1) node[pos=0.5, above left] {$\scriptstyle B_1$};
							\draw[fill=white] (0,0) circle (0.1);
							\draw[fill=gray] (1,0) circle (0.1);
							\draw[fill=white] (1,-1) circle (0.1);
							\draw[fill=gray] (2,-1) circle (0.1);
							\begin{scope}[shift={(1,-0.5)}]
								\begin{scope}[xscale=0.5]
									\draw[\myblue] (0,0) circle (0.8);
								\end{scope}
							\end{scope}
						\end{tikzpicture}
					\end{array} & \begin{array}{c}
						\begin{tikzpicture}[rotate=-45, scale=0.7, yscale=-1]
							\draw[postaction={decorate},
							decoration={markings, mark= at position 0.65 with {\arrow{stealth}}}] (0,0) -- (1,0) node[pos=0.5, below left] {$\scriptstyle A_1$};
							\draw[postaction={decorate},
							decoration={markings, mark= at position 0.65 with {\arrow{stealth}}}] (1,-1) -- (2,-1) node[pos=0.5, above right] {$\scriptstyle A_1$};
							\draw[postaction={decorate},
							decoration={markings, mark= at position 0.65 with {\arrow{stealth}}}] (1,0) -- (1,-1) node[pos=0.5, above left] {$\scriptstyle B_1$};
							\draw[fill=white] (0,0) circle (0.1);
							\draw[fill=gray] (1,0) circle (0.1);
							\draw[fill=white] (1,-1) circle (0.1);
							\draw[fill=gray] (2,-1) circle (0.1);
							\begin{scope}[shift={(1,-0.5)}]
								\begin{scope}[xscale=0.5]
									\draw[\myblue] (0,0) circle (0.8);
								\end{scope}
							\end{scope}
						\end{tikzpicture}
					\end{array}\\
				\end{array}
				\endgroup$};
			\draw[rounded corners, ultra thick, burgundy] (-5.5,2.1) -- (5.5,2.1) -- (5.5,-0.5) -- (-5.5,-0.5) -- cycle;
			%\node[right, burgundy] at (5.5,0.75) {\bf Major discrepancy};
		\end{tikzpicture}
	\end{array}
\end{equation}

\subsection{On non-mutant phases: crooked glasses}\label{sec:crooked}

Now we would like to move to the right lower quadrant of the $\zeta$-plane in fig.\ref{fig:new_phases}.
If we apply the mutation procedure to the black node of quiver \eqref{quiver} and follow all the procedures of sec.\ref{sec:mutation} we will again arrive to the same dual quiver \eqref{dual_quiver}  however the duality relation for the quiver dimensions and stability parameters would differ:
\begin{equation}\label{duality_rel_2}
	({\check d}^{\circ},{\check d}^{\bullet})=(d^{\circ},2d^{\circ}-d^{\bullet}),\quad ({\check \zeta}^{\circ},{\check \zeta}^{\bullet})=(2\zeta^{\bullet}+\zeta^{\circ},-\zeta^{\bullet})\,.
\end{equation}

And we observe that the cyclic phase of the dual quiver corresponds to a new phase of the quiver in question:
\begin{equation}\label{reg_1}
	2\zeta^{\bullet}+\zeta^{\circ}\geq 0,\quad \zeta^{\bullet}\leq 0\,,
\end{equation}
that corresponds to the nearest sector to the cyclic phase in the bottom left corner  in fig.\ref{fig:new_phases} depicted with the most intense violet color.

However if we apply the Seiberg duality directly we encounter the following problem.
The next to the lowest vacuum state $|\nhtile\rangle$ in the dual phase corresponds to a single black node in the atomic plot, that is given by the following dual dimension vector:
\begin{equation}
	(\check{d}^{\circ},\check{d}^{\bullet})=(0,1)\,.
\end{equation}
Applying \eqref{duality_rel_2} we arrive to \emph{negative} dimensions:
\begin{equation}
	(d^{\circ},d^{\bullet})=(0,-1)\,,
\end{equation}
that is an \emph{impossible} state.
Therefore we arrive to a conclusion that the Seiberg duality is inapplicable in this case in a naive straightforward way.

Let us bring up another observation supporting the conclusion made above.
As we were observing so far in the top right corner ($\zeta^{\circ}+\zeta^{\bullet}\geq 0$, $\zeta^{\bullet}\geq 0$) of the phase diagram in fig.\ref{fig:new_phases} in all the solutions $\langle R\rangle\neq 0$, $\langle S\rangle = 0$, whereas in the bottom left corner ($\zeta^{\circ}+\zeta^{\bullet}\leq 0$, $\zeta^{\circ}\leq 0$) $\langle R\rangle= 0$, $\langle S\rangle\neq 0$.
Therefore we could have nicknamed the phases belonging to the top left (bottom right) corner as $R$-dominated ($S$-dominated respectively).
Also let us remind that there is a natural $\IZ_2$-symmetry of vacuum equations \eqref{fixed point AB} permuting directions of all arrows and parameters $\zeta^{\circ}$ and $\zeta^{\bullet}$, so that the phase picture is symmetric with respect to the line $\zeta^{\circ}+\zeta^{\bullet}=0$.
This symmetry exchanges the roles of quiver morphisms $R$ and $S$ mapping the $R$-dominated phase to the $S$-dominated one and vice versa.
One might have expected that this picture of either $R$- or $S$-domination may be extended to the bottom left corner of the phase portrait as well.
However let us describe explicitly where the simplest $R$- and $S$-dominated solutions exist:
\begin{equation}
	\begin{aligned}
		&(d^{\circ},d^{\bullet})=(1,0),\; A_1=A_2=B_1=B_2=0,\; R=({\bf\color{burgundy}r}),\; S=0,\;|{\bf\color{burgundy}r}|=\sqrt{\zeta^\circ},\quad \zeta^\circ\geq 0\,,\\
		&(d^{\circ},d^{\bullet})=(0,1),\; A_1=A_2=B_1=B_2=0,\; R=0,\; S=({\bf\color{\myblue}s}),\;|{\bf\color{\myblue}s}|=\sqrt{-\zeta^\bullet},\quad \zeta^\bullet\leq 0\,,
	\end{aligned}
\end{equation}
Therefore both $R$- and $S$-dominated solutions \emph{exist simultaneously} in the bottom left corner ($\zeta^{\circ}\geq 0$, $\zeta^{\bullet}\leq 0$) of the phase portrait.
We find more of these examples in sec.\ref{sec:lot_of_examples}, implying that the simple solutions discussed above are not unique, and in general they form $R$- and $S$-dominated families \emph{overlapping} in the region ($\zeta^{\circ}\geq 0$, $\zeta^{\bullet}\leq 0$).

We conjecture that in this region there are multiple BPS state branches.
We will investigate this problem further elsewhere, however the presented data seems to be sufficient to distinguish at least two branches of $R$- and $S$-dominated solutions.

An appearance of multiple solution branches is another argument that the simple Seiberg duality is broken with this new phase since one expects that the number of solution branches is preserved by the duality.
The found solutions indicate as well that the quiver representation spaces are not cyclic, therefore these states could be neither associated with molten crystals.
However to distinguish them from the simple glass states we call them \emph{crooked glasses}.

Nevertheless, we \emph{conjecture} that the duality is not broken completely, and \emph{one} $R$-dominated branch of solutions in region \eqref{reg_1} is dual to a cyclic region of another quiver:
\begin{equation}\label{dual_quiver_2}
	\begin{array}{c}
		\begin{tikzpicture}
			\draw[postaction={decorate},
			decoration={markings, mark= at position 0.55 with {\arrow{stealth}}, mark= at position 0.65 with {\arrow{stealth}}}] (2,0) to[out=165, in=15] node[pos=0.5, above] {$\scriptstyle {\check A}_{1,2}$} (0,0);
			\draw[postaction={decorate},
			decoration={markings, mark= at position 0.55 with {\arrow{stealth}}, mark= at position 0.65 with {\arrow{stealth}}}] (0,0) to[out=345,in=195] node[pos=0.5, below] {$\scriptstyle  {\check B}_{1,2}$} (2,0);
			\draw[postaction={decorate},
			decoration={markings, mark= at position 0.55 with {\arrow{stealth}}}] (0,-2) to[out=105,in=255] node[pos=0.5, left] {$\scriptstyle R$} (0,0);
			\draw[postaction={decorate},
			decoration={markings, mark= at position 0.55 with {\arrow{stealth}}, mark= at position 0.65 with {\arrow{stealth}}}] (0,0) to[out=285,in=75] node[pos=0.5, right] {$\scriptstyle T_{1,2}$} (0,-2);
			\draw[fill=white] (0,0) circle (0.1);
			\draw[fill=gray] (2,0) circle (0.1);
			\begin{scope}[shift={(0,-2)}]
				\draw[fill=\myblue] (-0.1,-0.1) -- (-0.1,0.1) -- (0.1,0.1) -- (0.1,-0.1) -- cycle;
			\end{scope}
			\node[above left] at (-0.1,0.1) {$\scriptstyle {\check\zeta}_1,\;{\check d}_1$};
			\node[above right] at (2.1,0.1) {$\scriptstyle {\check\zeta}_2,\;{\check d}_2$};
		\end{tikzpicture}
	\end{array}\quad \begin{array}{c}
		{\check W}=\Tr\left({\check B}_1{\check A}_2{\check B}_2{\check A}_1-{\check B}_1{\check A}_1{\check B}_2{\check A}_2+T_1\check{A}_1\check{B}_2R+T_2\check{A}_1\check{B}_1R\right)\,,\\
		\\
		\begin{array}{c|c|c|c|c|c|c|c}
			\mbox{Fields}&{\check A}_1 & {\check A}_2 & {\check B}_1 & {\check B}_2 & R & T_1 & T_2\\
			\hline
			\mbox{Weights}&-\epsilon_1 & \epsilon_1 & -\epsilon_2 & \epsilon_2 & 0 & -\epsilon_2+\epsilon_1 & \epsilon_2+\epsilon_1
		\end{array}
	\end{array}
\end{equation}
Our conjecture is based on the following fact.
In sec.\ref{sec:d_1=1,d_2=1} we constructed \emph{two} solutions of dimensions $(d^{\circ}, d^{\bullet})=(1,1)$ valid in chamber \eqref{reg_1}.
The mutation formula \eqref{duality_rel_2} produces the following dual dimensions: $(\check{d}^{\circ}, \check{d}^{\bullet})=(1,1)$.
Therefore we expect that this solution branch is dual to a quiver theory having \emph{two} BPS solutions at level $(\check{d}^{\circ}, \check{d}^{\bullet})=(1,1)$.
Quiver \eqref{dual_quiver_2} is the seemingly simplest modification of \eqref{dual_quiver} having this property.
In the rest of this subsection we describe fixed points on the quiver \eqref{dual_quiver_2} representation moduli space in the cyclic chamber, identify those fixed points with \emph{skew} super-Young diagrams and discuss how our duality \emph{conjecture} works.

\begin{figure}[ht!]
	\begin{center}
		\begin{tikzpicture}
			\begin{scope}[scale=0.3]
				\foreach \x/\y/\z/\w in {0/0/1/0, 0/0/0/-1, 0/-1/1/-1, 1/0/1/-1, 0/-1/0/-2, 0/-2/1/-2, 1/-1/1/-2, 1/0/2/0, 1/-1/2/-1, 2/0/2/-1, 0/-2/0/-3, 0/-3/1/-3, 1/-2/1/-3, 1/-2/2/-2, 2/-1/2/-2, 2/0/3/0, 2/-1/3/-1, 3/0/3/-1, 3/0/4/0, 3/-1/4/-1, 4/0/4/-1, 1/-3/2/-2, 4/0/5/0, 4/-1/5/0}
				{
					\draw[thick] (\x,\y) -- (\z,\w);
				}
				\draw[fill=burgundy] (0,0) -- (1,0) -- (0,-1) -- cycle;
			\end{scope}
			%%%%%%%%%%%%%%%%%%%%%
			\begin{scope}[shift = {(4.5,0)}]
				\begin{scope}[shift={(-0.141421,0.141421)}]
					\begin{scope}[scale=0.565685]
						\foreach \x/\y/\z/\w in {0/-1/1/0, 0/-1/1/-1, 0/-1/1/-1, 1/0/1/-1, 0/-1/0/-2, 0/-2/1/-2, 1/-1/1/-2, 1/0/2/0, 1/-1/2/-1, 2/0/2/-1, 0/-2/0/-3, 0/-3/1/-3, 1/-2/1/-3, 1/-2/2/-2, 2/-1/2/-2, 2/0/3/0, 2/-1/3/-1, 3/0/3/-1, 3/0/4/0, 3/-1/4/-1, 4/0/4/-1, 1/-3/2/-2, 4/0/5/0, 4/-1/5/0}
						{
							\draw[thin, black!40!green] (\x,\y) -- (\z,\w);
						}
					\end{scope}
				\end{scope}
				\begin{scope}[rotate=-45, scale=0.4]
					\foreach \x/\y/\z/\w in {1/1/0/1, 1/-1/0/-1, 1/-1/2/-1, 1/0/1/-1, 1/0/1/1, 1/1/2/1, 2/-2/3/-2, 2/-1/2/-2, 2/-1/2/0, 2/0/3/0, 2/1/2/0, 2/1/2/2, 2/2/3/2, 3/-2/3/-1, 3/0/3/-1, 3/2/3/3, 3/3/4/3, 4/3/4/4}
					{
						\draw[postaction={decorate},decoration={markings,
							mark= at position 0.6 with {\arrow{stealth}}}] (\x,\y) -- (\z,\w);
					}
					\foreach \x/\y in {1/-1, 1/1, 2/-2, 2/0, 2/2, 3/-1, 3/3, 4/4}
					{
						\draw[fill=gray] (\x,\y) circle (0.2);
					}
					\foreach \x/\y in {0/1, 0/-1, 1/0, 2/-1, 2/1, 3/-2, 3/0, 3/2, 4/3}
					{
						\draw[fill=white] (\x,\y) circle (0.2);
					}
					\draw[thick, burgundy] (1,0) circle (0.5);
					\begin{scope}[shift={(0,1)}]
						\draw[ultra thick, orange] (-0.35,-0.35) -- (0.35,0.35) (0.35,-0.35) -- (-0.35,0.35);
					\end{scope}
					\begin{scope}[shift={(0,-1)}]
						\draw[ultra thick, orange] (-0.35,-0.35) -- (0.35,0.35) (0.35,-0.35) -- (-0.35,0.35);
					\end{scope}
				\end{scope}
			\end{scope}
			%%%%%%%%%%%%%%%%%%%%%
			\begin{scope}[shift = {(9,0)}]
				\begin{scope}[shift={(-0.141421,0.141421)}]
					\begin{scope}[scale=0.565685]
						\foreach \x/\y/\z/\w in {0/-1/1/0, 0/-1/1/-1, 0/-1/1/-1, 1/0/1/-1, 0/-1/0/-2, 0/-2/1/-2, 1/-1/1/-2, 1/0/2/0, 1/-1/2/-1, 2/0/2/-1, 0/-2/0/-3, 0/-3/1/-3, 1/-2/1/-3, 1/-2/2/-2, 2/-1/2/-2, 2/0/3/0, 2/-1/3/-1, 3/0/3/-1, 3/0/4/0, 3/-1/4/-1, 4/0/4/-1, 1/-3/2/-2, 4/0/5/0, 4/-1/5/0}
						{
							\draw[thin, black!40!green] (\x,\y) -- (\z,\w);
						}
					\end{scope}
				\end{scope}
				\begin{scope}[rotate=-45, scale=0.4]
					\begin{scope}[shift={(1,0)}]
					\foreach \x/\y/\z/\w in {0/0/1/0, 1/-1/2/-1, 1/1/2/1, 1/0/1/-1, 1/0/1/1, 2/-3/2/-2, 2/1/2/0, 2/1/2/2, 2/-1/2/-2, 2/-1/2/0, 2/2/3/2, 3/2/3/3, 2.5/1/2/0, 2.5/1/2/2}
					{
						\draw[postaction={decorate},decoration={markings,
							mark= at position 0.6 with {\arrow{stealth}}}] (\x,\y) -- (\z,\w);
					}
					\foreach \x/\y in {0/0, 1/-1, 1/1, 2/-2, 2/0, 2/2, 3/3}
					{
						\draw[fill=white] (\x,\y) circle (0.2);
					}
					\foreach \x/\y in {1/0, 2/-3, 2/1, 2/-1, 3/2, 2.5/1}
					{
						\draw[fill=gray] (\x,\y) circle (0.2);
					}
					\draw[thick, burgundy] (0,0) circle (0.5);
					\end{scope}
				\end{scope}
			\end{scope}
			%%%%%%%%%%%%%%%%%%%%%%%%%%%%%%%%%%%%%%%%%%%%%%%%%
			\begin{scope}[shift = {(6,-2.5)}]
				\draw[-stealth] (0,0) -- (-0.3,-0.3) node[below left] {$\scriptstyle \check B_1$};
				\draw[-stealth] (0,0) -- (0.3,-0.3) node[below right] {$\scriptstyle \check A_2$};
				\draw[-stealth] (0,0) -- (0.3,0.3) node[above right] {$\scriptstyle \check B_2$};
				\draw[-stealth] (0,0) -- (-0.3,0.3) node[above left] {$\scriptstyle \check A_1$};
			\end{scope}
			\begin{scope}[shift = {(10.5,-2.5)}]
				\draw[-stealth] (0,0) -- (-0.3,-0.3) node[below left] {$\scriptstyle B_2$};
				\draw[-stealth] (0,0) -- (0.3,-0.3) node[below right] {$\scriptstyle A_1$};
				\draw[-stealth] (0,0) -- (0.3,0.3) node[above right] {$\scriptstyle B_1$};
				\draw[-stealth] (0,0) -- (-0.3,0.3) node[above left] {$\scriptstyle A_2$};
			\end{scope}
			%%%%%%%%%%%%%%%%%%%%%%%%%%%%%%%%%%%%%%%%%%%%%%%%%%%%%%%%%
			\node[above right] at (0,0.4) {Diagram};
			\node[above right] at (4.5,0.4) {Cyclic of \eqref{dual_quiver_2}};
			\node[above right] at (9,0.4) {Crooked glass};
		\end{tikzpicture}
		\caption{Crooked glass example}\label{fig:crooked}
	\end{center}
\end{figure}
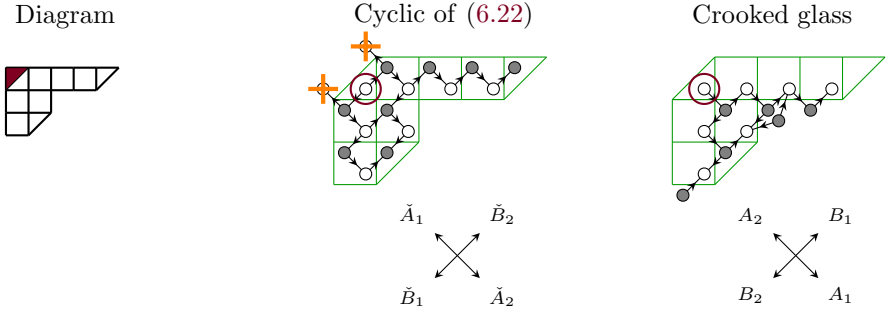

Fixed points on the moduli space of \eqref{dual_quiver_2} in the cyclic phase are enumerated by crystal atomic plots similar to the case of \eqref{dual_quiver}.
And the identification with the super-Young diagrams is also similar: a complete tile $\ntile$ is associated with a black-white atom pair and the upper half-tile $\nhtile$ is associated with a black atom.
The major difference now is that $\langle R\rangle\neq 0$ and we have to start growing a crystal from an atom of the white color.
Moreover, F-term relations:
\begin{equation}
	\check{A}_1\check{B}_2R=0,\quad \check{A}_1\check{B}_1R=0\,,
\end{equation}
prevent the crystal from growing along $\check{A}_1$ direction and forming another layers \cite{Galakhov:2021xum}, see fig.\ref{fig:crooked}.
If we associate it with a super-Young diagram then it is natural to associate it with a \emph{skew} super-Young diagram where we chip off a single triangular piece in the top left corner.

Let us enumerate few first levels of skew super-Young diagrams here:
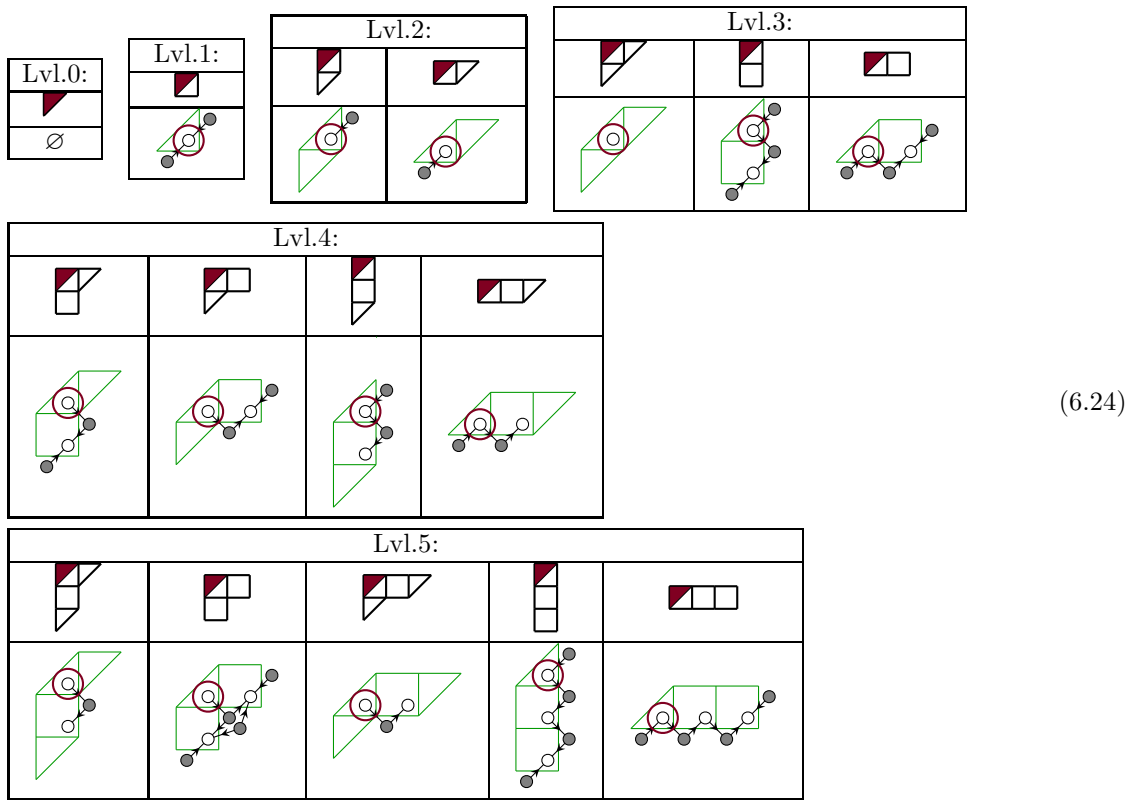
\begin{equation}
	\begin{aligned}
		&  \begin{array}{|c|}\hline\mbox{Lvl.0:}\\ \hline \begin{array}{c}
				\begin{tikzpicture}[scale=0.3]
					\foreach \x/\y/\z/\w in {0/0/1/0, 0/0/0/-1, 0/-1/1/0}
					{
						\draw[thick] (\x,\y) -- (\z,\w);
					}
					\draw[fill=burgundy] (0,0) -- (1,0) -- (0,-1) -- cycle;
				\end{tikzpicture}
			\end{array} \\ \hline \varnothing \\ \hline \end{array}\quad
		\begin{array}{|c|}
			\hline
			\mbox{Lvl.1:}\\
			\hline
			\begin{array}{c}
				\begin{tikzpicture}[scale=0.3]
					\foreach \x/\y/\z/\w in {0/0/1/0, 0/0/0/-1, 0/-1/1/-1, 1/0/1/-1}
					{
						\draw[thick] (\x,\y) -- (\z,\w);
					}
					\draw[fill=burgundy] (0,0) -- (1,0) -- (0,-1) -- cycle;
				\end{tikzpicture}
			\end{array}\\
			\hline
			\begin{array}{c}
				\begin{tikzpicture}
					\begin{scope}[shift={(-0.141421,0.141421)}]
						\begin{scope}[scale=0.565685]
							\foreach \x/\y/\z/\w in {1/0/0/-1, 0/-1/1/-1, 1/0/1/-1}
							{
								\draw[thin, black!40!green] (\x,\y) -- (\z,\w);
							}
						\end{scope}
					\end{scope}
					\begin{scope}[rotate=-45, scale=0.4]
						\begin{scope}[shift={(1,0)}]
							\foreach \x/\y/\z/\w in {0/1/0/0, 0/-1/0/0}
							{
								\draw[postaction={decorate},decoration={markings,
									mark= at position 0.6 with {\arrow{stealth}}}] (\x,\y) -- (\z,\w);
							}
							\foreach \x/\y in {0/0}
							{
								\draw[fill=white] (\x,\y) circle (0.2);
							}
							\foreach \x/\y in {0/1, 0/-1}
							{
								\draw[fill=gray] (\x,\y) circle (0.2);
							}
							\draw[thick, burgundy] (0,0) circle (0.5);
						\end{scope}
					\end{scope}
				\end{tikzpicture}
			\end{array}\\
			\hline
		\end{array}\quad
		\begin{array}{|c|c|}
			\hline
			\multicolumn{2}{|c|}{\mbox{Lvl.2:}}\\
			\hline
			\begin{array}{c}
				\begin{tikzpicture}[scale=0.3]
					\foreach \x/\y/\z/\w in {0/0/1/0, 0/0/0/-1, 0/-1/1/-1, 1/0/1/-1, 0/-1/0/-2, 0/-2/1/-1}
					{
						\draw[thick] (\x,\y) -- (\z,\w);
					}
					\draw[fill=burgundy] (0,0) -- (1,0) -- (0,-1) -- cycle;
				\end{tikzpicture}
			\end{array}&\begin{array}{c}
			\begin{tikzpicture}[scale=0.3]
				\foreach \x/\y/\z/\w in {0/0/1/0, 0/0/0/-1, 0/-1/1/-1, 1/0/1/-1, 1/0/2/0, 1/-1/2/0}
				{
					\draw[thick] (\x,\y) -- (\z,\w);
				}
				\draw[fill=burgundy] (0,0) -- (1,0) -- (0,-1) -- cycle;
			\end{tikzpicture}
		\end{array}\\
		\hline
		\begin{array}{c}
			\begin{tikzpicture}
				\begin{scope}[shift={(-0.141421,0.141421)}]
					\begin{scope}[scale=0.565685]
						\foreach \x/\y/\z/\w in {1/0/0/-1, 0/-1/1/-1, 1/0/1/-1, 0/-1/0/-2, 0/-2/1/-1}
						{
							\draw[thin, black!40!green] (\x,\y) -- (\z,\w);
						}
					\end{scope}
				\end{scope}
				\begin{scope}[rotate=-45, scale=0.4]
					\begin{scope}[shift={(1,0)}]
						\foreach \x/\y/\z/\w in {0/1/0/0}
						{
							\draw[postaction={decorate},decoration={markings,
								mark= at position 0.6 with {\arrow{stealth}}}] (\x,\y) -- (\z,\w);
						}
						\foreach \x/\y in {0/0}
						{
							\draw[fill=white] (\x,\y) circle (0.2);
						}
						\foreach \x/\y in {0/1}
						{
							\draw[fill=gray] (\x,\y) circle (0.2);
						}
						\draw[thick, burgundy] (0,0) circle (0.5);
					\end{scope}
				\end{scope}
			\end{tikzpicture}
		\end{array}& \begin{array}{c}
		\begin{tikzpicture}
			\begin{scope}[shift={(-0.141421,0.141421)}]
				\begin{scope}[scale=0.565685]
					\foreach \x/\y/\z/\w in {1/0/0/-1, 0/-1/1/-1, 1/0/1/-1, 1/0/2/0, 1/-1/2/0}
					{
						\draw[thin, black!40!green] (\x,\y) -- (\z,\w);
					}
				\end{scope}
			\end{scope}
			\begin{scope}[rotate=-45, scale=0.4]
				\begin{scope}[shift={(1,0)}]
					\foreach \x/\y/\z/\w in {0/-1/0/0}
					{
						\draw[postaction={decorate},decoration={markings,
							mark= at position 0.6 with {\arrow{stealth}}}] (\x,\y) -- (\z,\w);
					}
					\foreach \x/\y in {0/0}
					{
						\draw[fill=white] (\x,\y) circle (0.2);
					}
					\foreach \x/\y in {0/-1}
					{
						\draw[fill=gray] (\x,\y) circle (0.2);
					}
					\draw[thick, burgundy] (0,0) circle (0.5);
				\end{scope}
			\end{scope}
		\end{tikzpicture}
	\end{array} \\
		\hline
		\end{array}\quad \begin{array}{|c|c|c|}
		\hline
		\multicolumn{3}{|c|}{\mbox{Lvl.3:}}\\
		\hline
		\begin{array}{c}
			\begin{tikzpicture}[scale=0.3]
				\foreach \x/\y/\z/\w in {0/0/1/0, 0/0/0/-1, 0/-1/1/-1, 1/0/1/-1, 0/-1/0/-2, 0/-2/1/-1, 1/0/2/0, 1/-1/2/0}
				{
					\draw[thick] (\x,\y) -- (\z,\w);
				}
				\draw[fill=burgundy] (0,0) -- (1,0) -- (0,-1) -- cycle;
			\end{tikzpicture}
		\end{array} & \begin{array}{c}
			\begin{tikzpicture}[scale=0.3]
				\foreach \x/\y/\z/\w in {0/0/1/0, 0/0/0/-1, 0/-1/1/-1, 1/0/1/-1, 0/-1/0/-2, 0/-2/1/-2, 1/-1/1/-2}
				{
					\draw[thick] (\x,\y) -- (\z,\w);
				}
				\draw[fill=burgundy] (0,0) -- (1,0) -- (0,-1) -- cycle;
			\end{tikzpicture}
		\end{array} & \begin{array}{c}
			\begin{tikzpicture}[scale=0.3]
				\foreach \x/\y/\z/\w in {0/0/1/0, 0/0/0/-1, 0/-1/1/-1, 1/0/1/-1, 1/0/2/0, 1/-1/2/-1, 2/0/2/-1}
				{
					\draw[thick] (\x,\y) -- (\z,\w);
				}
				\draw[fill=burgundy] (0,0) -- (1,0) -- (0,-1) -- cycle;
			\end{tikzpicture}
		\end{array}\\
		\hline
		%%%%%%%%%%%%%%%%%%%%%%%%%%%%%%%%%%%%%%%%%%%%%%%%
		\begin{array}{c}
			\begin{tikzpicture}
				\begin{scope}[shift={(-0.141421,0.141421)}]
					\begin{scope}[scale=0.565685]
						\foreach \x/\y/\z/\w in {1/0/0/-1, 0/-1/1/-1, 1/0/1/-1, 0/-1/0/-2, 0/-2/1/-1, 1/0/2/0, 1/-1/2/0}
						{
							\draw[thin, black!40!green] (\x,\y) -- (\z,\w);
						}
					\end{scope}
				\end{scope}
				\begin{scope}[rotate=-45, scale=0.4]
					\begin{scope}[shift={(1,0)}]
						\foreach \x/\y in {0/0}
						{
							\draw[fill=white] (\x,\y) circle (0.2);
						}
						\draw[thick, burgundy] (0,0) circle (0.5);
					\end{scope}
				\end{scope}
			\end{tikzpicture}
		\end{array} &
		%%%%%%%%%%%%%%%%%%%%%%%%%%%%%%%%%%%%%%%%%%%%%%%%%
		\begin{array}{c}
			\begin{tikzpicture}
				\begin{scope}[shift={(-0.141421,0.141421)}]
					\begin{scope}[scale=0.565685]
						\foreach \x/\y/\z/\w in {1/0/0/-1, 0/-1/1/-1, 1/0/1/-1, 0/-1/0/-2, 0/-2/1/-2, 1/-1/1/-2}
						{
							\draw[thin, black!40!green] (\x,\y) -- (\z,\w);
						}
					\end{scope}
				\end{scope}
				\begin{scope}[rotate=-45, scale=0.4]
					\begin{scope}[shift={(1,0)}]
						\foreach \x/\y/\z/\w in {0/0/1/0, 0/1/0/0, 1/0/1/-1, 1/-2/1/-1}
						{
							\draw[postaction={decorate},decoration={markings,
								mark= at position 0.6 with {\arrow{stealth}}}] (\x,\y) -- (\z,\w);
						}
						\foreach \x/\y in {0/0, 1/-1}
						{
							\draw[fill=white] (\x,\y) circle (0.2);
						}
						\foreach \x/\y in {0/1, 1/0, 1/-2}
						{
							\draw[fill=gray] (\x,\y) circle (0.2);
						}
						\draw[thick, burgundy] (0,0) circle (0.5);
					\end{scope}
				\end{scope}
			\end{tikzpicture}
		\end{array}&
		%%%%%%%%%%%%%%%%%%%%%%%%%%%%%%%%%%%%%%%%%%%%%%%%
		\begin{array}{c}
			\begin{tikzpicture}
				\begin{scope}[shift={(-0.141421,0.141421)}]
					\begin{scope}[scale=0.565685]
						\foreach \x/\y/\z/\w in {1/0/0/-1, 0/-1/1/-1, 1/0/1/-1, 1/0/2/0, 1/-1/2/-1, 2/0/2/-1}
						{
							\draw[thin, black!40!green] (\x,\y) -- (\z,\w);
						}
					\end{scope}
				\end{scope}
				\begin{scope}[rotate=-45, scale=0.4]
					\begin{scope}[shift={(1,0)}]
						\foreach \x/\y/\z/\w in {0/0/1/0, 0/-1/0/0, 1/2/1/1, 1/0/1/1}
						{
							\draw[postaction={decorate},decoration={markings,
								mark= at position 0.6 with {\arrow{stealth}}}] (\x,\y) -- (\z,\w);
						}
						\foreach \x/\y in {0/0, 1/1}
						{
							\draw[fill=white] (\x,\y) circle (0.2);
						}
						\foreach \x/\y in {0/-1, 1/2, 1/0}
						{
							\draw[fill=gray] (\x,\y) circle (0.2);
						}
						\draw[thick, burgundy] (0,0) circle (0.5);
					\end{scope}
				\end{scope}
			\end{tikzpicture}
		\end{array}\\
		\hline
	\end{array}\\
	& \begin{array}{|c|c|c|c|}
		\hline
		\multicolumn{4}{|c|}{\mbox{Lvl.4:}}\\
		\hline
		\begin{array}{c}
			\begin{tikzpicture}[scale=0.3]
				\foreach \x/\y/\z/\w in {0/0/1/0, 0/0/0/-1, 0/-1/1/-1, 1/0/1/-1, 0/-1/0/-2, 0/-2/1/-2, 1/-1/1/-2, 1/0/2/0, 1/-1/2/0}
				{
					\draw[thick] (\x,\y) -- (\z,\w);
				}
				\draw[fill=burgundy] (0,0) -- (1,0) -- (0,-1) -- cycle;
			\end{tikzpicture}
		\end{array} & \begin{array}{c}
			\begin{tikzpicture}[scale=0.3]
				\foreach \x/\y/\z/\w in {0/0/1/0, 0/0/0/-1, 0/-1/1/-1, 1/0/1/-1, 1/0/2/0, 1/-1/2/-1, 2/0/2/-1, 0/-1/0/-2, 0/-2/1/-1}
				{
					\draw[thick] (\x,\y) -- (\z,\w);
				}
				\draw[fill=burgundy] (0,0) -- (1,0) -- (0,-1) -- cycle;
			\end{tikzpicture}
		\end{array} & \begin{array}{c}
			\begin{tikzpicture}[scale=0.3]
				\foreach \x/\y/\z/\w in {0/0/1/0, 0/0/0/-1, 0/-1/1/-1, 1/0/1/-1, 0/-1/0/-2, 0/-2/1/-2, 1/-1/1/-2, 0/-2/0/-3, 0/-3/1/-2}
				{
					\draw[thick] (\x,\y) -- (\z,\w);
				}
				\draw[fill=burgundy] (0,0) -- (1,0) -- (0,-1) -- cycle;
			\end{tikzpicture}
		\end{array} & \begin{array}{c}
			\begin{tikzpicture}[scale=0.3]
				\foreach \x/\y/\z/\w in {0/0/1/0, 0/0/0/-1, 0/-1/1/-1, 1/0/1/-1, 1/0/2/0, 1/-1/2/-1, 2/0/2/-1, 2/0/3/0, 2/-1/3/0}
				{
					\draw[thick] (\x,\y) -- (\z,\w);
				}
				\draw[fill=burgundy] (0,0) -- (1,0) -- (0,-1) -- cycle;
			\end{tikzpicture}
		\end{array}\\
		\hline
		%%%%%%%%%%%%%%%%%%%%%%%%%%%%%%%%%%%%%%%%%%%%%%%%%%%%%%%%
		\begin{array}{c}
			\begin{tikzpicture}
				\begin{scope}[shift={(-0.141421,0.141421)}]
					\begin{scope}[scale=0.565685]
						\foreach \x/\y/\z/\w in {1/0/0/-1, 0/-1/1/-1, 1/0/1/-1, 0/-1/0/-2, 0/-2/1/-2, 1/-1/1/-2, 1/0/2/0, 1/-1/2/0}
						{
							\draw[thin, black!40!green] (\x,\y) -- (\z,\w);
						}
					\end{scope}
				\end{scope}
				\begin{scope}[rotate=-45, scale=0.4]
					\begin{scope}[shift={(1,0)}]
						\foreach \x/\y/\z/\w in {0/0/1/0, 1/0/1/-1, 1/-2/1/-1}
						{
							\draw[postaction={decorate},decoration={markings,
								mark= at position 0.6 with {\arrow{stealth}}}] (\x,\y) -- (\z,\w);
						}
						\foreach \x/\y in {0/0, 1/-1}
						{
							\draw[fill=white] (\x,\y) circle (0.2);
						}
						\foreach \x/\y in {1/0, 1/-2}
						{
							\draw[fill=gray] (\x,\y) circle (0.2);
						}
						\draw[thick, burgundy] (0,0) circle (0.5);
					\end{scope}
				\end{scope}
			\end{tikzpicture}
		\end{array}&
		%%%%%%%%%%%%%%%%%%%%%%%%%%%%%%%%%%%%%%%%%%%%%%%%%%%%%%
		\begin{array}{c}
			\begin{tikzpicture}
				\begin{scope}[shift={(-0.141421,0.141421)}]
					\begin{scope}[scale=0.565685]
						\foreach \x/\y/\z/\w in {1/0/0/-1, 0/-1/1/-1, 1/0/1/-1, 1/0/2/0, 1/-1/2/-1, 2/0/2/-1, 0/-1/0/-2, 0/-2/1/-1}
						{
							\draw[thin, black!40!green] (\x,\y) -- (\z,\w);
						}
					\end{scope}
				\end{scope}
				\begin{scope}[rotate=-45, scale=0.4]
					\begin{scope}[shift={(1,0)}]
						\foreach \x/\y/\z/\w in {0/0/1/0, 1/2/1/1, 1/0/1/1}
						{
							\draw[postaction={decorate},decoration={markings,
								mark= at position 0.6 with {\arrow{stealth}}}] (\x,\y) -- (\z,\w);
						}
						\foreach \x/\y in {0/0, 1/1}
						{
							\draw[fill=white] (\x,\y) circle (0.2);
						}
						\foreach \x/\y in {1/2, 1/0}
						{
							\draw[fill=gray] (\x,\y) circle (0.2);
						}
						\draw[thick, burgundy] (0,0) circle (0.5);
					\end{scope}
				\end{scope}
			\end{tikzpicture}
		\end{array}&
		%%%%%%%%%%%%%%%%%%%%%%%%%%%%%%%%%%%%%%%%%%%%%%%%%%%%%%%%%%%%%
		\begin{array}{c}
			\begin{tikzpicture}
				\begin{scope}[shift={(-0.141421,0.141421)}]
					\begin{scope}[scale=0.565685]
						\foreach \x/\y/\z/\w in {1/0/0/-1, 0/-1/1/-1, 1/0/1/-1, 0/-1/0/-2, 0/-2/1/-2, 1/-1/1/-2, 0/-2/0/-3, 0/-3/1/-2, }
						{
							\draw[thin, black!40!green] (\x,\y) -- (\z,\w);
						}
					\end{scope}
				\end{scope}
				\begin{scope}[rotate=-45, scale=0.4]
					\begin{scope}[shift={(1,0)}]
						\foreach \x/\y/\z/\w in {0/0/1/0, 0/1/0/0, 1/0/1/-1}
						{
							\draw[postaction={decorate},decoration={markings,
								mark= at position 0.6 with {\arrow{stealth}}}] (\x,\y) -- (\z,\w);
						}
						\foreach \x/\y in {0/0, 1/-1}
						{
							\draw[fill=white] (\x,\y) circle (0.2);
						}
						\foreach \x/\y in {0/1, 1/0}
						{
							\draw[fill=gray] (\x,\y) circle (0.2);
						}
						\draw[thick, burgundy] (0,0) circle (0.5);
					\end{scope}
				\end{scope}
			\end{tikzpicture}
		\end{array}&
		%%%%%%%%%%%%%%%%%%%%%%%%%%%%%%%%%%%%%%%%%%%%%%%%%%%%
		\begin{array}{c}
			\begin{tikzpicture}
				\begin{scope}[shift={(-0.141421,0.141421)}]
					\begin{scope}[scale=0.565685]
						\foreach \x/\y/\z/\w in {1/0/0/-1, 0/-1/1/-1, 1/0/1/-1, 1/0/2/0, 1/-1/2/-1, 2/0/2/-1, 2/0/3/0, 2/-1/3/0}
						{
							\draw[thin, black!40!green] (\x,\y) -- (\z,\w);
						}
					\end{scope}
				\end{scope}
				\begin{scope}[rotate=-45, scale=0.4]
					\begin{scope}[shift={(1,0)}]
						\foreach \x/\y/\z/\w in {0/0/1/0, 0/-1/0/0, 1/0/1/1}
						{
							\draw[postaction={decorate},decoration={markings,
								mark= at position 0.6 with {\arrow{stealth}}}] (\x,\y) -- (\z,\w);
						}
						\foreach \x/\y in {0/0, 1/1}
						{
							\draw[fill=white] (\x,\y) circle (0.2);
						}
						\foreach \x/\y in {0/-1, 1/0}
						{
							\draw[fill=gray] (\x,\y) circle (0.2);
						}
						\draw[thick, burgundy] (0,0) circle (0.5);
					\end{scope}
				\end{scope}
			\end{tikzpicture}
		\end{array}\\
		\hline
	\end{array}\\
	&
	\begin{array}{|c|c|c|c|c|}
	\hline
	\multicolumn{5}{|c|}{\mbox{Lvl.5:}}\\
	\hline
	 \begin{array}{c}
		\begin{tikzpicture}[scale=0.3]
			\foreach \x/\y/\z/\w in {0/0/1/0, 0/0/0/-1, 0/-1/1/-1, 1/0/1/-1, 0/-1/0/-2, 0/-2/1/-2, 1/-1/1/-2, 1/0/2/0, 1/-1/2/0, 0/-2/0/-3, 0/-3/1/-2}
			{
				\draw[thick] (\x,\y) -- (\z,\w);
			}
			\draw[fill=burgundy] (0,0) -- (1,0) -- (0,-1) -- cycle;
		\end{tikzpicture}
	\end{array} & \begin{array}{c}
		\begin{tikzpicture}[scale=0.3]
			\foreach \x/\y/\z/\w in {0/0/1/0, 0/0/0/-1, 0/-1/1/-1, 1/0/1/-1, 0/-1/0/-2, 0/-2/1/-2, 1/-1/1/-2, 1/0/2/0, 1/-1/2/-1, 2/0/2/-1}
			{
				\draw[thick] (\x,\y) -- (\z,\w);
			}
			\draw[fill=burgundy] (0,0) -- (1,0) -- (0,-1) -- cycle;
		\end{tikzpicture}
	\end{array} & \begin{array}{c}
		\begin{tikzpicture}[scale=0.3]
			\foreach \x/\y/\z/\w in {0/0/1/0, 0/0/0/-1, 0/-1/1/-1, 1/0/1/-1, 1/0/2/0, 1/-1/2/-1, 2/0/2/-1, 0/-1/0/-2, 0/-2/1/-1, 2/0/3/0, 2/-1/3/0}
			{
				\draw[thick] (\x,\y) -- (\z,\w);
			}
			\draw[fill=burgundy] (0,0) -- (1,0) -- (0,-1) -- cycle;
		\end{tikzpicture}
	\end{array} & \begin{array}{c}
		\begin{tikzpicture}[scale=0.3]
			\foreach \x/\y/\z/\w in {0/0/1/0, 0/0/0/-1, 0/-1/1/-1, 1/0/1/-1, 0/-1/0/-2, 0/-2/1/-2, 1/-1/1/-2, 0/-2/0/-3, 0/-3/1/-3, 1/-2/1/-3}
			{
				\draw[thick] (\x,\y) -- (\z,\w);
			}
			\draw[fill=burgundy] (0,0) -- (1,0) -- (0,-1) -- cycle;
		\end{tikzpicture}
	\end{array} & \begin{array}{c}
		\begin{tikzpicture}[scale=0.3]
			\foreach \x/\y/\z/\w in {0/0/1/0, 0/0/0/-1, 0/-1/1/-1, 1/0/1/-1, 1/0/2/0, 1/-1/2/-1, 2/0/2/-1, 2/0/3/0, 2/-1/3/-1, 3/0/3/-1}
			{
				\draw[thick] (\x,\y) -- (\z,\w);
			}
			\draw[fill=burgundy] (0,0) -- (1,0) -- (0,-1) -- cycle;
		\end{tikzpicture}
	\end{array}\\
	\hline
	%%%%%%%%%%%%%%%%%%%%%%%%%%%%%%%%%%%%%%%%%%%%%%%%%%%%%%%%%%%%%%%%%%
	\begin{array}{c}
		\begin{tikzpicture}
			\begin{scope}[shift={(-0.141421,0.141421)}]
				\begin{scope}[scale=0.565685]
					\foreach \x/\y/\z/\w in {1/0/0/-1, 0/-1/1/-1, 1/0/1/-1, 0/-1/0/-2, 0/-2/1/-2, 1/-1/1/-2, 1/0/2/0, 1/-1/2/0, 0/-2/0/-3, 0/-3/1/-2}
					{
						\draw[thin, black!40!green] (\x,\y) -- (\z,\w);
					}
				\end{scope}
			\end{scope}
			\begin{scope}[rotate=-45, scale=0.4]
				\begin{scope}[shift={(1,0)}]
					\foreach \x/\y/\z/\w in {0/0/1/0, 1/0/1/-1}
					{
						\draw[postaction={decorate},decoration={markings,
							mark= at position 0.6 with {\arrow{stealth}}}] (\x,\y) -- (\z,\w);
					}
					\foreach \x/\y in {0/0, 1/-1}
					{
						\draw[fill=white] (\x,\y) circle (0.2);
					}
					\foreach \x/\y in {1/0}
					{
						\draw[fill=gray] (\x,\y) circle (0.2);
					}
					\draw[thick, burgundy] (0,0) circle (0.5);
				\end{scope}
			\end{scope}
		\end{tikzpicture}
	\end{array}&
	%%%%%%%%%%%%%%%%%%%%%%%%%%%%%%%%%%%%%%%%%%%%%%%%%%%%%%%%%%%%%%%%%
	\begin{array}{c}
		\begin{tikzpicture}
			\begin{scope}[shift={(-0.141421,0.141421)}]
				\begin{scope}[scale=0.565685]
					\foreach \x/\y/\z/\w in {1/0/0/-1, 0/-1/1/-1, 1/0/1/-1, 0/-1/0/-2, 0/-2/1/-2, 1/-1/1/-2, 1/0/2/0, 1/-1/2/-1, 2/0/2/-1}
					{
						\draw[thin, black!40!green] (\x,\y) -- (\z,\w);
					}
				\end{scope}
			\end{scope}
			\begin{scope}[rotate=-45, scale=0.4]
				\begin{scope}[shift={(1,0)}]
					\foreach \x/\y/\z/\w in {0/0/1/0, 1/0/1/-1, 1/0/1/1, 1/-2/1/-1, 1/2/1/1, 1.5/0/1/-1, 1.5/0/1/1}
					{
						\draw[postaction={decorate},decoration={markings,
							mark= at position 0.6 with {\arrow{stealth}}}] (\x,\y) -- (\z,\w);
					}
					\foreach \x/\y in {0/0, 1/-1, 1/1}
					{
						\draw[fill=white] (\x,\y) circle (0.2);
					}
					\foreach \x/\y in {1/0, 1/-2, 1/2, 1.5/0}
					{
						\draw[fill=gray] (\x,\y) circle (0.2);
					}
					\draw[thick, burgundy] (0,0) circle (0.5);
				\end{scope}
			\end{scope}
		\end{tikzpicture}
	\end{array}&
	%%%%%%%%%%%%%%%%%%%%%%%%%%%%%%%%%%%%%%%%%%%%%%%%%%%%%%%%%%%%%%%%%%%%%%
	\begin{array}{c}
		\begin{tikzpicture}
			\begin{scope}[shift={(-0.141421,0.141421)}]
				\begin{scope}[scale=0.565685]
					\foreach \x/\y/\z/\w in {1/0/0/-1, 0/-1/1/-1, 1/0/1/-1, 1/0/2/0, 1/-1/2/-1, 2/0/2/-1, 0/-1/0/-2, 0/-2/1/-1, 2/0/3/0, 2/-1/3/0}
					{
						\draw[thin, black!40!green] (\x,\y) -- (\z,\w);
					}
				\end{scope}
			\end{scope}
			\begin{scope}[rotate=-45, scale=0.4]
				\begin{scope}[shift={(1,0)}]
					\foreach \x/\y/\z/\w in {0/0/1/0, 1/0/1/1}
					{
						\draw[postaction={decorate},decoration={markings,
							mark= at position 0.6 with {\arrow{stealth}}}] (\x,\y) -- (\z,\w);
					}
					\foreach \x/\y in {0/0, 1/1}
					{
						\draw[fill=white] (\x,\y) circle (0.2);
					}
					\foreach \x/\y in {1/0}
					{
						\draw[fill=gray] (\x,\y) circle (0.2);
					}
					\draw[thick, burgundy] (0,0) circle (0.5);
				\end{scope}
			\end{scope}
		\end{tikzpicture}
	\end{array}&
	%%%%%%%%%%%%%%%%%%%%%%%%%%%%%%%%%%%%%%%%%%%%%%%%%%%%%%%%%%%
	\begin{array}{c}
		\begin{tikzpicture}
			\begin{scope}[shift={(-0.141421,0.141421)}]
				\begin{scope}[scale=0.565685]
					\foreach \x/\y/\z/\w in {1/0/0/-1, 0/-1/1/-1, 1/0/1/-1, 0/-1/0/-2, 0/-2/1/-2, 1/-1/1/-2, 0/-2/0/-3, 0/-3/1/-3, 1/-2/1/-3}
					{
						\draw[thin, black!40!green] (\x,\y) -- (\z,\w);
					}
				\end{scope}
			\end{scope}
			\begin{scope}[rotate=-45, scale=0.4]
				\begin{scope}[shift={(1,0)}]
					\foreach \x/\y/\z/\w in {0/0/1/0, 0/1/0/0, 1/-1/2/-1, 1/0/1/-1, 2/-1/2/-2, 2/-3/2/-2}
					{
						\draw[postaction={decorate},decoration={markings,
							mark= at position 0.6 with {\arrow{stealth}}}] (\x,\y) -- (\z,\w);
					}
					\foreach \x/\y in {0/0, 1/-1, 2/-2}
					{
						\draw[fill=white] (\x,\y) circle (0.2);
					}
					\foreach \x/\y in {0/1, 1/0, 2/-1, 2/-3}
					{
						\draw[fill=gray] (\x,\y) circle (0.2);
					}
					\draw[thick, burgundy] (0,0) circle (0.5);
				\end{scope}
			\end{scope}
		\end{tikzpicture}
	\end{array}&
	\begin{array}{c}
		\begin{tikzpicture}
			\begin{scope}[shift={(-0.141421,0.141421)}]
				\begin{scope}[scale=0.565685]
					\foreach \x/\y/\z/\w in {1/0/0/-1, 0/-1/1/-1, 1/0/1/-1, 1/0/2/0, 1/-1/2/-1, 2/0/2/-1, 2/0/3/0, 2/-1/3/-1, 3/0/3/-1}
					{
						\draw[thin, black!40!green] (\x,\y) -- (\z,\w);
					}
				\end{scope}
			\end{scope}
			\begin{scope}[rotate=-45, scale=0.4]
				\begin{scope}[shift={(1,0)}]
					\foreach \x/\y/\z/\w in {0/0/1/0, 0/-1/0/0, 1/1/2/1, 1/0/1/1, 2/3/2/2, 2/1/2/2}
					{
						\draw[postaction={decorate},decoration={markings,
							mark= at position 0.6 with {\arrow{stealth}}}] (\x,\y) -- (\z,\w);
					}
					\foreach \x/\y in {0/0, 1/1, 2/2}
					{
						\draw[fill=white] (\x,\y) circle (0.2);
					}
					\foreach \x/\y in {0/-1, 1/0, 2/3, 2/1}
					{
						\draw[fill=gray] (\x,\y) circle (0.2);
					}
					\draw[thick, burgundy] (0,0) circle (0.5);
				\end{scope}
			\end{scope}
		\end{tikzpicture}
	\end{array}\\
	\hline
	\end{array}
	\end{aligned}
\end{equation}

Having an atomic structure plot of a dual fixed point for \eqref{duality_rel_2} we \emph{propose} the following algorithm to acquire an atomic structure of a fixed point for theory \eqref{quiver} in phase \eqref{reg_1}.
Each white atom at position $z$ is translated to a complex of atoms: a white atom at position $z$ and two black atoms at positions $z\pm \epsilon_2$.
Furthermore each black atom at position $z$ is translated to a subtraction of a black atom at the same position $z$.
$A_1$, $B_1$ and $B_2$ arrows are restored uniquely as all the possible directed edges of length 1 in the resulting picture.

Finally, we might encounter a situation when at some position $z$ there are two black atoms.
In this case the $GL(2,\IC)$ subgroup of the gauge group commutes with the equivariant action.
We may further use the gauge symmetry on the solutions to fix this freedom.
We choose to do it in the following way: we choose one of the vectors associated to a pair of black atoms to belong to the image of map $A_1$, whereas the other atom belongs to its co-kernel.
Diagrammatically this situation may be represented by a picture where both atoms are connected via $B_{1,2}$-edges to neighbor white atoms, however just one of the black atoms is connected by the $A_1$-edge.
We treat this situation as the atomic structure plot is sitting simultaneously on different sheets of a plane cover.

To calculate the matrix coefficients in the representation we will not need an exact solution to \eqref{fixed point AB}, rather just a point in its $G_{\IC}$-orbit (see sec.\ref{sec:integrals}).
To acquire such a point we used to choose the non-zero elements to be given just by unit elements.
In the case of black atom doubling we have to choose one of the matrix elements to be negative as we have seen in sec.\ref{sec:double_black}.

We could schematize these rules in the following way:
\begin{equation}
	\check{\begin{array}{c}
		\begin{tikzpicture}[scale=0.4]
			\draw[fill=white] (0,0) circle (0.2);
		\end{tikzpicture}
	\end{array}}=\begin{array}{c}
\begin{tikzpicture}[scale=0.4]
\draw[postaction={decorate},decoration={markings,
	mark= at position 0.6 with {\arrow{stealth}}}] (0,0) -- (0,1);
\draw[postaction={decorate},decoration={markings,
	mark= at position 0.6 with {\arrow{stealth}}}] (0,0) -- (0,-1);
\draw[fill=white] (0,0) circle (0.2);
\draw[fill=gray] (0,1) circle (0.2) (0,-1) circle (0.2);
\end{tikzpicture}
\end{array},\quad \check{\begin{array}{c}
	\begin{tikzpicture}[scale=0.4]
		\draw[fill=gray] (0,0) circle (0.2);
	\end{tikzpicture}
\end{array}}=(-1)\begin{array}{c}
\begin{tikzpicture}[scale=0.4]
	\draw[fill=gray] (0,0) circle (0.2);
\end{tikzpicture}
\end{array},\quad \left(\begin{array}{c}
\begin{tikzpicture}[scale=0.4]
	\draw[postaction={decorate},decoration={markings,
		mark= at position 0.6 with {\arrow{stealth}}}] (0,0) -- (1.5,0) node[pos=0.5, above] {$\scriptstyle A_1$};
	\draw[fill=white] (0,0) circle (0.2);
	\draw[fill=gray] (1.5,0) circle (0.2) ;
\end{tikzpicture}
\end{array},\; \begin{array}{c}
\begin{tikzpicture}[scale=0.4]
	\draw[postaction={decorate},decoration={markings,
		mark= at position 0.6 with {\arrow{stealth}}}] (0,0) -- (0,1.5) node[pos=0.5, left] {$\scriptstyle B_1$};
	\draw[fill=white] (0,0) circle (0.2);
	\draw[fill=gray] (0,1.5) circle (0.2) ;
\end{tikzpicture}
\end{array},\; \begin{array}{c}
\begin{tikzpicture}[scale=0.4]
	\draw[postaction={decorate},decoration={markings,
		mark= at position 0.6 with {\arrow{stealth}}}] (0,0) -- (0,-1.5) node[pos=0.5, left] {$\scriptstyle B_2$};
	\draw[fill=white] (0,0) circle (0.2);
	\draw[fill=gray] (0,-1.5) circle (0.2) ;
\end{tikzpicture}
\end{array} \right),\quad \begin{array}{c}
\begin{tikzpicture}[scale=0.4]
	\draw[postaction={decorate},decoration={markings,
		mark= at position 0.6 with {\arrow{stealth}}}] (0,0) -- (0,1);
	\draw[postaction={decorate},decoration={markings,
		mark= at position 0.6 with {\arrow{stealth}}}] (0,0) -- (0,-1);
	\draw[postaction={decorate},decoration={markings,
		mark= at position 0.6 with {\arrow{stealth}}}] (-1,0) -- (0,0);
	\draw[fill=gray] (0,0) circle (0.2);
	\draw[fill=white] (-1,0) circle (0.2) (0,1) circle (0.2) (0,-1) circle (0.2);
	\node[right] at (0,0) {$\scriptstyle 2$};
\end{tikzpicture}
\end{array} \; \to \;
\begin{array}{c}
	\begin{tikzpicture}[scale=0.8]
		\draw[postaction={decorate},decoration={markings,
			mark= at position 0.6 with {\arrow{stealth}}}] (0,0) -- (0,1) node[pos=0.5,left] {$\scriptstyle 1 $};
		\draw[postaction={decorate},decoration={markings,
			mark= at position 0.6 with {\arrow{stealth}}}] (0,0) -- (0,-1) node[pos=0.5,left] {$\scriptstyle 1 $};
		\draw[postaction={decorate},decoration={markings,
			mark= at position 0.6 with {\arrow{stealth}}}] (0.5,0) -- (0,1) node[pos=0.5,right] {$\scriptstyle 1 $};
		\draw[postaction={decorate},decoration={markings,
			mark= at position 0.6 with {\arrow{stealth}}}] (0.5,0) -- (0,-1) node[pos=0.5,right] {$\scriptstyle -1 $};
		\draw[postaction={decorate},decoration={markings,
			mark= at position 0.6 with {\arrow{stealth}}}] (-1,0) -- (0,0) node[pos=0.2,above] {$\scriptstyle 1 $};
		\draw[fill=gray] (0,0) circle (0.1) (0.5,0) circle (0.1);
		\draw[fill=white] (-1,0) circle (0.1) (0,1) circle (0.1) (0,-1) circle (0.1);
	\end{tikzpicture}
\end{array}
\end{equation}
Examples of applying this recipe could be found in fig.\ref{fig:crooked} and in \eqref{crooked_ex}.

%%%%%%%%%%%%%%%%%%%%%%%%%%%%%%%%%%%%%%%%%%%%%%%%%%%%%%%%%%%%%%%%%%%
%%%%%%%%%%%%%%%%%%%%%%%%%%%%%%%%%%%%%%%%%%%%%%%%%%%%%%%%%%%%%%%%%%%
%%%%%%%%%%%%%%%%%%%%%%%%%%%%%%%%%%%%%%%%%%%%%%%%%%%%%%%%%%%%%%%%%%%
%%%%%%%%%%%%%%%%%%%%%%%%%%%%%%%%%%%%%%%%%%%%%%%%%%%%%%%%%%%%%%%%%%%

\section{Algebras across the walls}\label{sec:wc_algebra}

\subsection{Representation form integrals over moduli spaces}\label{sec:integrals}

In this subsection we remind some steps of constructing quiver Yangian representations on quiver moduli space fixed points (see details in \cite{Galakhov:2020vyb,Galakhov:2021vbo,Galakhov:2023mak}) in the cyclic phase and extend this notion to other phases.

Since when working with equivariant cohomologies localization shrinks all the system properties to tangent vectors in a neighborhood of a fixed point we would like to start with a modified version of the Euler class for a graded linear space proposed in \cite{Galakhov:2020vyb}.
Suppose the vectors of the linear space $\CN$ are graded with equivariant weights $w_i$ some of which may be zeroes:
\begin{equation}
	\CN=\bigoplus\lm_i \IC|w_i\rangle\,.
\end{equation}
We define the Euler class for this space as the following function:
\begin{equation}\label{Eul}
	{\rm Eul}\;\CN:=(-1)^{\left\lfloor\frac{1}{2}\#\{i:\,w_i=0\}\right\rfloor}\prod\lm_{i:\,w_i\neq 0}w_i\,,
\end{equation}
where in the exponent we take a floor integer part of a half of a total number of vectors with a zero weight, and in the product we put all the non-zero weights.

Now we consider a quiver representation moduli space.
In this subsection we tend to define all the structures in general for a generic quiver $\fQ$ with a set of nodes $\fQ_0$ and a dimension vector $\vec{d}=\{d_a\}_{a\in\fQ_0}$, and a stability parameter vector $\vec{\zeta}=\{\zeta_a\}_{a\in\fQ_0}$.
Let us define the quiver representation moduli space as a set of solutions to the D-term equations \eqref{D-term} modulo the action of the gauge group:
\begin{equation}
	G_{\IR}=\prod\lm_{a\in\fQ_0}U(d_a)\,.
\end{equation}
This space depends explicitly on the stability parameters $\vec\zeta$ appearing in the right hand side of the D-term equations.
The Narashiman-Shishadri-Hitchin-Kobayashi correspondence \cite{donaldson1983new, nakajima1999lectures} allows one to identify this space with a space of stable quiver representations modulo the complexified gauge group:
\begin{equation}
	G_{\IC}=\prod\lm_{a\in\fQ_0}GL(d_a,\IC)\,.
\end{equation}
This alternative definition has certain practical advantages and disadvantages.
On one hand the space of stable quiver matrices is a complex algebraic variety admitting an application of all the modern algebraic geometry methods.
On the other hand the stability condition is rather abstract and makes the role of the stability parameters $\vec\zeta$ more obscure than simply parameters in equations.

Therefore for the purposes of this paper we adopt a mixed definition of the quiver representation moduli space as the following set of points:
\begin{equation}
	\mathscr{M}(\vec\zeta,\vec d):=\left\{\{G_{\IC}\mbox{-orbit of quiver matrices containing a solution to D-term equations \eqref{D-term}}\}/G_{\IC}\right\}
\end{equation}

Having two points $p\in \mathscr{M}(\vec\zeta,\vec d)$ and $p'\in \mathscr{M}(\vec\zeta,\vec d')$ we define a homomorphism of quiver representations as a singular gauge transform such that the following diagram commutes for all the quiver morphisms:
\begin{equation}
	\tau_a\in{\rm Hom}_{\IC}(V_a(p),V_a(p')),\quad\begin{array}{c}
	\begin{tikzpicture}
		\node(A) at (0,0) {$V_a(p)$};
		\node(B) at (2.5,0) {$V_b(p)$};
		\node(C) at (0,-1) {$V_a(p')$};
		\node(D) at (2.5,-1) {$V_b(p')$};
		\path (A) edge[-stealth] node[above] {$\scriptstyle \phi_{ba}(p)$} (B) (C) edge[-stealth] node[above] {$\scriptstyle \phi_{ba}(p')$} (D) (A) edge[-stealth] node[left] {$\scriptstyle \tau_a$} (C) (B) edge[-stealth] node[right] {$\scriptstyle \tau_b$} (D);
	\end{tikzpicture}
	\end{array}\;\Rightarrow\;\tau_b\cdot\phi_{ba}(p)=\phi_{ba}(p')\cdot\tau_a\,,
\end{equation}
where $V_a$ and $\phi_{ba}$ are respective vector spaces and quiver matrices associated to quiver nodes and arrows, $a,b\in\fQ_0$.

We call an \emph{incidence locus} $\CI(\vec\zeta,\vec d,\vec d')\subset \mathscr{M}(\vec\zeta,\vec d)\times \mathscr{M}(\vec\zeta,\vec d')$ a variety of point pairs where a homomorphism exists.

In our model we have a natural equivariant action induced by the flavor symmetry.
This symmetry leaves on $\mathscr{M}(\vec\zeta,\vec d)$ isolated fixed points.
Fixed points on $\CI(\vec\zeta,\vec d,\vec d')$ are enumerated by pairs of points $(p,p')$ with $p\in\mathscr{M}(\zeta,\vec d)$, $p'\in\mathscr{M}(\zeta,\vec d')$ where a homomorphism exists.
We denote corresponding tangent spaces to fixed points as:
\begin{equation}
	\CT(p)\left[\mathscr{M}(\vec\zeta,\vec d)\right],\quad \CT(p,p')\left[\CI(\vec\zeta,\vec d,\vec d')\right]\subset\CT(p)\left[\mathscr{M}(\vec\zeta,\vec d)\right]\oplus \CT(p')\left[\mathscr{M}(\vec\zeta,\vec d')\right]\,.
\end{equation}
Both these spaces are graded vector spaces, therefore definition \eqref{Eul} is applicable for them.

In sec.\ref{sec:phases} we have observed that some families of fixed points on quiver representation moduli spaces of various dimensions admit \emph{partition ordering}.
Let us introduce an ordering function:
\begin{equation}
	\fo_{\vec\zeta}: \quad \lambda \; \longrightarrow\; \left\{ \mbox{Set of fixed points on }\mathscr{M}\right\}\,,
\end{equation}
that maps a particular (skew) super-partition to a particular fixed point on the quiver representation moduli space.
This function depends explicitly peace-wise on the stability parameters $\vec\zeta$ as $\mathscr{M}$ does.
Therefore for each phase chamber in fig.\ref{fig:new_phases} we have to define its own ordering function\footnote{We should stress that this mapping is not injective either: as we have seen in sec.\ref{sec:ex_cycle} a single atomic structure plot may be represented by \emph{different} super-Young diagrams in different phases.}.
Moreover the ordering function $\fo_{\vec\zeta}$ defines the quiver dimensions \eqref{dimensions} of the resulting variety  $\mathscr{M}(\vec \zeta,\vec d)$.

To calculate matrix elements for the raising and lowering operators these operations are considered as a version of Hecke modifications on ADHM data associated to a quiver \cite{Galakhov:2020vyb}, or a Fourier-Mukai transform on two moduli spaces $\mathscr{M}({\vec \zeta},{\vec d})$ and $\mathscr{M}({\vec \zeta},{\vec d}')$ with a kernel given by the structure sheaf of the incidence locus $\CI$.
The latter action on the cohomologies boils down to the canonical pullback-pushforward construction \cite{Rapcak:2020ueh,nakajima1999lectures,Rapcak:2018nsl}.
After applying the Berline-Vergne-Atiyah-Bott integration localization formula we acquire a result that is simply a ratio of Euler classes of $\CT(p)\left[\mathscr{M}({\vec d},{\vec \zeta})\right]$ and $\CT(p,p')\left[\CI({\vec \zeta},{\vec d},{\vec d}')\right]$.

In various phase chambers of parameters $\vec\zeta$ having constructed a system of super-Young diagrams and an ordering function $\fo_{\vec\zeta}$ we define representation matrices:
\begin{equation}\label{rep_phase}
	{\bf E}_{\lambda,\lambda+a}:=\frac{{\rm Eul}\left\{\CT\left(\fo_{\vec\zeta}(\lambda)\right)\left[\mathscr{M}\right]\right\}}{{\rm Eul}\left\{\CT\left(\fo_{\vec\zeta}(\lambda+a),\fo_{\vec\zeta}(\lambda)\right)\left[\CI\right]\right\}},\quad {\bf F}_{\lambda,\lambda-a}:=\frac{{\rm Eul}\left\{\CT\left(\fo_{\vec\zeta}(\lambda)\right)\left[\mathscr{M}\right]\right\}}{{\rm Eul}\left\{\CT\left(\fo_{\vec\zeta}(\lambda),\fo_{\vec\zeta}(\lambda-a)\right)\left[\CI\right]\right\}}\,.
\end{equation}
Now we are in a position to check if the defined matrices form a representation of any quiver Yangian, in other words if they satisfy \eqref{hysteresis}.

\subsection{Simple glasses}\label{sec:res_sg}

We check that defined in the previous subsection matrix elements \emph{do satisfy}  \eqref{hysteresis} for all the phases we call simple glasses in the sector $\zeta^{\circ}\leq 0$, $\zeta^{\circ}+\zeta^{\bullet}\geq 0$ defined in sec.\ref{sec:simp_glasses}.
This observation indicates that BPS vacua for all glass-$n$ phases are isomorphic to the Fock module of affine super-Yangian $\mathsf{Y}(\widehat{\fg\fl}_{1|1})$  with the algebra action given by \eqref{rep_phase}.

However as we find different glass chambers may be distinguished by vacuum Cartan eigenvalues:
\begingroup
\renewcommand{\arraystretch}{2}
\begin{equation}\label{mut_alg}
	\begin{array}{c|c|c}
		\mbox{Phase} & \psi_{\varnothing}^{\circ}(z) & \psi_{\varnothing}^{\bullet}(z)\\
		\hline
		\mbox{Crystal} &\dfrac{1}{z} & z+\epsilon_1\\
		\hline
		\mbox{Glass-1} &-\dfrac{1}{(\epsilon_1^2-\epsilon_2^2)^2}\cdot\dfrac{1}{z} & z+\epsilon_1\\
		\mbox{Glass-2} &\dfrac{1}{(\epsilon_1^2-\epsilon_2^2)^4(9\epsilon_1^2-\epsilon_2^2)^2}\cdot\dfrac{1}{z} & -\dfrac{1}{(\epsilon_1^2-\epsilon_2^2)^2}\cdot\left(z+\epsilon_1\right)\\
		\ldots & \ldots & \ldots\\
		\mbox{Glass-}n & \upsilon_n\dfrac{1}{z} & \upsilon_{n-1}(z+\epsilon_1)
	\end{array}
\end{equation}
\endgroup
where
\begin{equation}
	\upsilon_k=(-1)^k\prod\lm_{j=1}^k\left((2j-1)^2\epsilon_1^2-\epsilon_2^2\right)^{-2(k-j+1)}\,.
\end{equation}

In other words, comparing \eqref{mut_alg} and \eqref{psi_rep} we find that in the phase of glass-$n$ (we may think of the cyclic phase as a glass-$0$ phase) we have the following parameters defining the representation:
\begin{equation}
	c_{\circ}=\upsilon_n,\quad c_{\bullet}=\upsilon_{n-1}\,.
\end{equation}

This observation indicates that the mutation supports not only Seiberg duality of the fixed points on different sides of a marginal stability wall, rather it supports even equivariant loci flowing through these points.
Moreover we observe that the duality may be raised indeed to the level of the incidence loci as it was proposed in \cite{Galakhov:2021xum}.
However the variation of the Cartan elements from a phase to a phase, rather than $c_{\circ}=c_{\bullet}=1$ for all simple glasses, might seem surprising from this point of view.

To explain this phenomenon let us remind first the structure of vacuum charge functions representing the Cartan eigenvalues on the top module vector $|\varnothing\rangle$.
For a chosen quiver gauge node $a\in \fQ_0$ we have \cite{Galakhov:2021xum}:
\begin{equation}\label{zero_charge}
	\psi_{\varnothing}^{(a)}(z)=\prod\lm_{\gamma\in\{a\to a\}}\left(-\frac{1}{w_{\gamma}}\right)\times\frac{\prod\lm_{\beta\in\{ a\to\ff\}}\left(z+w_{\beta}\right)}{\prod\lm_{\alpha\in\{\ff\to a\}}\left(z-w_{\alpha}\right)}\,,
\end{equation}
where $\ff$ is the framing node, and $w_{\alpha,\beta,\gamma}$ are equivariant weights of the corresponding arrows.

Then we note that in the glass-1 phase the numerical pre-factor in the Cartan eigenvalue coincides exactly with a product over looped arrow weights appeared after the mutation transform in \eqref{mut_1}:
\begin{equation}
	\upsilon_1\sim\frac{1}{\left(\epsilon_1^2-\epsilon_2^2\right)^2}=\frac{1}{\prod\lm_{i,j=1}^2w\left[H_{i,j}\right]}\,.
\end{equation}
Therefore despite the F-term constraint allows us to integrate over these fields (imposes an explicit constraint \eqref{du_morph_2}) these fields contribute on the quantum level.

Unfortunately, at the moment we are unable to explain higher multipliers $\upsilon_{n\geq 2}$ for the following reason.
Clearly, they appear as a result of dualizing operators presented by mixtures of dynamical and non-dynamical fields, like $\check{A}_k H_{i,j} \check{B}_l $.
However since $H_{i,j}$ is charged adjointly this duality is a version of more complicated Kutasov duality \cite{Kutasov:1995ss} we are not considering in this paper.

\begin{comment}

\bigskip

{\footnotesize
As an example of further development of the subject we present a version of
the structure functions that describe hysteresis rule according to \eqref{hysteresis} depend on the phase:
\begingroup
\renewcommand{\arraystretch}{2}
\begin{equation}\label{mut_alg}
	\begin{array}{c|c|c|c}
		\mbox{Phase} & \varphi_{\circ, \circ}(z) & \varphi_{\bullet, \bullet}(z) & \varphi_{\bullet, \circ}(z)  \\
		\hline
		\mbox{Crystal} &-1 & -1 & \dfrac{(z - \epsilon_2)(z + \epsilon_2)}{(z - \epsilon_1)(z + \epsilon_1)} \\
		\hline
		\mbox{Glass-1} &-1 & \dfrac{z - 2 \epsilon_1}{z + 2 \epsilon_1} &   \dfrac{(z - \epsilon_2)(z + \epsilon_2)}{(z - \epsilon_1) (z + \epsilon_1)} \cdot \dfrac{(-1)}{(z - \epsilon_1)^2 (z - 3 \epsilon_1)} \\
	\end{array}
\end{equation}
\endgroup
\noindent
and the other function is restored via the rule
\begin{equation}
	 \varphi_{\circ, \bullet}(z) = \dfrac{1}{\varphi_{\bullet, \circ}(-z)}
\end{equation}
This is, however, only an attempt, perhaps, not fully correct -- thus we present it in the footnotesize.
In its present form it implies the breakdown of the $\mathsf{Y}(\widehat{\fg\fl}_{1|1})$ algebra, what is hardly true.

\bigskip
}
\end{comment}

\subsection{Crooked glass branch}

As we have discussed in sec.\ref{sec:crooked} a structure of the BPS spectrum becomes rather involved.
Nevertheless we managed to separate a branch of $R$-dominated solutions in the next to the boundary region \eqref{reg_1} and constructed the corresponding ordering function $\fo_{\vec\zeta}$ associating the fixed points of this branch with skew super-partitions.

We apply to this branch of $\fo_{\vec\zeta}$  our construction of the algebra representation \eqref{rep_phase}.
As the result of this construction we find that the representation matrices $\bf E$ and $\bf F$ satisfy hysteresis relations \eqref{hysteresis} if the vacuum charge functions defining Cartan eigenvalues on the module top vector $|\varnothing\rangle$ read:
\begin{equation}\label{crooked_psi}
	\psi_\varnothing^{\circ}(z)=-\frac{1}{z},\quad \psi_\varnothing^{\bullet}(z)=\frac{1}{(\epsilon_1^2-\epsilon_2^2)^2}\frac{(z-\epsilon_2)(z+\epsilon_2)}{z-\epsilon_1}\,.
\end{equation}
The fact that generator $e^{\bullet}$ adds the lower left triangle tile $\nhhtile$ that corresponds in the dual phase to a white atom the eigenvalue $\psi_\varnothing^{\bullet}(z)$ is given by  expression \eqref{zero_charge} for dual quiver \eqref{dual_quiver_2} we predicted.
Even overall normalization $\upsilon_1$ of this factor appears due to the same effect of integration over non-dynamical fields as in sec.\ref{sec:res_sg}.

However as we see $\psi_\varnothing^{\circ}(z)$ also has a pole.
That brings us to a necessity to modify the predicted dual quiver \eqref{dual_quiver_2} by adding another arrow $P$ responsible for the pole in $\psi_\varnothing^{\circ}(z)$ that connects the black node with the framing node (here we emphasize new elements in comparison to \eqref{dual_quiver_2}):
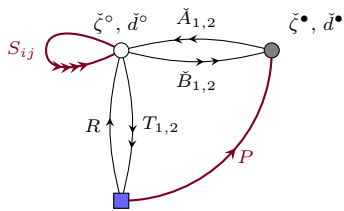
\begin{equation}
\begin{array}{c}
	\begin{tikzpicture}
		\begin{scope}[rotate=180]
			\draw[thick, burgundy, postaction={decorate},
			decoration={markings, mark= at position 0.65 with {\arrow{stealth}}, mark= at position 0.7 with {\arrow{stealth}},mark= at position 0.75 with {\arrow{stealth}}, mark= at position 0.8 with {\arrow{stealth}}}] (0,0) to[out=330,in=270] (1,0) to[out=90,in=30] (0,0);
		\end{scope}
		\node[left,burgundy] at (-1,0) {$\scriptstyle S_{ij}$};
		\draw[postaction={decorate},
		decoration={markings, mark= at position 0.55 with {\arrow{stealth}}, mark= at position 0.65 with {\arrow{stealth}}}] (2,0) to[out=165, in=15] node[pos=0.5, above] {$\scriptstyle {\check A}_{1,2}$} (0,0);
		\draw[postaction={decorate},
		decoration={markings, mark= at position 0.55 with {\arrow{stealth}}, mark= at position 0.65 with {\arrow{stealth}}}] (0,0) to[out=345,in=195] node[pos=0.5, below] {$\scriptstyle  {\check B}_{1,2}$} (2,0);
		\draw[postaction={decorate},
		decoration={markings, mark= at position 0.55 with {\arrow{stealth}}}] (0,-2) to[out=105,in=255] node[pos=0.5, left] {$\scriptstyle R$} (0,0);
		\draw[postaction={decorate},
		decoration={markings, mark= at position 0.55 with {\arrow{stealth}}, mark= at position 0.65 with {\arrow{stealth}}}] (0,0) to[out=285,in=75] node[pos=0.5, right] {$\scriptstyle T_{1,2}$} (0,-2);
		\draw[thick, burgundy, postaction={decorate},
		decoration={markings, mark= at position 0.55 with {\arrow{stealth}}}] (0,-2) to[out=0,in=270] node[pos=0.5, right] {$\scriptstyle P$} (2,0);
		\draw[fill=white] (0,0) circle (0.1);
		\draw[fill=gray] (2,0) circle (0.1);
		\begin{scope}[shift={(0,-2)}]
			\draw[fill=\myblue] (-0.1,-0.1) -- (-0.1,0.1) -- (0.1,0.1) -- (0.1,-0.1) -- cycle;
		\end{scope}
		\node[above] at (0,0.1) {$\scriptstyle {\check\zeta}^{\circ},\;{\check d}^{\circ}$};
		\node[above right] at (2.1,0.1) {$\scriptstyle {\check\zeta}^{\bullet},\;{\check d}^{\bullet}$};
	\end{tikzpicture}
\end{array}\; \begin{array}{c}
	{\check W}=\Tr\big({\check B}_1{\check A}_2{\check B}_2{\check A}_1-{\check B}_1{\check A}_1{\check B}_2{\check A}_2+\\
	+T_1\check{A}_1\check{B}_2R+T_2\check{A}_1\check{B}_1R+S_{11}S_{22}+S_{12}S_{21}\big)\,,\\
	\\
	\scalebox{0.9}{$
	\begin{array}{c|c|c|c|c|c|c|c|c|c}
		\mbox{Fields}&{\check A}_1 & {\check A}_2 & {\check B}_1 & {\check B}_2 & R & T_1 & T_2 & S_{ij}& P\\
		\hline
		\mbox{Weights}&-\epsilon_1 & \epsilon_1 & -\epsilon_2 & \epsilon_2 & 0 & -\epsilon_2+\epsilon_1 & \epsilon_2+\epsilon_1& \pm\epsilon_1\pm\epsilon_2 &-\epsilon_1
	\end{array}$}
\end{array}
\end{equation}
However after modification the dual quiver has now \emph{two} growth points for crystals in the cyclic chamber.
According to a discussion in \cite{Galakhov:2022uyu} one may treat this situation as a simple model for tensor products on quiver Yangian modules.
So that it is natural to treat vectors in a module with \eqref{crooked_psi} as elements of a tensor product $|\lambda\rangle\otimes|\tilde \lambda\rangle$, where $\lambda$, $\tilde\lambda$ are, in general, partitions.
And so far as skew super partitions we have enumerated only vectors of form $\lambda\otimes\varnothing$.
It might be natural to try to identify \emph{additional} vectors in the tensor product module with other solution branches we noticed in sec.\ref{sec:crooked}.
However, even if this is the case, we are unaware of how to describe the action of duality explicitly in this setting.
We will return to this problem elsewhere.

We should remind here that the numbers of poles and zeroes in the charge functions \eqref{crooked_psi} define a net characteristic of the affine Yangian algebras called \emph{shifts} \cite{Galakhov:2021xum}.
So as the simple glasses \eqref{mut_alg} all have the same shifts coinciding with those of the crystal (cyclic) phase, crooked glasses \eqref{crooked_psi} account for different shifts.

%%%%%%%%%%%%%%%%%%%%%%%%%%%%%%%%%%%%%%%%%%%%%%%%%%%%%%%%%%%%%%%%%%%
%%%%%%%%%%%%%%%%%%%%%%%%%%%%%%%%%%%%%%%%%%%%%%%%%%%%%%%%%%%%%%%%%%%
%%%%%%%%%%%%%%%%%%%%%%%%%%%%%%%%%%%%%%%%%%%%%%%%%%%%%%%%%%%%%%%%%%%
%%%%%%%%%%%%%%%%%%%%%%%%%%%%%%%%%%%%%%%%%%%%%%%%%%%%%%%%%%%%%%%%%%%

\section{Open problems}
\begin{comment}
The main goal of this paper was to expose the variety of glass phases at generic values of the parameters $\zeta$.
BPS states in this case are described by {\it atomic structures}, which do not reduce to any kind of
(super)Young diagrams.
Still they are often??? related to them by mutation transformations of the underlying quivers.
\end{comment}
The main goal of this paper was to expose the effects of moving between phases in the phase portrait on the BPS algebraic structure defined canonically solely in the cyclic chamber.
BPS states in this case are described by {\it atomic structures}, which differ intrinsically from the molten crystals, and therefore do not reduce to any kind of
(super)Young diagrams.
Still we are able to relate those concepts in some cases by mutation (Seiberg duality) transformations of the underlying quivers.
In those cases we were able to present a related algebraic structure and to show that BPS states in those new phases also form representations of the same affine super-Yangian $\mathsf{Y}(\widehat{\fg\fl}_{1|1})$.

Description in terms of Young diagrams implies that Yangian generators act by gluing or subtracting boxes,
and no boxes can appear in any other place but at the boundary of the crystal.
This is what is expressed by the image of the {\it molten crystal}.
Also in this case each box (atom) is connected by a sequence of arrows to the origin,
and this is designated by the name {\it cyclic}.
In glass phases both points are violated: there are atoms which can not be reached along the arrows
from the origin -- some arrows point in the opposite direction.

As concluding remarks for this paper we propose a list of open problems that might be interesting to investigate further:
\begin{itemize}

\item{} One might try to calculate explicitly the quantities ${\bf E}$ and ${\bf F}$ in eq.(\ref{rep}) in all glass phases,
and an analogue of hook formulas might be found in terms of atomic structures, not just (super)Young diagrams.
In principle, an obstacle for this formulation is that the quiver varieties we arrive at turn out to be singular.
However on the other hand a setting of 2d atomic structures seems simpler than 3d partitions of Macmahon modules \cite{Galakhov:2023mak, Morozov:2023vra, Morozov:2022ndt}.

\item{} The variety of glass phases requires a more thorough and illuminating classification,
perhaps, as a kind of a bundle over that of the (super)Young diagrams.
However, the exact role of Young diagrams in the description of {\it glass} phases remains obscure
and even questionable.

\item{} Commutation relations are the same in all phases, what makes them (the algebraic structure) a kind of invariant
of BPS algebras, preserved by the wall crossing.
%For ordinary crystal representations they are also the same, and
For ordinary Lie algebras this is formulated in terms of {\it co-multiplication},
and probably this is also straightforward in crystal representations.
In glass phases co-multiplication seems to become more sophisticated.
Wall crossing is a kind of Weyl reflection in the root system --
but far more complicated,
%should be considered as a more sophisticated kind of comultiplication,
which remains to be carefully defined and investigated.

%[{\bf\color{burgundy}Doubtful:} I wouldn't say this it is a good analogy.
%Wall-crossing looks more like automorphisms, say, Weyl reflections of the root systems.
%Probably for $\mathsf{Y}(\fs\fl_n)$ they will be literally Weyl reflections.]

\item{} Our examples in this paper were restricted to the first seemingly simple yet interesting case of $\mathsf{Y}(\widehat{\fg\fl}_{1|1})$,
where the pattern of glass phases is already quite rich.
Some of the relevant mutations for more general  $\mathsf{Y}(\widehat{\fg\fl}_{m|n})$ are described in Appendix \ref{sec:Y(gl(m|n))}.
Those mutations represent known morphisms of the underlying $\widehat{\fg\fl}_{m|n}$ Dynkin diagram modifying the corresponding quiver as well.
Yet it is expected \cite{BM} those morphisms are promoted to automorphisms of the very  $\mathsf{Y}(\widehat{\fg\fl}_{m|n})$.
Still a detailed analysis of the phases and effects of the wall-crossing on the BPS algebra remain to be done.

\end{itemize}

%%%%%%%%%%%%%%%%%%%%%%%%%%%%%%%%%%%%%%%%%%%%%%%%%%%%%%%%%%%%%%%%%%%
%%%%%%%%%%%%%%%%%%%%%%%%%%%%%%%%%%%%%%%%%%%%%%%%%%%%%%%%%%%%%%%%%%%
%%%%%%%%%%%%%%%%%%%%%%%%%%%%%%%%%%%%%%%%%%%%%%%%%%%%%%%%%%%%%%%%%%%
%%%%%%%%%%%%%%%%%%%%%%%%%%%%%%%%%%%%%%%%%%%%%%%%%%%%%%%%%%%%%%%%%%%

\section*{Acknowledgments}
We would like to thank Alexey Litvinov for useful comments on the draft.
Our work is partly supported by grants RFBR 21-51-46010 ST\_a (D.G., A.M., N.T.), by the grants of the Foundation for the Advancement of Theoretical Physics and Mathematics “BASIS” (A.M., N.T.). This research was also partly supported by the Ministry of Science and Higher Education of the Russian Federation, agreement 075-15-2022-289 date 06/04/2022 (D.G., N.T.).

%%%%%%%%%%%%%%%%%%%%%%%%%%%%%%%%%%%%%%%%%%%%%%%%%%%%%%%%%%%%%%%%%%%
%%%%%%%%%%%%%%%%%%%%%%%%%%%%%%%%%%%%%%%%%%%%%%%%%%%%%%%%%%%%%%%%%%%
%%%%%%%%%%%%%%%%%%%%%%%%%%%%%%%%%%%%%%%%%%%%%%%%%%%%%%%%%%%%%%%%%%%
%%%%%%%%%%%%%%%%%%%%%%%%%%%%%%%%%%%%%%%%%%%%%%%%%%%%%%%%%%%%%%%%%%%

\appendix

\section{Braid mutation action in $\mathsf{Y}(\widehat{\fg\fl}_{m|n})$}\label{sec:Y(gl(m|n))}

For affine algebra $\widehat{\fg\fl}_{m|n}$ there are different choices of Dynkin diagrams.
It is specified by a signature (cf.\ \cite{Nagao:2009rq,BM}):
\begin{equation}
\Sigma_{m,n}:\quad \{1,2,\ldots,m+n \}\longrightarrow \{+1,-1\},\;\mbox{so that}\;\#(+1)=m,\;\#(-1)=n \;,
\end{equation}
where in the following we consider indices modulo $m+n$.

Different choices of signatures $\Sigma_{m,n}$ and $\Sigma_{m,n}'$ for the same pair $(m|n)$ may seem to produce different algebras.
However those signatures are related by a permutation that can be lifted to a braid group action of isomorphisms of  quantum toroidal algebras \cite{BM}.
In this section we would like to show that an elementary permutation of two neighboring opposite spins is induced by a mutation.

First let us remind the identification pattern between signatures and Dynkin/quiver diagrams in the following table:
\begin{equation}\renewcommand{\arraystretch}{1.3}
\begin{array}{c|c|c}
	\mbox{spin arrangement} & \sigma_i\sigma_{i+1}=1& \sigma_i\sigma_{i+1}=-1\\
	\hline
	\IZ_2\mbox{-parity} &\mbox{Even} & \mbox{Odd} \\
	\hline
	\mbox{Dynkin node} & \begin{array}{c}
		\begin{tikzpicture}
			\draw[ultra thick] (-0.5,0) -- (0.5,0);
			\draw[fill=white] (0,0) circle (0.15);
		\end{tikzpicture}
	\end{array}& \begin{array}{c}
		\begin{tikzpicture}
			\draw[ultra thick] (-0.5,0) -- (0.5,0);
			\draw[fill=white] (0,0) circle (0.15);
			\draw (-0.2,-0.2) -- (0.2,0.2) (0.2,-0.2) -- (-0.2,0.2);
		\end{tikzpicture}
	\end{array}\\
	\hline
	\mbox{quiver node} & \begin{array}{c}
		\begin{tikzpicture}
			\draw (0,0) circle (0.15);
			\draw[-stealth] (-1,0.05) -- (-0.141421,0.05) node[pos=0.2,above] {$A_i$};
			\draw[stealth-] (-1,-0.05) -- (-0.141421,-0.05) node[pos=0.2,below] {$B_i$};
			\draw[stealth-] (1,0.05) -- (0.141421,0.05) node[pos=0.2,above] {$A_{i+1}$};
			\draw[-stealth] (1,-0.05) -- (0.141421,-0.05) node[pos=0.2,below] {$B_{i+1}$};
			\draw[postaction={decorate},decoration={markings,
				mark= at position 0.85 with {\arrow{stealth}}}] (-0.05,0.141421) to[out=120,in=180] node[pos=0.7,above left] {$C_i$} (0,0.7) to[out=0,in=60] (0.05,0.141421);
			\draw[white] (-0.1,1.1) -- (0.1,1.1);
		\end{tikzpicture}
	\end{array}& \begin{array}{c}
		\begin{tikzpicture}
			\draw (0,0) circle (0.15);
			\draw[-stealth] (-1,0.05) -- (-0.141421,0.05) node[pos=0.2,above] {$A_i$};
			\draw[stealth-] (-1,-0.05) -- (-0.141421,-0.05) node[pos=0.2,below] {$B_i$};
			\draw[stealth-] (1,0.05) -- (0.141421,0.05) node[pos=0.2,above] {$A_{i+1}$};
			\draw[-stealth] (1,-0.05) -- (0.141421,-0.05) node[pos=0.2,below] {$B_{i+1}$};
			\draw[white] (-0.1,1.1) -- (0.1,1.1);
		\end{tikzpicture}
	\end{array}\\
	\hline
	\mbox{superpotential }\delta W_i & \sigma_i\,\Tr\, C_i\left(B_{i+1}A_{i+1}-A_{i}B_{i}\right)& -\sigma_i\,\Tr\,B_{i+1}A_{i+1}A_iB_i\\
\end{array}
\end{equation}

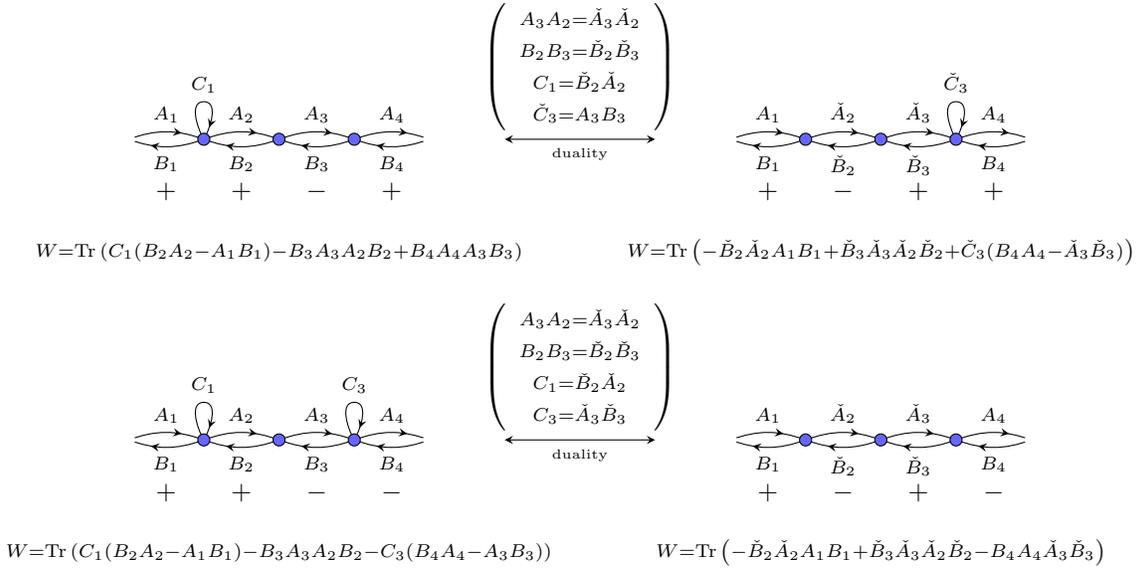
\begin{figure}[ht!]
	\begin{center}
		\begin{tikzpicture}
			\begin{scope}
				\begin{scope}
					\draw[postaction={decorate},
					decoration={markings, mark= at position 0.7 with {\arrow{stealth}}}] (-2,0) to[out=20,in=160] node[pos=0.5,above] {$\scriptstyle A_1$} (-1,0);
					\draw[postaction={decorate},
					decoration={markings, mark= at position 0.7 with {\arrow{stealth}}}] (-1,0) to[out=200,in=340] node[pos=0.5,below] {$\scriptstyle B_1$} (-2,0);
				\end{scope}
				%%%%%%
				\begin{scope}[shift={(1,0)}]
					\draw[postaction={decorate},
					decoration={markings, mark= at position 0.7 with {\arrow{stealth}}}] (-2,0) to[out=20,in=160] node[pos=0.5,above] {$\scriptstyle A_2$} (-1,0);
					\draw[postaction={decorate},
					decoration={markings, mark= at position 0.7 with {\arrow{stealth}}}] (-1,0) to[out=200,in=340] node[pos=0.5,below] {$\scriptstyle B_2$} (-2,0);
				\end{scope}
				%%%%%%
				\begin{scope}[shift={(2,0)}]
					\draw[postaction={decorate},
					decoration={markings, mark= at position 0.7 with {\arrow{stealth}}}] (-2,0) to[out=20,in=160] node[pos=0.5,above] {$\scriptstyle A_3$} (-1,0);
					\draw[postaction={decorate},
					decoration={markings, mark= at position 0.7 with {\arrow{stealth}}}] (-1,0) to[out=200,in=340] node[pos=0.5,below] {$\scriptstyle B_3$} (-2,0);
				\end{scope}
				%%%%%%
				\begin{scope}[shift={(3,0)}]
					\draw[postaction={decorate},
					decoration={markings, mark= at position 0.7 with {\arrow{stealth}}}] (-2,0) to[out=20,in=160] node[pos=0.5,above] {$\scriptstyle A_4$} (-1,0);
					\draw[postaction={decorate},
					decoration={markings, mark= at position 0.7 with {\arrow{stealth}}}] (-1,0) to[out=200,in=340] node[pos=0.5,below] {$\scriptstyle B_4$} (-2,0);
				\end{scope}
				%%%%%%
				\begin{scope}[shift={(-1,0)}]
					\draw[postaction={decorate},
					decoration={markings, mark= at position 0.85 with {\arrow{stealth}}}] (0,0) to[out=120,in=180] (0,0.5) to[out=0,in=60] (0,0);
					\node[above] at (0,0.5) {$\scriptstyle C_1$};
				\end{scope}
				\draw[white, fill=white] (-2,0) circle (0.08) (2,0) circle (0.08);
				\draw[fill=\myblue] (-1,0) circle (0.08) (0,0) circle (0.08) (1,0) circle (0.08);
				\node at (-1.5,-0.7) {$+$};
				\node at (-0.5,-0.7) {$+$};
				\node at (0.5,-0.7) {$-$};
				\node at (1.5,-0.7) {$+$};
				\node at (0,-1.5) {$\scriptstyle W=\Tr\left(C_1(B_2A_2-A_1B_1)-B_3A_3A_2B_2+B_4A_4A_3B_3\right)$};
			\end{scope}
			%%%%%%%%%%%%%%%%%%%%%%%%%%%%%%%%%
			\begin{scope}[shift={(8,0)}]
				\begin{scope}
					\draw[postaction={decorate},
					decoration={markings, mark= at position 0.7 with {\arrow{stealth}}}] (-2,0) to[out=20,in=160] node[pos=0.5,above] {$\scriptstyle A_1$} (-1,0);
					\draw[postaction={decorate},
					decoration={markings, mark= at position 0.7 with {\arrow{stealth}}}] (-1,0) to[out=200,in=340] node[pos=0.5,below] {$\scriptstyle B_1$} (-2,0);
				\end{scope}
				%%%%%%
				\begin{scope}[shift={(1,0)}]
					\draw[postaction={decorate},
					decoration={markings, mark= at position 0.7 with {\arrow{stealth}}}] (-2,0) to[out=20,in=160] node[pos=0.5,above] {$\scriptstyle \check A_2$} (-1,0);
					\draw[postaction={decorate},
					decoration={markings, mark= at position 0.7 with {\arrow{stealth}}}] (-1,0) to[out=200,in=340] node[pos=0.5,below] {$\scriptstyle \check B_2$} (-2,0);
				\end{scope}
				%%%%%%
				\begin{scope}[shift={(2,0)}]
					\draw[postaction={decorate},
					decoration={markings, mark= at position 0.7 with {\arrow{stealth}}}] (-2,0) to[out=20,in=160] node[pos=0.5,above] {$\scriptstyle \check A_3$} (-1,0);
					\draw[postaction={decorate},
					decoration={markings, mark= at position 0.7 with {\arrow{stealth}}}] (-1,0) to[out=200,in=340] node[pos=0.5,below] {$\scriptstyle \check B_3$} (-2,0);
				\end{scope}
				%%%%%%
				\begin{scope}[shift={(3,0)}]
					\draw[postaction={decorate},
					decoration={markings, mark= at position 0.7 with {\arrow{stealth}}}] (-2,0) to[out=20,in=160] node[pos=0.5,above] {$\scriptstyle A_4$} (-1,0);
					\draw[postaction={decorate},
					decoration={markings, mark= at position 0.7 with {\arrow{stealth}}}] (-1,0) to[out=200,in=340] node[pos=0.5,below] {$\scriptstyle B_4$} (-2,0);
				\end{scope}
				%%%%%%
				\begin{scope}[shift={(1,0)}]
					\draw[postaction={decorate},
					decoration={markings, mark= at position 0.85 with {\arrow{stealth}}}] (0,0) to[out=120,in=180] (0,0.5) to[out=0,in=60] (0,0);
					\node[above] at (0,0.5) {$\scriptstyle \check C_3$};
				\end{scope}
				\draw[white, fill=white] (-2,0) circle (0.08) (2,0) circle (0.08);
				\draw[fill=\myblue] (-1,0) circle (0.08) (0,0) circle (0.08) (1,0) circle (0.08);
				\node at (-1.5,-0.7) {$+$};
				\node at (-0.5,-0.7) {$-$};
				\node at (0.5,-0.7) {$+$};
				\node at (1.5,-0.7) {$+$};
				\node at (0,-1.5) {$\scriptstyle W=\Tr\left(-\check B_2\check A_2 A_1B_1+\check B_3\check A_3 \check A_2\check B_2+\check C_3(B_4A_4-\check A_3\check B_3)\right)$};
			\end{scope}
			\draw[stealth-stealth] (3,0) -- (5,0) node[pos=0.5,below] {\tiny duality} node[pos=0.5,above] {$\left(\begin{array}{c}
					\scriptstyle A_3A_2=\check A_3\check A_2\\
					\scriptstyle B_2B_3=\check B_2\check B_3\\
					\scriptstyle C_1=\check B_2\check A_2\\
					\scriptstyle \check C_3=A_3B_3
				\end{array}\right)$};
			%%%%%%%%%%%%%%%%%%%%%%%%%%%%%%%%%%%%%%%
			%%%%%%%%%%%%%%%%%%%%%%%%%%%%%%%%%%%%%%%
			%%%%%%%%%%%%%%%%%%%%%%%%%%%%%%%%%%%%%%%
			\begin{scope}[shift={(0,-4)}]
			\begin{scope}
				\begin{scope}
					\draw[postaction={decorate},
					decoration={markings, mark= at position 0.7 with {\arrow{stealth}}}] (-2,0) to[out=20,in=160] node[pos=0.5,above] {$\scriptstyle A_1$} (-1,0);
					\draw[postaction={decorate},
					decoration={markings, mark= at position 0.7 with {\arrow{stealth}}}] (-1,0) to[out=200,in=340] node[pos=0.5,below] {$\scriptstyle B_1$} (-2,0);
				\end{scope}
				%%%%%%
				\begin{scope}[shift={(1,0)}]
					\draw[postaction={decorate},
					decoration={markings, mark= at position 0.7 with {\arrow{stealth}}}] (-2,0) to[out=20,in=160] node[pos=0.5,above] {$\scriptstyle A_2$} (-1,0);
					\draw[postaction={decorate},
					decoration={markings, mark= at position 0.7 with {\arrow{stealth}}}] (-1,0) to[out=200,in=340] node[pos=0.5,below] {$\scriptstyle B_2$} (-2,0);
				\end{scope}
				%%%%%%
				\begin{scope}[shift={(2,0)}]
					\draw[postaction={decorate},
					decoration={markings, mark= at position 0.7 with {\arrow{stealth}}}] (-2,0) to[out=20,in=160] node[pos=0.5,above] {$\scriptstyle A_3$} (-1,0);
					\draw[postaction={decorate},
					decoration={markings, mark= at position 0.7 with {\arrow{stealth}}}] (-1,0) to[out=200,in=340] node[pos=0.5,below] {$\scriptstyle B_3$} (-2,0);
				\end{scope}
				%%%%%%
				\begin{scope}[shift={(3,0)}]
					\draw[postaction={decorate},
					decoration={markings, mark= at position 0.7 with {\arrow{stealth}}}] (-2,0) to[out=20,in=160] node[pos=0.5,above] {$\scriptstyle A_4$} (-1,0);
					\draw[postaction={decorate},
					decoration={markings, mark= at position 0.7 with {\arrow{stealth}}}] (-1,0) to[out=200,in=340] node[pos=0.5,below] {$\scriptstyle B_4$} (-2,0);
				\end{scope}
				%%%%%%
				\begin{scope}[shift={(-1,0)}]
					\draw[postaction={decorate},
					decoration={markings, mark= at position 0.85 with {\arrow{stealth}}}] (0,0) to[out=120,in=180] (0,0.5) to[out=0,in=60] (0,0);
					\node[above] at (0,0.5) {$\scriptstyle C_1$};
				\end{scope}
				\begin{scope}[shift={(1,0)}]
					\draw[postaction={decorate},
					decoration={markings, mark= at position 0.85 with {\arrow{stealth}}}] (0,0) to[out=120,in=180] (0,0.5) to[out=0,in=60] (0,0);
					\node[above] at (0,0.5) {$\scriptstyle C_3$};
				\end{scope}
				\draw[white, fill=white] (-2,0) circle (0.08) (2,0) circle (0.08);
				\draw[fill=\myblue] (-1,0) circle (0.08) (0,0) circle (0.08) (1,0) circle (0.08);
				\node at (-1.5,-0.7) {$+$};
				\node at (-0.5,-0.7) {$+$};
				\node at (0.5,-0.7) {$-$};
				\node at (1.5,-0.7) {$-$};
				\node at (0,-1.5) {$\scriptstyle W=\Tr\left(C_1(B_2A_2-A_1B_1)-B_3A_3A_2B_2-C_3(B_4A_4-A_3B_3)\right)$};
			\end{scope}
			%%%%%%%%%%%%%%%%%%%%%%%%%%%%%%%%%
			\begin{scope}[shift={(8,0)}]
				\begin{scope}
					\draw[postaction={decorate},
					decoration={markings, mark= at position 0.7 with {\arrow{stealth}}}] (-2,0) to[out=20,in=160] node[pos=0.5,above] {$\scriptstyle A_1$} (-1,0);
					\draw[postaction={decorate},
					decoration={markings, mark= at position 0.7 with {\arrow{stealth}}}] (-1,0) to[out=200,in=340] node[pos=0.5,below] {$\scriptstyle B_1$} (-2,0);
				\end{scope}
				%%%%%%
				\begin{scope}[shift={(1,0)}]
					\draw[postaction={decorate},
					decoration={markings, mark= at position 0.7 with {\arrow{stealth}}}] (-2,0) to[out=20,in=160] node[pos=0.5,above] {$\scriptstyle \check A_2$} (-1,0);
					\draw[postaction={decorate},
					decoration={markings, mark= at position 0.7 with {\arrow{stealth}}}] (-1,0) to[out=200,in=340] node[pos=0.5,below] {$\scriptstyle \check B_2$} (-2,0);
				\end{scope}
				%%%%%%
				\begin{scope}[shift={(2,0)}]
					\draw[postaction={decorate},
					decoration={markings, mark= at position 0.7 with {\arrow{stealth}}}] (-2,0) to[out=20,in=160] node[pos=0.5,above] {$\scriptstyle \check A_3$} (-1,0);
					\draw[postaction={decorate},
					decoration={markings, mark= at position 0.7 with {\arrow{stealth}}}] (-1,0) to[out=200,in=340] node[pos=0.5,below] {$\scriptstyle \check B_3$} (-2,0);
				\end{scope}
				%%%%%%
				\begin{scope}[shift={(3,0)}]
					\draw[postaction={decorate},
					decoration={markings, mark= at position 0.7 with {\arrow{stealth}}}] (-2,0) to[out=20,in=160] node[pos=0.5,above] {$\scriptstyle A_4$} (-1,0);
					\draw[postaction={decorate},
					decoration={markings, mark= at position 0.7 with {\arrow{stealth}}}] (-1,0) to[out=200,in=340] node[pos=0.5,below] {$\scriptstyle B_4$} (-2,0);
				\end{scope}
				%%%%%%
				\draw[white, fill=white] (-2,0) circle (0.08) (2,0) circle (0.08);
				\draw[fill=\myblue] (-1,0) circle (0.08) (0,0) circle (0.08) (1,0) circle (0.08);
				\node at (-1.5,-0.7) {$+$};
				\node at (-0.5,-0.7) {$-$};
				\node at (0.5,-0.7) {$+$};
				\node at (1.5,-0.7) {$-$};
				\node at (0,-1.5) {$\scriptstyle W=\Tr\left(-\check B_2\check A_2 A_1B_1+\check B_3\check A_3 \check A_2\check B_2-B_4A_4\check A_3\check B_3\right)$};
			\end{scope}
			\draw[stealth-stealth] (3,0) -- (5,0) node[pos=0.5,below] {\tiny duality} node[pos=0.5,above] {$\left(\begin{array}{c}
					\scriptstyle A_3A_2=\check A_3\check A_2\\
					\scriptstyle B_2B_3=\check B_2\check B_3\\
					\scriptstyle C_1=\check B_2\check A_2\\
					\scriptstyle C_3=\check A_3\check B_3
				\end{array}\right)$};
			\end{scope}
		\end{tikzpicture}
		\caption{Elementary dualities on $\widehat{\fg\fl}_{m|n}$ diagrams.} \label{fig:dualities}
	\end{center}
\end{figure}

In this setting we observe two distinct mutation actions, see fig.\ref{fig:dualities}, all the remaining cases may be derived from these two by symmetries of the problem.

%%%%%%%%%%%%%%%%%%%%%%%%%%%%%%%%%%%%%%%%%%%%%%%%%%%%%%%%%%%%%%%%%%%
%%%%%%%%%%%%%%%%%%%%%%%%%%%%%%%%%%%%%%%%%%%%%%%%%%%%%%%%%%%%%%%%%%%
%%%%%%%%%%%%%%%%%%%%%%%%%%%%%%%%%%%%%%%%%%%%%%%%%%%%%%%%%%%%%%%%%%%
%%%%%%%%%%%%%%%%%%%%%%%%%%%%%%%%%%%%%%%%%%%%%%%%%%%%%%%%%%%%%%%%%%%

\bibliographystyle{utphys}
\bibliography{biblio}

\end{document}